\documentclass[twocolumn,aps,prb,longbibliography,superscriptaddress]{revtex4-2}
\usepackage[pdftex]{graphics}
\usepackage{graphicx,bm}
\graphicspath{{figures/}}
\usepackage{xcolor}
\usepackage{amsmath, amsfonts,amssymb} 
\usepackage{physics}
\usepackage{mathtools,dsfont}
\usepackage{ulem}

\begin{document}
\title{Three-dimensional spin-orbital liquids}
\author{Anna Sandberg}
\affiliation{Department of Physics, Stockholm University, AlbaNova University Center, SE-106 91 Stockholm, Sweden}
\author{Lukas R\o dland}
\affiliation{Max Planck Institute for the Science of Light, Staudtstrasse 2, 91058 Erlangen, Germany}
\author{Maria Hermanns}
\affiliation{Department of Physics, Stockholm University, AlbaNova University Center, SE-106 91 Stockholm, Sweden}
\date{\today}

\begin{abstract}
Spin–orbital liquids provide an exactly solvable route to three-dimensional $\mathbb Z_2$ quantum spin liquids beyond the original Kitaev setting.
Built from higher-dimensional Clifford-algebra representations, spin-orbital Hamiltonians can be realized on both three- and four-coordinated lattices, giving rise to phases with $\nu=3$ and $\nu=2$ itinerant Majorana flavors, respectively. 
We demonstrate that these models host a rich set of gapless Majorana metals, characterized, in particular, by topological Fermi surfaces, nodal lines, and Weyl semimetal phases. 
We analyze the stability of these structures under physically motivated perturbations and identify generic splitting patterns and topological transitions driven by symmetry breaking and flavor mixing. 
This yields a unified organizing framework for three-dimensional Majorana metals in fractionalized spin liquids.
\end{abstract}
\maketitle

\section{Introduction}\label{sec:Introduction}

Quantum spin liquids (QSLs) represent one of the most intriguing paradigms in modern condensed matter physics, embodying states of matter where strong quantum fluctuations prevent conventional magnetic ordering even at zero temperature. 
Initially conceived in the context of frustrated magnetism and resonating valence bond physics \cite{anderson1973resonating}, QSLs have since emerged as fertile ground for exploring fractionalization, long-range entanglement, and emergent gauge structures \cite{wen2002quantum,balents2010spin,broholm2020quantum}.
A central challenge is to understand how distinct microscopic mechanisms can conspire to stabilize QSL behavior, classify the resulting phases, and determine their properties and experimental signatures \cite{knolle2019field,zhou2017quantum,savary_quantum_2016}. 
Within this broader context, simplified toy models provide invaluable reference points \cite{kitaev2003fault, levin_string-net_2005,kitaev2006anyons,moessner2001resonating}, as they allow us to rigorously establish the existence of QSL phases and may thus serve as a cornerstone for the exploration of more realistic and complex settings.

The Kitaev honeycomb model \cite{kitaev2006anyons} is arguably one of the most important toy models for QSLs. 
It is one of the rare examples where the existence of QSL phases can be shown exactly, since the low-energy properties of the model can be determined exactly by representing the interacting spins in terms of Majorana fermions hopping in an emergent, static gauge field \cite{baskaran2007exact}. 
The emergent gauge field turns out to be gapped---a feature that holds generically even for the various generalizations discussed later---which provides the stability of the QSL phases against generic perturbation. 
The relative simplicity of the model also allows computation of several experimental probes, such as nuclear resonance, Raman scattering, or non-linear response features \cite{knolle2014dynamics,knolle2014raman,kanega2021linear}.

It was soon realized that the Kitaev model can be generalized in several distinct ways. 
This was an important insight because it allows us to form a more comprehensive picture of the behavior of spin liquids beyond the Kitaev honeycomb model. 
One fruitful generalization amounts to placing the Kitaev interaction on other three-coordinated lattices \cite{yao2007exact,yang2007mosaic}, in particular, three-dimensional (3D) ones, where exactly solvable toy models offer a rare opportunity to study QSLs in an analytically controlled setting \cite{mandal2009exactly,hermanns2014quantum}. Depending on the lattice geometry and/or symmetry, the itinerant Majorana fermions may form Fermi surfaces (FSs), nodal lines, or Weyl points (WPs), often carrying nontrivial topological charges protected by time-reversal or crystalline symmetries \cite{obrien_classification_2016}.
These three-dimensional Kitaev spin liquids (KSLs) serve as valuable platforms for exploring the interplay of emergent gauge fields, crystalline symmetry, and topological band theory in a setting that remains exactly tractable \cite{eschmann2019thermodynamics,yamada2017crystalline}.

Another fruitful route to generalizing the Kitaev model focuses instead on the algebraic structure underlying its exact solvability. 
In particular, the use of Pauli matrices realizing a Clifford algebra plays a central role in the fractionalization into Majorana fermions.
By enlarging the on-site Hilbert space and replacing Pauli matrices with higher-dimensional representations of the Clifford algebra, one obtains a broad class of exactly solvable models whose ground states realize a variety of QSL phases \cite{yao2009algebraic,ryu_three-dimensional_2009,wu2009gamma,chua2011exact, nakai_time-reversal_2012}.
Within this broader framework, a particularly interesting subclass is formed by \emph{spin–orbital liquids} (SOLs) \cite{wang2009spin-orbital,corboz2012spin-orbital,chulliparambil_microscopic_2020}.
 These models combine spin and orbital degrees of freedom in such a way that the enlarged local Hilbert space admits multiple itinerant Majorana flavors, leading to rich patterns of fractionalization and emergent gauge structure.
To date, SOLs have been studied predominantly in two spatial dimensions, most notably on honeycomb and square lattices, where they were shown to realize all topological phases of Kitaev’s 16-fold way \cite{chulliparambil_microscopic_2020,chulliparambil_flux_2021}.

Relatively few works have explored SOLs in three dimensions. 
The two-flavor diamond-lattice model introduced by Ryu \cite{ryu_three-dimensional_2009} constitutes the first example of a three-dimensional SOL, already illustrating that the interplay of higher dimensionality and multiple Majorana flavors leads to qualitatively new band structures and topological phenomena. 
Building on this insight, the present work aims to systematically combine the SOL framework with genuinely three-dimensional lattice geometries, thereby uncovering a much broader landscape of three-dimensional spin–orbital liquids and their characteristic low-energy excitations.
Whereas Refs.~\cite{chulliparambil_flux_2021, ryu_three-dimensional_2009} elucidate intriguing features of specific lattice realizations (in two and three dimensions, respectively), our work extends these ideas toward a classification of generic SOL properties realizable across a broad range of three-dimensional lattices. 
We focus our discussion  on a set of lattices---both three- and four-coordinated ones---that serve as representative examples of such generic behavior in 3D SOLs. 
For each lattice, we determine the low-energy properties and the robust nodal structures of the itinerant Majorana fermions. 
We also study the response to a variety of physically motivated perturbations.

\paragraph*{Outline of this paper}
The remainder of the manuscript is organized as follows: In Sec.~\ref{sec:SOLs}, we review SOLs and how to solve them. 
Sec.~\ref{sec:Symmetries} discusses symmetries, how they are implemented, and most importantly, how they can be broken. 
Sec.~\ref{sec:Tricoordinated} contains our results for the three-coordinated lattices, while Sec.~\ref{sec:Fourcoordinated} is concerned with the four-coordinated ones. 
We conclude in Sec.~\ref{sec:Conclusions} and show how our results can aid future research in 3D QSLs. 
To keep the main manuscript accessible, we discuss all the more technical parts in the various appendices. 

\section{Spin-orbital liquid models}\label{sec:SOLs}

The Kitaev honeycomb model demonstrates that bond-dependent Ising interactions can give rise to an exactly solvable QSL, in which spins fractionalize into itinerant Majorana fermions coupled to a static $\mathbb{Z}_2$ gauge field. 
A natural question is how far this construction can be generalized while retaining exact solvability. 
One systematic route is to enlarge the local Hilbert space by replacing the Pauli matrices with higher-dimensional representations of the Clifford algebra.

The key idea behind these generalizations is simple: the structure that makes the Kitaev model solvable can be preserved if one replaces Pauli operators $\sigma^\gamma$ by $\Gamma$-matrices---higher-dimensional representations of the Clifford algebra.
The resulting models continue to map onto free Majorana fermions moving in a static $\mathbb{Z}_2$ gauge field, but exhibit multiple flavors of itinerant Majorana fermions. 
The number of flavors, $\nu$, depends on the coordination number of the underlying lattice. 
For instance, when considering the four-dimensional representation of the Clifford algebra---the focus of our work---the number of Majorana flavors is $\nu=3$ on three-coordinated lattices, while four-coordinated ones harbor only $\nu=2$ Majorana flavors.
This multiplicity enriches the possible band structures and topological features compared to the single-flavor Kitaev model.

\subsection{The model}\label{subsec:ModelSOLS}
To generalize Kitaev's construction, we replace the Pauli matrices with a set of $\Gamma$ matrices, which are higher-dimensional representations of the Clifford algebra,
\begin{align}
\{\Gamma^\alpha,\Gamma^\beta\} = 2 \delta^{\alpha\beta},
\end{align}
where $\delta$ denotes the Kronecker delta. In the $2^{q+1}$-dimensional representation, this set contains $2q+3$ distinct matrices of size $2^{q+1} \times 2^{q+1}$.
Note that $q=0$ corresponds to the original Kitaev model, with ($2\times 2$)-dimensional Pauli matrices. 
How to generalize the Hamiltonian is \emph{not} unique if the number of nearest-neighbor (NN) bonds (or coordination number) $\gamma_m$ is smaller than the number of available $\Gamma$-matrices, $2q+3$. 
In the literature, different variations have been proposed, e.g.,  \cite{yao2009algebraic,ryu_three-dimensional_2009,wu2009gamma,chua2011exact, nakai_time-reversal_2012,chulliparambil_microscopic_2020}. 

We choose to follow the approach of Refs.~\cite{chulliparambil_microscopic_2020,chulliparambil_flux_2021}, since it allows us to  treat lattices with different coordination numbers in a systematic way:  
 \begin{equation}
     \mathcal H_J^{(\nu)}=-\sum_{\langle ij\rangle_\gamma}J_\gamma \left( \Gamma_i^\gamma \Gamma_j^\gamma + \sum_{\beta=\gamma_m+1}^{2q+3}\Gamma_i^{\gamma\beta}\Gamma_j^{\gamma\beta} \right)\label{eq:SeifertHamiltonian}.
 \end{equation}
The sum here goes over all nearest neighbor bonds $\langle ij\rangle_\gamma$, where $\gamma$ denotes the bond type. That is, if we for example have a three-coordinated lattice, $\gamma=1,2,3$ denote the three different types of bonds. We also allow interactions on different bond types to have different coupling strengths, $J_\gamma$. The first term looks completely equivalent to the original Kitaev model, placing a $\Gamma^\gamma_i\Gamma^\gamma_j$ Ising interaction on the nearest neighbor bonds of type $\gamma$. 
 This leaves $2q+3-\gamma_m$ $\Gamma$ matrices that one needs to take care of. 
 For each of these matrices, we use the commutator
 \begin{align}
     \Gamma^{\alpha\beta}&=\frac{i}{2}[\Gamma^\alpha,\Gamma^\beta], 
 \end{align}
to define the remaining bond-interaction terms of \eqref{eq:SeifertHamiltonian}. 
The reasoning behind this particular construction is the following: When representing the $\Gamma$ matrices by $2q+4$ Majorana fermions ($b^1\ldots b^{2q+3},\, c)$ via 
$   \Gamma^\alpha = i b^\alpha c, $
we can combine the first $\gamma_m$ $b$-Majoranas to form (static) bond operators, $\hat{u}^\gamma_{ij} = i b_i^\gamma b_j^\gamma$, whereas the remaining $\nu=2q+4-\gamma_m$ Majoranas are itinerant and form the band structure \cite{chulliparambil_microscopic_2020,chua2011exact}.
This is completely analogous to the original Kitaev model, albeit with a larger number of itinerant Majorana fermions. 
For instance, the original Kitaev model is constructed using Pauli matrices, which are two-dimensional, and is then represented by four Majorana fermions. Out of these, $\gamma_m=3$ are assigned to the bonds, i.e., the gauge sector, leaving $\nu=1$ itinerant Majorana fermion.

Since the total number of Majoranas $2q+4$ is always even, the number of itinerant Majoranas $\nu$  is odd for lattices with odd coordination number, while it is even for those with even coordination number. 
Note that in two spatial dimensions, $\nu$ has an additional physical significance: it corresponds to the total Chern number of the gapped state obtained when breaking TR symmetry \cite{chulliparambil_flux_2021}.

\subsection{Spin-orbital liquids for $q=1$}\label{subsec:q=1SOLS}
Having introduced the general structure of spin–orbital models, we now focus on the smallest nontrivial case, $q=1$, which is also the most relevant for realistic materials. 
Higher values of $q$ require increasingly large on-site degeneracies---corresponding to eight or more local degrees of freedom---which are exceedingly difficult to realize or stabilize experimentally. 
In contrast, the $q=1$ model acts on a four-dimensional Hilbert space, naturally interpreted as a combination of spin and orbital degrees of freedom. 
Despite being relatively simple, it already showcases all the general features that occur in KSLs when enlarging the number of itinerant Majorana flavors. 

The four-dimensional local Hilbert space may, in principle, also arise from a spin-$\frac{3}{2}$ moment or from two coupled spin-$\frac{1}{2}$ degrees of freedom, rather than the coupled spin and orbital degree of freedom that we assume here. 
The latter interpretation is more flexible and phenomenologically appealing, as it permits distinct symmetry actions on the spin and orbital sectors.
In particular, if one were to adopt either of the first two viewpoints, the symmetry-allowed perturbations would take forms that differ from those employed in the rest of this work. 

For a spin-orbital interpretation, a suitable representation of the five $\Gamma$ matrices is given by
 \begin{align}
     \Gamma^i &=-\sigma^y\otimes \tau^i, & \Gamma^4&=\sigma^x \otimes \mathds{1}, & \Gamma^5&=-\sigma^z \otimes \mathds{1}, \label{eq:SpinOrbitRepresentationSeifert}
\end{align}
where we identify the indices $i=1,2,3$ with $i=x,y,z$, a labeling we use throughout this work, and $\sigma/\tau$ are the Pauli matrices acting on the spin/orbital subspace, respectively. 
It is beneficial to introduce a fourth $\tau$ matrix as $\tau^w = \mathds 1$.

Substituting \eqref{eq:SpinOrbitRepresentationSeifert} into \eqref{eq:SeifertHamiltonian} results in a Hamiltonian which depends on the coordination number, $\gamma_m$, of the lattice. 
For three-coordinated lattices, it takes the form
\begin{equation}
    \mathcal H_J^{(3)}=-\sum_{\langle ij\rangle_\gamma}J_\gamma (\bm{\sigma}_i \cdot \bm{\sigma}_j)\otimes \tau_i^\gamma \tau_j^\gamma, \label{eq:SpinOrbitHam3}
\end{equation}
where the summation runs over all nearest neighbor bonds $\langle i j\rangle_\gamma$ for all bond-types $\gamma$. 
Due to the term $(\bm{\sigma}_i \cdot \bm{\sigma}_j)$, which is invariant under arbitrary rotations in spin space, the model has an emergent $\mathrm{SO}(3)$ symmetry, which---as we will see later---manifests itself in $\nu=3$ identical, itinerant Majoranas. 
For four-coordinated lattices, on the other hand, the spin part only contains the $x$ and $y$ components, as $\Gamma^4$ is now associated with a bond:
\begin{equation}
   \mathcal H_J^{(2)}=-\sum_{\langle ij\rangle_\gamma}J_\gamma(\sigma_i^x\sigma_j^x+\sigma_i^y\sigma_j^y)\otimes (\tau_i^\gamma \tau_j^\gamma).\label{eq:SpinOrbitHam2}
\end{equation}
This reduces the emergent rotation symmetry to $\mathrm{SO}(2)$, corresponding to $\nu=2$ Majorana flavors. 
Note that we can also place the Hamiltonian on a five-coordinated lattice. 
In this case, however, the choice of $\Gamma$ matrix representation in Eq. \eqref{eq:SpinOrbitRepresentationSeifert} is not ideal, since it obscures, rather than highlights, the symmetries of the model.

 \subsection{How to solve the model}\label{subsec:SolvingSOLS}
 
 \subsubsection{Majorana representation}\label{subsubsec:MajoranaSOLS}
To solve the model, we express the four-dimensional $\Gamma$ matrices in terms of six Majorana fermions at each site:
 \begin{align}\label{eq:GammaToMajorana}
     \Gamma_j^\alpha = i b_j^\alpha c_j, \text{ for } \alpha = 1,\dots, 5.
 \end{align}
Substituting this into the Hamiltonian \eqref{eq:SeifertHamiltonian} and using the anti-commutation relations of Majorana fermions to simplify the expressions, gives the following result for the spin-orbital Hamiltonian with $\nu =6-\gamma_m$:
  \begin{align}\label{eq:MajoranaHamGeneral}
\tilde{\mathcal H}_J^{(\nu)}=\sum_{\langle ij\rangle_\gamma}J_\gamma \hat u^\gamma_{ij} \left( i c_i c_j + \sum_{\beta=\gamma_m+1}^{5}i b_i^\beta b_j^\beta \right),
 \end{align} 
 where the bond operators are defined as $\hat{u}^\gamma_{ij} = i b_i^\gamma b_j^\gamma$.

The subsequent discussion is in complete analogy to Kitaev's original discussion. 
First, we can note that the bond operators commute both with the Hamiltonian and with each other. 
Thus, they are independent conserved quantities. 
Since the Majorana anti-commutation relations ensure that the bond operators square to the identity operator, $\left({\hat{u}^\gamma}_{ij}\right)^2 = \mathds 1$, they can be interpreted as emergent $\mathbb{Z}_2$ gauge fields, living on the bonds, with eigenvalues $\pm 1$. 
We can now replace the bond operators with their respective eigenvalues $\hat u^\gamma_{ij} \rightarrow u^\gamma_{ij}= \pm 1$, which turns the Hamiltonian \eqref{eq:MajoranaHamGeneral} into one of \emph{non-interacting} Majorana fermions hopping in a \emph{static} $\mathbb{Z}_2$ gauge field configuration. 
The remaining step is to identify, which one of the extensive number of different $u$-configurations gives the lowest energy. 

\subsubsection{Physical and unphysical states}
\label{subsubsec:enlargedHilbertSpaceSOLS}
Before discussing how to determine the ground state flux configurations, we notice that when representing the $\Gamma$ matrices in terms of Majorana fermions, we have enlarged the local Hilbert space from 4 (the dimension of the $\Gamma$ matrices) to 8 (6 Majoranas correspond to three complex fermionic modes, which all can be occupied or empty). 
We also note that the right-hand-side of Eq. \eqref{eq:GammaToMajorana} does not obey the Clifford algebra for an arbitrary state in the enlarged Hilbert space---only those that obey 
\begin{align}\label{eq:physical_states}
 D_j |\text{phys}\rangle & = |\text{phys}\rangle \, \,  \forall \,  j \quad \mbox{ with }\nonumber\\
  D_j &=i b^1_jb^2_jb^3_jb^4_j b^5_j c_j. 
\end{align}
This local gauge constraint defines the physical subspace of our enlarged Hilbert space. 
The bond operators, $\hat u$, do not commute with the gauge constraint. 
While we want to fix the gauge, i.e., pick a set of eigenvalues of the $\hat u$'s, to simplify our calculations, we afterwards need to ensure that the ground state (and excited states) we computed have a non-vanishing weight in the physical sector. 
The latter is determined by the fermion parity, see Ref.~\cite{pedrocchi2011physical} for the original Kitaev honeycomb model. 
In particular, only one of the parity sectors---which one depends on the systems geometry and size---corresponds to physical states, while the other only contains unphysical states. 
For gapless states---which is the generic scenario in 3D SOLs as we see later---we can always adjust the parity without energy cost (at least in the thermodynamic limit), to ensure that the resulting state has a non-vanishing weight in the physical subspace and, thus, corresponds to a physical state.

\subsubsection{$\mathbb{Z}_2$ flux}\label{subsubsec:Z2flux}
While the bond operators themselves are not physical observables, we can use them to define Wilson loop operators for any closed loop $p$ (with length $N$) as 
\begin{align}
    \hat W_p=(-i)^N \prod_{j\in p} \hat u_{j,j+1},
    \label{wilson loop}
\end{align}
which do correspond to observables. 
The Wilson loop operators commute both with the Hamiltonian and among themselves, and---since we are only considering lattices with even-length plaquettes---have eigenvalues $\pm 1$. 
Thus, for any closed loop---in particular any plaquette $p$ in the lattice---we can interpret $\hat W_p$ as the operator measuring the absence (eigenvalue $+1$) or presence (eigenvalue $-1$) of a $\pi$-flux on this plaquette. 
The $\mathbb{Z}_2$ gauge flux excitations associated with nontrivial---i.e. $-1$---eigenvalues of the plaquette operators are commonly referred to as \emph{visons}.

In certain cases, the ground-state flux configuration---i.e. the flux configuration that allows for the lowest energy ground state for the corresponding Majorana Hamiltonian---can be determined using Lieb’s theorem \cite{lieb_flux_1994}.
If the lattice contains a plaquette that is invariant under a mirror symmetry which intersects only bonds (but not sites), then the ground state resides in the flux sector where that plaquette carries zero flux when its length is $4n+2$, and $\pi$ flux when its length is $4n$. 
For the lattices considered here, it implies that loops of length 6 or 10 want to carry 0 flux, while loops of length 8 want to carry $\pi$ flux. 
In our study, only the hyperhexagon and the layered honeycomb possess such mirror symmetries, and neither possesses enough to fix the ground state flux sector unambiguously.

In cases where the Lieb theorem cannot be applied, one is forced to resolve the question using numerical simulations along the lines of Refs. \cite{nasu2014finite}. 
Previous studies of 3D generalizations of the Kitaev model, however, strongly indicate that Lieb's theorem predicts the correct flux sector, even in cases where it cannot be applied \cite{eschmann2020thermodynamic}. 
The only known counterexamples are lattices where the flux sector is geometrically frustrated, e.g., the lattices (9,3)a \cite{mishchenko2020chiral} and (8,3)c \cite{eschmann2019thermodynamics}. 
None of the lattices studied in this work exhibits geometric frustration in the gauge sector. 

Since the unperturbed SOLs on the three-coordinated lattices are copies of the Kitaev model, the numerical simulations of Ref.~\cite{eschmann2020thermodynamic} ensure that the Lieb flux configuration is the ground state configuration in the vicinity of the unperturbed SOL.
It would be interesting to see whether perturbations can drive the system into other flux sectors. 
This is, however, left for future studies, since it requires rather heavy numerical simulations.

For SOLs on the two four-coordinated lattices, we have no prior reference point for determining the ground state flux sector, although the layered honeycomb lattice possesses mirror planes that determine half of the relevant Wilson loops to have eigenvalues $+1$. 
Therefore, we performed numerical simulations for small sizes of the flux unit cells for the unperturbed model. 
These indicate that the Lieb configuration is the correct ground state sector.
Therefore, we will use the Lieb flux sector to determine the behavior of the Majorana fermions. 
The analysis of whether large perturbation strength stabilizes other flux sectors is left for future numerical studies. 
Note that we can enforce the Lieb flux sector to be the ground state by adding additional terms to the Hamiltonian of the form  
\begin{align}\label{eq:fluxHam}
     - \sum_p E_p\hat W_p, 
\end{align}
where $E_p$ is a sufficiently large positive (negative) energy on all fundamental plaquettes \footnote{Fundamental plaquettes are those that cannot be built up by smaller plaquettes.} $p$ of length $4n+2$ ($4n$).

\subsubsection{Thermal stability}
 An important distinction between two- and three-dimensional realizations of Kitaev-type models arises at finite temperature.
In two dimensions, visons are point-like excitations with a finite energy cost and therefore proliferate at any nonzero temperature, destroying the QSL.
In three dimensions, by contrast, visons necessarily form extended loop-like objects due to the volume constraints imposed by the lattice geometry; see Appendix~\ref{subsec:Loops_volume_constraints}. 
As the energy cost of loops scales with their length, three-dimensional models admit a finite-temperature regime in which flux excitations remain dilute, leading to thermal stability of the QSL phase \cite{nasu2014vaporization}.

\subsubsection{Explicit Majorana descriptions}
For the benefit of the reader, we bring the Hamiltonian \eqref{eq:MajoranaHamGeneral} in a more explicit and symmetric form, both for the four-coordinated lattices with $\nu=2$ and the three-coordinated ones with $\nu=3$. 
 For the case $\nu=2$, the Majorana Hamiltonian---in a fixed gauge---can be written as
  \begin{align}
      \tilde{\mathcal H}_J^{(2)}=\sum_{\substack{\langle ij\rangle_\gamma}}J_\gamma u^\gamma_{ij} \left(   i c_i^x c_j^x + i c_i^y c_j^y  \right),
 \end{align}
where the Majorana operators have been relabeled as $(b^5,c)\rightarrow (c^x,c^y)$ and $\gamma = x,y,z,w$ denotes the four nearest neighbor bonds. 
Because the two flavors appear in exactly the same way on all bonds, the model is simply two identical copies of the Kitaev model with $\nu = 1$.
In fact, the model is invariant under a $\mathrm{SO}(2)$ symmetry rotating the Majoranas $(c^x,c^y)$, showing the equivalence of the flavors.

For three-coordinated lattices with $\nu=3$, we relabel the Majorana operators $b^5\rightarrow c^x$, $c\rightarrow c^y$, $b^4\rightarrow c^z$, to bring the Hamiltonian in its most symmetric form: 
 \begin{align}\label{eq:tricoordHam}
      \tilde{\mathcal H}_J^{(3)}=\sum_{\langle ij\rangle_\gamma}J_\gamma u^\gamma_{ij} \left(   i c_i^x c_j^x + i c_i^y c_j^y + i c_i^z c_j^z \right).
 \end{align}
Since all three flavors couple in the same way on every bond, the model corresponds to three identical copies of the Kitaev model. 
The Hamiltonian is invariant under $\mathrm{SO}(3)$ rotations of the itinerant Majorana modes $(c^x,c^y,c^z)$, reflecting the complete equivalence of the three itinerant Majorana flavors. 
In the next section, we will discuss general symmetries of the model and how to break them. 
In particular, we will be interested in different ways to break the $\mathrm{SO}(\nu)$ symmetry.

\section{Symmetries and how to break them}\label{sec:Symmetries}
In this section, we summarize the symmetries of the unperturbed spin-orbital Hamiltonian and introduce the perturbations that preserve exact solvability.

\subsection{Symmetries }\label{subsec:Symmetries}
In order to classify the possible low-energy properties of the spin-orbital system, we need to determine the structure and symmetries of the Majorana Hamiltonian, \eqref{eq:MajoranaHamGeneral}. 
When rewriting the spin-orbital Hamiltonian in terms of Majorana fermions, we introduce a particle-hole  `symmetry' of the Majorana Bloch Hamiltonian. 
For each energy $E(\mathbf{k})$ at momentum $\mathbf{k}$, there exists a partner $-E(-\mathbf{k})$. 
This particle-hole symmetry is actually not a symmetry of the system (most importantly, it cannot be broken), but rather a redundancy. 
Rephrasing the problem into Majoranas---each of which is half of a fermionic mode---amounts to counting every degree of freedom twice. 
This double-counting is usually resolved by either only considering the upper half of the bands, or only those modes with $E\geq 0$ \cite{kitaev2006anyons}. \footnote{Band degeneracies (former choice) or zero modes (latter choice) need to be handled carefully.}

Another important symmetry in our model is time-reversal (TR). 
As mentioned earlier, its explicit implementation depends on the interpretation of the local Hilbert space. 
We choose to follow the conventions of Ref.~\cite{chulliparambil_flux_2021}, so  TR symmetry is given by
\begin{align}
    T &= \left(i \sigma^y\otimes\mathds{1}\right)\kappa,
\end{align}
with $\kappa$ denoting complex conjugation. 
When rewritten in the Majorana representation, $T$ can be shown to square to 1. 
This---together with particle-hole symmetry---places the unperturbed Majorana Hamiltonian in the Altland-Zirnbauer class BDI. \footnote{In \cite{ryu_three-dimensional_2009, nakai_time-reversal_2012}, an alternative TR symmetry was considered for the $\nu=2$ case, involving a combination of $T$ with a rotation of the spin degrees of freedom. In the Majorana representation, this changes the symmetry class to DIII, yielding a $\mathbb{Z}_2$ invariant in two dimensions and a $\mathbb{Z}$ invariant in three dimensions. Such a symmetry construction is not possible for the $\nu=3$ model.}

The remaining symmetries fall into two classes.
First, there is a global $\mathrm{SO}(\nu)$ symmetry associated with rotations in flavor space, reflecting the equivalence of the $\nu$ itinerant Majorana fermions.
Second, there are lattice symmetries—such as inversion, rotations, and mirror symmetries—which depend on the specific lattice under consideration.

In this work, we are interested in adding perturbations that mix the different Majorana flavors and therefore break the $\mathrm{SO}(\nu)$ flavor symmetry. 
At the same time, we largely restrict ourselves to perturbations that respect the underlying lattice symmetries, as these strongly constrain the allowed low-energy structures. The explicit construction of such perturbations is given in Sec.~\ref{subsec:pertubations}.

\subsection{Projective symmetry group}\label{subsec:PSG}
Having identified the relevant symmetries of the spin-orbital Hamiltonian, we now discuss how these symmetries constrain the gauge-fixed Majorana Hamiltonian. 
A subtlety arises because physical states must be invariant under physical symmetries, whereas the actual calculations are performed in a gauge-fixed representation. 
Consequently, the eigenstates of the gauge-fixed Hamiltonian need not transform linearly under a physical symmetry; instead, the symmetry action may have to be supplemented by a gauge transformation to return to the chosen gauge. 
Symmetries realized in this combined way are said to be implemented projectively, and the set of all such implementations forms what Wen termed the projective symmetry group (PSG)\cite{wen2002quantum}.

The essential idea can be illustrated most transparently using TR symmetry.
In the spin–orbital representation, TR flips the spin degrees of freedom while leaving the orbitals invariant. 
In the Majorana language, this corresponds to all Majoranas being even under TR. 
However, TR also flips the $\mathbb{Z}_2$ bond operators, $\hat u_{ij}^\gamma$, which means that applying TR takes us from the chosen gauge into a different gauge sector. 
To determine the constraints on the Bloch Hamiltonian, we must therefore augment TR by a gauge transformation that restores the original gauge.

For a bipartite lattice, such a transformation can be chosen as 
\begin{align}
    c^\alpha_j&\rightarrow -c^\alpha_j \, \mbox{ for }j\in B,& c^\alpha_j&\rightarrow c^\alpha_j\, \mbox{ for }j\in A, 
\end{align}
with $\alpha$ denoting the itinerant flavors.
If the sublattices are compatible with lattice translations---meaning that every translation maps A to A and B to B---then this gauge transformation is momentum independent, and TR relates $H(\mathbf{k})$ to $H(-\mathbf{k})$.
Combining TR and particle-hole symmetry, thus gives us constraints on the Hamiltonian at fixed $\mathbf{k}$. 

In contrast, if the sublattice structure is not compatible with all translation vectors, the gauge transformation acquires a nontrivial momentum dependence. 
One then finds that TR relates $H(\mathbf{k})$ to $H(-\mathbf{k}+\mathbf{k_0})$, where $\mathbf{k}_0$ is half a reciprocal lattice vector, i.e.  satisfying $2\mathbf{k}_0\equiv 0$. 
In this case, the combination of particle-hole and TR does not constrain the Hamiltonian at a single momentum point; instead, it relates the Hamiltonian at pairs of momenta. 
We refer to this as a projective implementation of TR.

Whether a symmetry is implemented trivially or projectively has profound consequences for the allowed nodal structures. 
This is already familiar from the two-dimensional Kitaev model: a trivial implementation of TR stabilizes Dirac cones, whereas a projective implementation allows extended Fermi lines~\cite{yang2007mosaic}. 
The same dichotomy appears in three dimensions and plays a central role in determining the zero-mode structure of SOLs on different lattices \cite{obrien_classification_2016}.

\subsection{Perturbation terms}\label{subsec:pertubations}
We now introduce the perturbation terms relevant for the remainder of this manuscript.  
Throughout, we restrict ourselves to quadratic perturbations that preserve exact solvability; that is, we only consider terms that commute with all plaquette operators.  
As shown in Ref.~\cite{chulliparambil_flux_2021}, any such solvable perturbation can be written as a hopping process between site $j$ and site $j+l$ of the form
\begin{align}\label{SolvPert}
    \tilde{\mathcal H}' = 
    f^{\alpha\beta}_{j,j+l}\,
    i\, c^\alpha_j 
    \left[\prod_{\langle jk\rangle \in \mathcal{L}} \hat u_{jk}\right]
    c^\beta_{j+l},
\end{align}
where the product of bond operators is taken along an arbitrary path $\mathcal{L}$ connecting sites $j$ and $j+l$. The a priori arbitrary couplings $f^{\alpha\beta}_{j,j+l}$ are, as we will see later in this section,  constrained by symmetries.
Although longer-range terms are formally allowed, in practice it suffices to consider on-site, nearest-neighbor, and next-nearest-neighbor (NNN) perturbations, as these already capture all phenomena relevant to the following sections.

\begin{table*}[t]
	\centering
	\caption{Overview of the possible QSL behavior for the SOL on three-coordinated lattices under perturbations. }
	
	\label{tab:summary_tricoord_tight}
	\begin{tabular}{p{3.7cm} | p{3.1cm}| p{4.7cm} |p{3.0cm}| p{2.8cm}}
		\hline\hline
		\textbf{Lattice} &
		\textbf{Unperturbed} &
		\textbf{TR-preserving (\(K,\Gamma,\Gamma'\))} &
		\textbf{TR-breaking (\(\mathbf{h}\))} &
		\textbf{TR-breaking (\(\kappa\))} \\
		\hline
		Hyperoctagon \((10,3)a\) &
		Top. FS &
		Top. FS $\leftrightarrow$ trivial FS $\leftrightarrow$ gapped &
		(Nested) FS  &
		Deformed top. FS \\
		\hline
		Hyperhoneycomb \((10,3)b\) &
		Threefold degenerate nodal line &
		Split nodal lines $\leftrightarrow$ gapped &
		Tubular  FS &
		WPs \\
		\hline
		Hyperhexagon \((8,3)b\) &
		WPs &
		WPs (split / shifted) $\leftrightarrow$ gapped &
		Top. FS  &
		WPs \\
		\hline\hline
	\end{tabular}
\end{table*}   
\paragraph{On-site perturbations}
For on-site perturbations, solvability reduces the number of potential terms dramatically: since no bonds are traversed, 
Eq. \eqref{SolvPert} reduces to 
\begin{align}
             \tilde{\mathcal H}' = \sum_{ j} f^{\alpha\beta}_{j} i c^\alpha_j c^\beta_{j}.
\end{align}
For the three-coordinated lattices, which host three itinerant Majorana flavors $c^x, c^y, c^z$, there are only three independent terms, $h_z := f^{xy}$, $h_y:=f^{zx}$, and $h_x: = f^{yz}$.
The Majorana Hamiltonian takes the form
\begin{align}
\tilde{\mathcal H}^{(3)}_h=\sum_j(h_x ic_j^y c_j^z+h_y ic_j^zc_j^x+ h_zic_j^x c_j^y).
\end{align}
In the spin-orbital representation, this becomes 
\begin{align}
    \mathcal H_h^{(3)} = - \boldsymbol{h}\sum_i \boldsymbol{\sigma}_i \otimes \mathds{1},
\end{align}
so the perturbation acts as a coupling of the spins to an external magnetic field. 
Since the on-site perturbation mixes the Majorana flavors, it not only breaks TR symmetry but also the $\mathrm{SO}(3)$ symmetry among the flavors since $\boldsymbol{h}\neq 0$ picks out a preferred direction in flavor space.

The four-coordinated models are quite similar, but now there are only two itinerant Majorana flavors, $c^x$, and $c^y$. 
Consequently, the allowed on-site perturbations are reduced to a single independent term, $h_z:=f^{xy}$, leading to 
\begin{align}
             \tilde{\mathcal H}_h^{(2)} = h_z \sum_j i c_j^x c_j^y, 
\end{align}
which, in the spin-orbital language, corresponds to
\begin{align}
    \mathcal H_h^{(2)} =  -h_z \sum \sigma_i^z \otimes \mathds{1}.
\end{align}
This term acts as a magnetic field in the $\hat z$-direction,  breaking TR symmetry. 
As in the $\nu = 3$ case, this term mixes the two itinerant Majorana flavors and hence breaks the $\mathrm{SO}(2)$ flavor symmetry.

\paragraph{Nearest-neighbor perturbations}
For NN interactions,  Eq.~\eqref{SolvPert} reduces to
\begin{align}
         \tilde{\mathcal H}'_{NN} = \sum_{\langle ij \rangle_\gamma} f^{\alpha\beta}_{ij} i c^\alpha_i \hat u_{ij}c^\beta_{j}.
\end{align}
Following Ref.~\cite{chulliparambil_flux_2021}, we impose lattice symmetries to restrict the allowed structure of the perturbation terms. 
For the three-coordinated lattices, a combined threefold rotation symmetry acting on both the spin and orbital sectors enforces isotropy of the perturbations: all symmetry-equivalent bonds must carry the same couplings.
Although the lattices we consider differ from those in Ref.~\cite{chulliparambil_flux_2021}, the resulting symmetry-allowed perturbations for $\nu=3$ are exactly the same. 
In contrast, for the $\nu=2$ models, we allow the NN couplings to differ between the four bond types whenever the lattice does not render them symmetry-equivalent.

For the three-coordinated lattices, the NN perturbations take the form 
\begin{align}
    \mathcal{\tilde{H}}^{(3)}_{K\Gamma\Gamma'} &= \sum_{\langle ij \rangle_\gamma}  -i u_{ij}\left( -Kc^\gamma_ic^\gamma_j + \Gamma (c^{\alpha}c^{\beta}+c^{\beta}c^{\alpha})\right.\nonumber\\
    &+ \left. \Gamma' (c^{\gamma}c^{\alpha}+c_i^{\alpha}c_j^{\gamma}+c_i^{\gamma}c_j^{\beta}+c_i^{\beta}c_j^{\gamma})\right),
\end{align}
and, in the spin-orbital representation, 
\begin{align}
    \mathcal{H}^{(3)}_{K\Gamma\Gamma'} &= \sum_{\langle ij \rangle_\gamma} \left( -K\sigma^\gamma_i\sigma^\gamma_j + \Gamma (\sigma_i^{\alpha}\sigma_j^{\beta}+\sigma_i^{\beta}\sigma_j^{\alpha})\right. \nonumber\\
    &\left.+ \Gamma' (\sigma_i^{\gamma}\sigma_j^{\alpha}+\sigma_i^{\alpha}\sigma_j^{\gamma}+\sigma_i^{\gamma}\sigma_j^{\beta}+\sigma_i^{\beta}\sigma_j^{\gamma})\right)\otimes\tau_i^{\gamma}\tau_j^{\gamma},
\end{align}
where $(\alpha,\beta,\gamma)= (y,z,x)$, $(z,x,y)$ and $(x,y,z)$ on $x$, $y$, and $z$ bonds, respectively. 
These terms correspond, in the spin–orbital language, to the familiar Kitaev–Heisenberg–$\Gamma$–$\Gamma'$ model. 
They break the $\mathrm{SO}(3)$ Majorana flavor symmetry explicitly while preserving all lattice symmetries imposed in the construction.

The situation for the four-coordinated lattices is slightly simpler, since only two itinerant Majorana flavors are present. As in Ref.~\cite{chulliparambil_flux_2021}, we consider only terms that mix these two flavors, i.e., those where only $f^{xy}_{ij} = f^{yx}_{ij}$ are non-zero. 
However, unlike Ref.~\cite{chulliparambil_flux_2021}, we only consider lattices where the four different bonds are not all related by symmetry. 
Therefore, we do not impose equivalence of the perturbation terms between the different bond types, but only those that are symmetry-equivalent. 
This leads to perturbations of the form
\begin{align} \label{eq:NN_pert_fourcoord}
   \mathcal{H}^{(2)}_{\bar \Gamma}&= \sum_{\langle ij \rangle_\gamma} \Bar{\Gamma}_{\gamma} (\sigma^x_i \sigma^y_j + \sigma^y_i \sigma^x_j )\otimes\tau^\gamma_i\tau^\gamma_j\nonumber\\
   \tilde{\mathcal{H}}^{(2)}_{\bar \Gamma}&=  \sum_{\langle ij \rangle_\gamma} -i u_{ij} \Bar{\Gamma}_{\gamma} \left( c_i^x c_j^y + c_i^y c_j^x \right) .
\end{align}
where the coefficients $\Bar{\Gamma}_{\gamma}$ may be different for symmetry-inequivalent bond types.

\paragraph{Next-nearest-neighbor perturbations}

NNN perturbations involve a string of operators over three consecutively connected sites $i, j, k$. 
Such terms necessarily break TR symmetry, but their effect on the Majorana bands differs qualitatively from that of the on-site TR symmetry-breaking perturbations discussed above.  
To streamline the analysis and limit the number of parameters, we adopt the same structure as in the original Kitaev model:
\begin{align}
    \tilde{\mathcal H}_\kappa^{(\nu)}
    = \kappa
      \sum_{\langle ijk \rangle_{\alpha\beta}}
        \epsilon^{\alpha\beta\gamma}\,
        u_{ij}^\alpha u_{jk}^\beta
        \left(
            i c_i c_k
            + \sum_{\delta = \gamma_m + 1}^{2q+3}
              i\, b_i^\delta b_k^\delta
        \right), 
    \label{eq:NNN_maj}
\end{align}
where ${\langle ijk \rangle_{\alpha\beta}}$ denotes the string of three sites  $i$, $j$, $k$, with bond $\langle ij\rangle$ being of type $\alpha$ and  bond $\langle jk\rangle$ of type $\beta$, and $\epsilon ^{\alpha\beta\gamma}$ denotes the  Levi-Civita tensor. 
In particular, the NNN perturbation is identical for the different Majorana flavors. 
In the original Kitaev model this term arises  from the effect of an external magnetic field projected to the ground-state flux sector \cite{kitaev2006anyons}. 
In the present SOL context,  a perturbative analysis analogous to the Kitaev case becomes substantially more involved and generically generates interaction terms. We therefore introduce the on-site term discussed above and the NNN term below as simple representative perturbations that capture how generic time-reversal-breaking effects enter the Majorana Hamiltonian. 
The NNN term are identical to those used in Ref.~\cite{chulliparambil_flux_2021}. 
For the $\nu = 3$ model, Eq.~\eqref{eq:NNN_maj} reduces to
\begin{align}
    \tilde{\mathcal H}_\kappa^{(3)}
    = \kappa \sum_{\langle ijk\rangle_{\alpha\beta}}
        \epsilon^{\alpha\beta\gamma}
        u^\alpha_{ij} u^\beta_{jk}
        \left(
            i c_i^x c_k^x
          + i c_i^y c_k^y
          + i c_i^z c_k^z
        \right),
    \label{eq:NNN_tricord}
\end{align}
while for the most natural generalization to the  $\nu = 2$ model is given by 
\begin{align}\label{eq:NNN_fourcoord1}
    \tilde{\mathcal H}_\kappa^{(2)}
    = \kappa \sum_{\langle ijk\rangle_{\alpha\beta}}
        \epsilon^{\alpha\beta\gamma\delta}\,
        u^\alpha_{ij} u^\beta_{jk}
        \left(
            i c_i^x c_k^x
          + i c_i^y c_k^y
        \right).
\end{align}
The motivation behind and further details on Eq.~\eqref{eq:NNN_fourcoord1} can be found in Sect.~\ref{subsec:Fourcoordinated_general}. 

\section{Three-coordinated lattices}\label{sec:Tricoordinated}
We will start our analysis of three-dimensional SOLs by analyzing Hamiltonian \eqref{eq:SeifertHamiltonian} and possible perturbations on three three-coordinated lattices. 
The three lattices are known from the generalization of the original Kitaev model to three spatial dimensions: the hyperoctagon, hyperhoneycomb, and hyperhexagon lattice \cite{obrien_classification_2016}.  
We chose these lattices for our analysis, since they are relatively simple and showcase the three main possibilities for the gapless (or nodal) structures: The zero modes can form two-dimensional FSs, a one-dimensional nodal line, or zero-dimensional WPs. 
Since (for three-coordinated lattices) the unperturbed SOL, \eqref{eq:SeifertHamiltonian}, corresponds to three copies of the corresponding KSL, we can use the intuition of the latter to understand the properties of the SOL. 
This will give us an easy introduction to 3D SOL before tackling the four-coordinated lattice, which has not been discussed before. 
In each case, we will give a short introduction to the lattice and its symmetries, review the properties and the band structure of the unperturbed SOL, as well as discuss the effect of perturbations. 
A summary of the dominating behavior for all three lattices can be found in Table~\ref{tab:summary_tricoord_tight}.

\subsection{General comments}\label{subsec:TRIgeneral}

\subsubsection{Perturbations}\label{subsubsec:TRIperturbations}
We start with a general discussion on the different effects that the perturbations---introduced in Sec.~\ref{subsec:pertubations}---can have on the low-energy behavior of the SOL. 
The TR-breaking NNN term is identical for all the flavors. Thus, it is the only perturbation that does not affect the flavor degeneracy, or in other words, the $\mathrm{SO}(3)$ symmetry of the SOL.
This term is constructed to be identical to the TR-breaking (due to an external magnetic field) of the original KSL. 
Hence, its effect is to produce three degenerate copies of the corresponding TR broken KSL. 

Breaking the $\mathrm{SO}(3)$ symmetry requires the onsite TR breaking term or one of the NN terms. 
The on-site TR breaking term  
\begin{align}
\tilde{\mathcal H}^{(3)}_{\bm h}&=\sum_i\Big(h_x ic_i^y c_i^z+h_y ic_i^zc_i^x+ h_zic_i^x c_i^y\Big)
\end{align}
acts as a chemical potential for suitably rotated vectors in the flavor space. 
While the band structure of one of the rotated Majorana flavors is left completely unchanged, another gets shifted upwards by a constant energy $+|\bm h|$,  while the band of the third is shifted downward by $-|\bm h|$. 

Of the remaining NN terms, the K perturbation is the simplest. 
It does not mix different Majorana flavors, but only splits their degeneracy by changing one of the Ising couplings $J_\gamma\rightarrow J_\gamma+K$ for the $\gamma$-Majorana flavor: 
\begin{align}
    \mathcal{\tilde{H}}^{(3)}_{K} = \sum_{\langle ij\rangle_\gamma} -i u_{ij}\Big( -Kc^\gamma_ic^\gamma_j\Big), 
\end{align}
i.e., each Majorana flavor is perturbed away from the isotropic point, but in a symmetric fashion so that the full model stays isotropic.
The perturbation terms for $\Gamma$ and $\Gamma'$ correspond to two distinct ways of mixing the Majorana flavors, again keeping the full model isotropic: 
\begin{align}
    \mathcal{\tilde{H}}^{(3)}_{\Gamma} &=\Gamma  \sum_{\langle ij\rangle_\gamma} -i u_{ij}\Big(c_i^{\alpha}c_j^{\beta}+c_i^{\beta}c_j^{\alpha}\Big)\\
   \mathcal{\tilde{H}}^{(3)}_{\Gamma'} &= \Gamma'  \sum_{\langle ij\rangle_\gamma} -i u_{ij} \Big(c_i^{\gamma}c_j^{\alpha}+c_i^{\alpha}c_j^{\gamma}+c_i^{\gamma}c_j^{\beta}+c_i^{\beta}c_j^{\gamma})\Big),
\end{align}
 where $(\alpha,\beta,\gamma) = (y,z,x), (z,x,y),$ and $(x,y,z)$ for an $x$, $y$ and $z$ bond, respectively.

 \paragraph{Special case: $\Gamma=\Gamma'$ with $K=0$}
For the special case $\Gamma=\Gamma'$ and $K=0$, we can bring the perturbed Hamiltonian into a simpler form using a global spin-rotation introduced in  Ref.~\cite{chaloupka2015hidden}. 
We rotate to a new coordinate frame $(\tilde X, \tilde Y, \tilde Z)$ that is related to the old one by the rotation 
\begin{align}
R_1 &=    
    \begin{pmatrix}
      -\frac{1}{\sqrt{2}}& -\frac{1}{\sqrt{6}}& \frac{1}{\sqrt{3}}\\
      \frac{1}{\sqrt{2}}&-\frac{1}{\sqrt{6}} & \frac{1}{\sqrt{3}}\\
      0& \sqrt{\frac{2}{3}} & \frac{1}{\sqrt{3}} \\
    \end{pmatrix} & \mbox{with } 
    \begin{pmatrix}
        \tilde X \\ \tilde Y\\ \tilde Z\\ 
    \end{pmatrix}
    &= R_1
    \begin{pmatrix}
         X \\  Y\\  Z\\ 
    \end{pmatrix}. 
\end{align}
Let us now consider the full Hamiltonian for $\Gamma=\Gamma'$, $K=0$ on the bond $\langle ij\rangle_\gamma$, given by  
\begin{align}
    -J\left(\bm \sigma_i\cdot \bm \sigma_j \right) \otimes \tau_i^\gamma \tau_j^\gamma 
    + 
    \Gamma \sum_{\alpha<\beta} \left( \sigma_i^\alpha \sigma_j^\beta+\sigma_i^\beta \sigma_j^\alpha\right)\otimes \tau_i^\gamma\tau_j^\gamma, 
\end{align}
where $(\alpha,\beta)$  runs over all combinations $(x,y)$, $(x,z)$ and $(y,z)$. 
The first part is invariant under any spin rotation, but the second part transforms to 
\begin{align}
    \Gamma \left( -\sigma_i^x \sigma_j^x -\sigma_i^y\sigma_j^y+2 \sigma_i^z\sigma_j^z \right)\otimes \tau_i^\gamma\tau_j^\gamma. 
\end{align}
Thus, the full Hamiltonian can be brought into the simple form 
\begin{align}
    -\left[(J+\Gamma)(\sigma_i^x \sigma_j^x +\sigma_i^y \sigma_j^y) + (J-2\Gamma)\sigma_i^z\sigma_j^z\right]  \otimes \tau_i^\gamma \tau_j^\gamma. 
\end{align}
In other words, on the line $\Gamma=\Gamma'$, the only effect of the perturbations is to rescale the coupling constants for $x$/$y$ Majorana flavor to $J+\Gamma$  and that of the $z$ flavor to $J-2\Gamma$.
Two points in the phase diagram, namely $\Gamma=J/2$ ($\Gamma=-J$), are special, as one (two) of the Majorana flavors drops out of the Hamiltonian completely, thus creating two (four) completely flat Majorana bands at zero energy.

\begin{figure}
    \centering
    \includegraphics[width=0.5\linewidth]{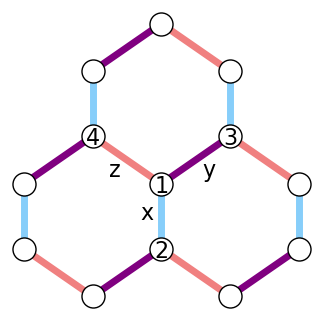}
    \caption{Local choice of sublattice for the Klein rotation. }
    \label{fig:Klein rotation}
\end{figure}

\paragraph{Special line $\Gamma=\Gamma'=0$}
If only the $K$ perturbation is non-zero, we can again employ a spin rotation---the so-called Klein rotation---to bring the Hamiltonian into a simpler form. 
We first choose a 4-site sublattice according to Fig.~\ref{fig:Klein rotation} \cite{khaliullin2003theory}. 
As shown in Ref.~\cite{kimchi_kitaev-heisenberg_2014}, this can be done for any graph whose fundamental loops contain a number of $x$/$y$/$z$ bonds that are either all even or all odd.
This condition is fulfilled for both the layered honeycomb and the chiral square octagon lattice discussed below. 

The rotation will be different on the four sublattices, namely: 
\begin{align}
    \mbox{site 1: }& (X,Y,Z)\rightarrow (\tilde X,\tilde Y, \tilde Z) \nonumber \\
    \mbox{site 2: }& (X,Y,Z)\rightarrow (\tilde X,-\tilde Y,- \tilde Z) \nonumber\\
    \mbox{site 3: }& (X,Y,Z)\rightarrow (-\tilde X,\tilde Y, -\tilde Z) \nonumber\\
    \mbox{site 4: }& (X,Y,Z)\rightarrow (-\tilde X,-\tilde Y, \tilde Z).
\end{align}
Since the rotation is different on the different sublattices, the unperturbed Hamiltonian is not invariant under this transformation. 
For instance, for an $x$-bond, the rotated $J$ term reads: 
\begin{align}
    -\left[J(\sigma_i^x \sigma_j^x) -J (\sigma_i^y \sigma_j^y +\sigma_i^z\sigma_j^z)\right]  \otimes \tau_i^x \tau_j^x. 
\end{align}
whereas the $K$ term transforms to 
\begin{align}
    -K(\sigma_i^x \sigma_j^x)  \otimes \tau_i^x \tau_j^x. 
\end{align}
The other bond types transform similarly, and thus, the full Hamiltonian is transformed to 
\begin{align}
    -J (\bm \sigma_i\cdot \bm \sigma_j)\otimes \tau_i^\gamma \tau_j^\gamma
    - (K+ 2J) \sigma_i^\gamma \sigma_j^\gamma \otimes \tau_i^\gamma \tau_j^\gamma, 
\end{align}
on a bond $\langle ij \rangle_\gamma$. 
At the special point $K=-2J$, the model again looks like the unperturbed one, albeit at coupling $-J$. 
For our discussion, the point $K=-J$ is of more interest. 
For this value, the coupling constant $J_\gamma$ for the $c^\gamma$ Majorana flavor vanishes, $J_\gamma=0$ . 
For our three lattices of interest, this implies that the Majoranas can only hop along disconnected 1D chains, instead of a truly 3D lattice.\footnote{Note, however, that the resulting 1D chains are different for each of the Majorana flavors.}
Consequently, flat zero energy bands appear at this point in parameter space.

\subsubsection{Structure of perturbed Hamiltonian in momentum space}\label{subsubsec:TRImomentum}
In order to simplify the notation---as well as avoid the display of large matrices---later on, we now discuss the general structure of the perturbed Hamiltonian in momentum space and define certain sub-blocks. 
Doing a Fourier transform and combining all the Majoranas into a single spinor $\psi$ 
\begin{align}
    \psi(\mathbf{k}) &= (c_{\mathbf{k},1}^x, \ldots,c_{\mathbf{k},d}^x,c_{\mathbf{k},1}^y, \ldots, c_{\mathbf{k},d}^z)
\end{align}
where $d$ denotes the size of the unit cell for a given lattice, we find the general matrix representation of the perturbed SOL as 
\begin{align}
		\tilde{\mathcal{H}}^{(3)}&=\sum_{\mathbf{k}}
		\psi(-\mathbf{k})^\dagger 
		H(\mathbf{k})
	\psi(\mathbf{k}), 
\end{align}
which can be expressed in terms of $d\times d$ matrices $M_{\alpha\beta}$ as 
\begin{align}
	\begin{aligned}
		H(\mathbf{k})=
		\begin{pmatrix}
			&M_{xx}(\mathbf{k}) &M_{xy}(\mathbf{k}) &M_{xz}(\mathbf{k}) \\
			&M_{xy}^\dagger(\mathbf{k}) &M_{yy}(\mathbf{k}) &M_{yz}(\mathbf{k}) \\
			&M_{xz}^\dagger(\mathbf{k}) &M_{yz}^\dagger(\mathbf{k}) &M_{zz}(\mathbf{k})
		\end{pmatrix}.
	\end{aligned}
	\label{eq:perturbedH}
\end{align}

The diagonal terms $M_{\alpha\alpha}(\mathbf{k})$ are identical to the Hamiltonian of the original Kitaev model on this lattice \cite{obrien_classification_2016}, with the modification that in $M_{\alpha\alpha}$ the $J_\alpha$ Ising coupling is replaced by $J_\alpha+K$. 
The perturbations $\Gamma$ and $\Gamma'$, as well as the on-site TR breaking terms, appear only in the off-diagonal blocks, i.e., $M_{\alpha\beta}$ with $\alpha\neq \beta$. 
To keep the subsequent discussion accessible, explicit matrix expressions for \eqref{eq:perturbedH} are given in  Appendix~\ref{subsec:explicitHamiltonians} for each of the three-coordinated lattices we discuss.

\subsection{Hyperoctagon}\label{subsec:hyperoctagon}
\subsubsection{The lattice and symmetries}\label{subsubsec:hyperoctagonLattice}
\begin{figure}[t]
    \centering
    \includegraphics[width=0.8\linewidth]{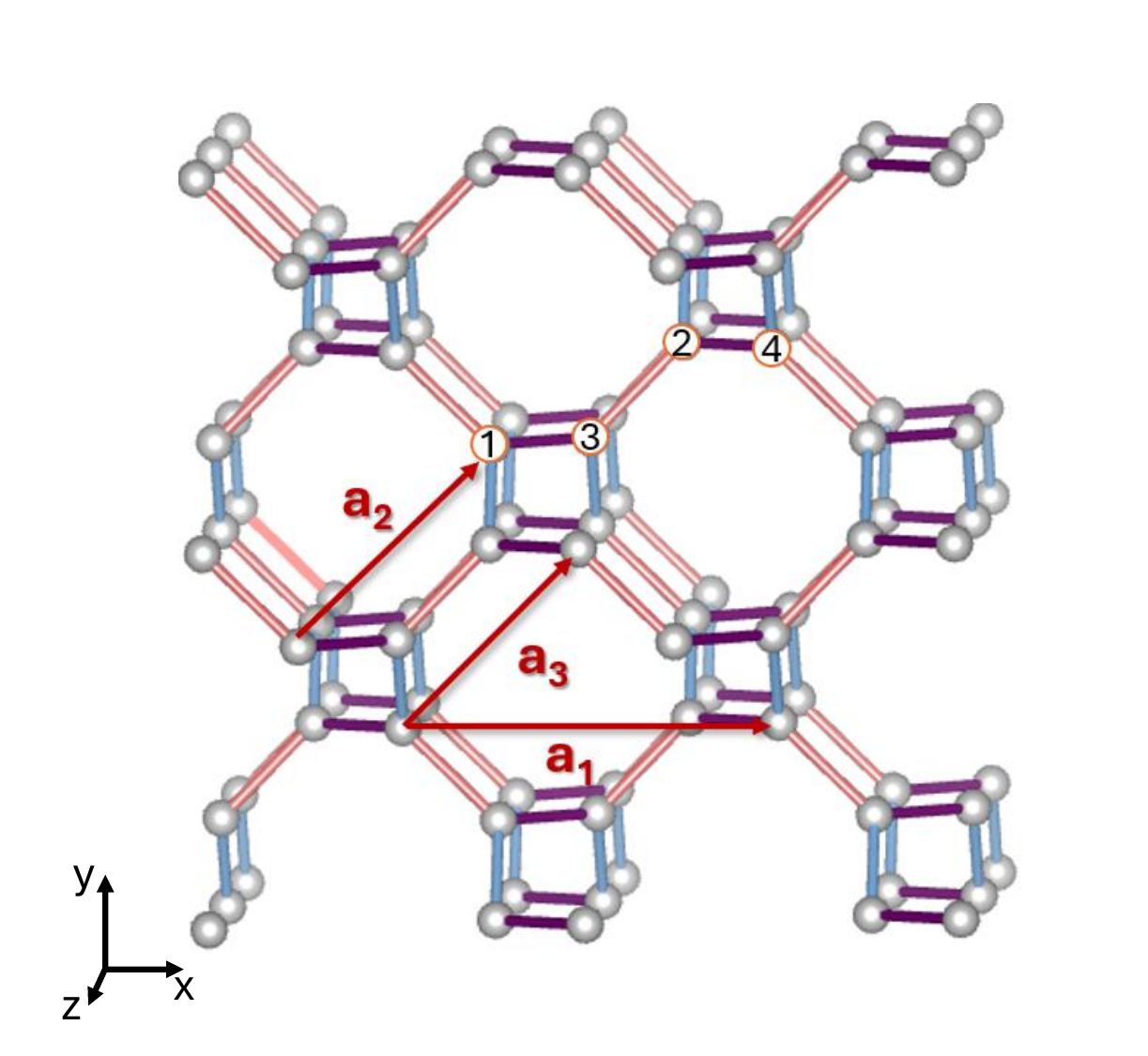}
    \caption{Hyperoctagon  lattice with unit cell and translation vectors.  Colors blue/purple/pink denote  $x$/$y$/$z$-bonds.}
    \label{fig:10_3A_lattice}
\end{figure}

\begin{figure*}[t]
    \centering
    \includegraphics[width=\linewidth]{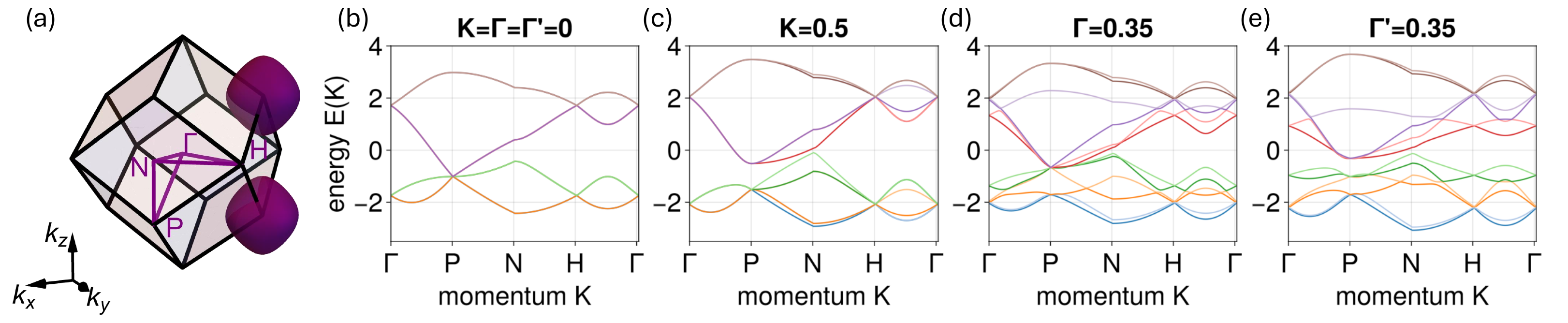}
    \caption{(a) BZ and FSs for the unperturbed SOL on the hyperoctagon lattice. (b)-(e) High-symmetry line plots for the unperturbed model, as well as for switching on one of the perturbations.  }
    \label{fig:10_3A_BZ_perturbations}
\end{figure*}

We begin our discussion with the most symmetric three-coordinated lattice, the hyperoctagon, or (10,3)a lattice.
It can be visualized as a square-octagon lattice, where the squares are replaced by co-rotating spirals, see Fig.~\ref{fig:10_3A_lattice}.

The lattice has a four-site unit cell with positions  \cite{obrien_classification_2016}
\begin{align}
    \mathbf{r}_1&=\left(\frac{1}{8},\frac{1}{8},\frac{1}{8} \right),& 
    \mathbf{r}_2&=\left(\frac{5}{8},\frac{3}{8},-\frac{1}{8}, \right)\nonumber\\
    \mathbf{r}_3&=\left(\frac{3}{8},\frac{1}{8},-\frac{1}{8} \right),&
    \mathbf{r}_4&=\left(\frac{7}{8},\frac{3}{8},\frac{1}{8} \right),  
\end{align}
with lattice translation vectors
\begin{align}
    \mathbf{a}_1&=(1,0,0), &
    \mathbf{a}_2&=\left(\frac{1}{2},\frac{1}{2},-\frac{1}{2} \right),&
    \mathbf{a}_3&=\left(\frac{1}{2},\frac{1}{2},\frac{1}{2} \right). 
\end{align}
and corresponding reciprocal lattice vectors
\begin{align}
    \mathbf{q}_1&=2\pi (1,-1,0),&  \mathbf{q}_2&=2\pi (0,1,-1), & \mathbf{q}_3&=2\pi (0,1,1).
\end{align}
    
The lattice possesses several symmetries, most important for the Majorana band structure are (i)  threefold rotation (around, e.g., the $[1,1,1]$ axis) and (ii) 2-fold screw rotations (around, e.g., $[1,0,0]$) that relate the different bonds to each other.  
It is a bipartite lattice, but lattice translation symmetries map the two sublattices onto each other---a feature that automatically implies that TR symmetry is implemented projectively and, thus,  connects states at momentum $\mathbf{k}$ to those at $\mathbf{-k}+\mathbf{k_0}$, where $\mathbf{k_0}= (-\mathbf{q_2}+\mathbf{q_3})/2$ for the hyperoctagon lattice. 
This projective implementation of TR symmetry has important consequences for the nature of the Majorana metal as explained below. 

\subsubsection{Unperturbed spin-orbital liquid} \label{subsubsec:unperturbed_Hyperoctagon}
For the unperturbed SOL on the hyperoctagon lattice, the momentum Hamiltonian for each of the Majorana flavors is identical to the one of the original Kitaev model, and is given by 
\begin{align}
        M(\bm{k})=
        \begin{pmatrix}
            &0 &-iA_2 &-iJ_y&-i A_1 \\
            &iA_2^* &0 &-iJ_z &iJ_y  \\
            &iJ_y  &iJ_z  &0 &-iA_3 \\
            &i A_1^* &-iJ_y &iA_3^* &0
        \end{pmatrix}, 
\end{align}
where $A_1=J_z e^{-2\pi i k_1}$,  $A_2=J_x e^{-2\pi i k_2}$, and $A_3=J_xe^{-2\pi i k_3}$.
The original KSL exhibits Majorana FSs, which sit at the corners of the Brillouin zone (BZ), see Fig.~\ref{fig:10_3A_BZ_perturbations} (a). 
FSs are inherently stable against generic perturbations that are quadratic in the Majorana, which can only deform the FS \cite{hermanns2014quantum}. 
Thus, no infinitesimal perturbation can gap it. \footnote{Quartic perturbations, on the other hand, can gap the Majorana FS of the isotropic system to nodal lines, which in turn are protected by TR \cite{hermanns2015spin}.} 

In addition to the general stability of FSs, the Majorana FS of the KSL possesses an additional topological protection.  
As shown in \cite{bradlyn2016beyond}, the combination of 3-fold rotation and 2-fold screw rotation protects a threefold degeneracy at the high-symmetry point $P$. 
This degeneracy is linked to a non-zero Chern number of $\pm 2$, and thus, turns the surrounding Majorana FS into a topological FS.  
Consequently, the KSL on the hyperoctagon lattice cannot be gapped at all by (quadratic) perturbations that retain threefold rotation symmetry. 
Since the unperturbed SOL consists of three identical Majorana flavors, each flavor contributes an identical copy of the FS structure, leading to a threefold degeneracy throughout the band structure, a 9-fold degeneracy at the P point and a total charge of the FS of $3\times 2=6$, see Figure~\ref{fig:10_3A_BZ_perturbations} (b).

\subsubsection{Effect of perturbations}\label{subsubsec:perturbed_Hyperoctagon}
    
The topological charge of the FS is stable against any perturbation: while infinitesimal perturbations (even those that obey threefold rotation symmetry) split the 9-fold degeneracy at the P point, the topological charge of the FS only changes when WPs move into or out of the surface. 
This requires a finite perturbation strength.

\begin{figure}
    \centering
    \includegraphics[width=\linewidth]{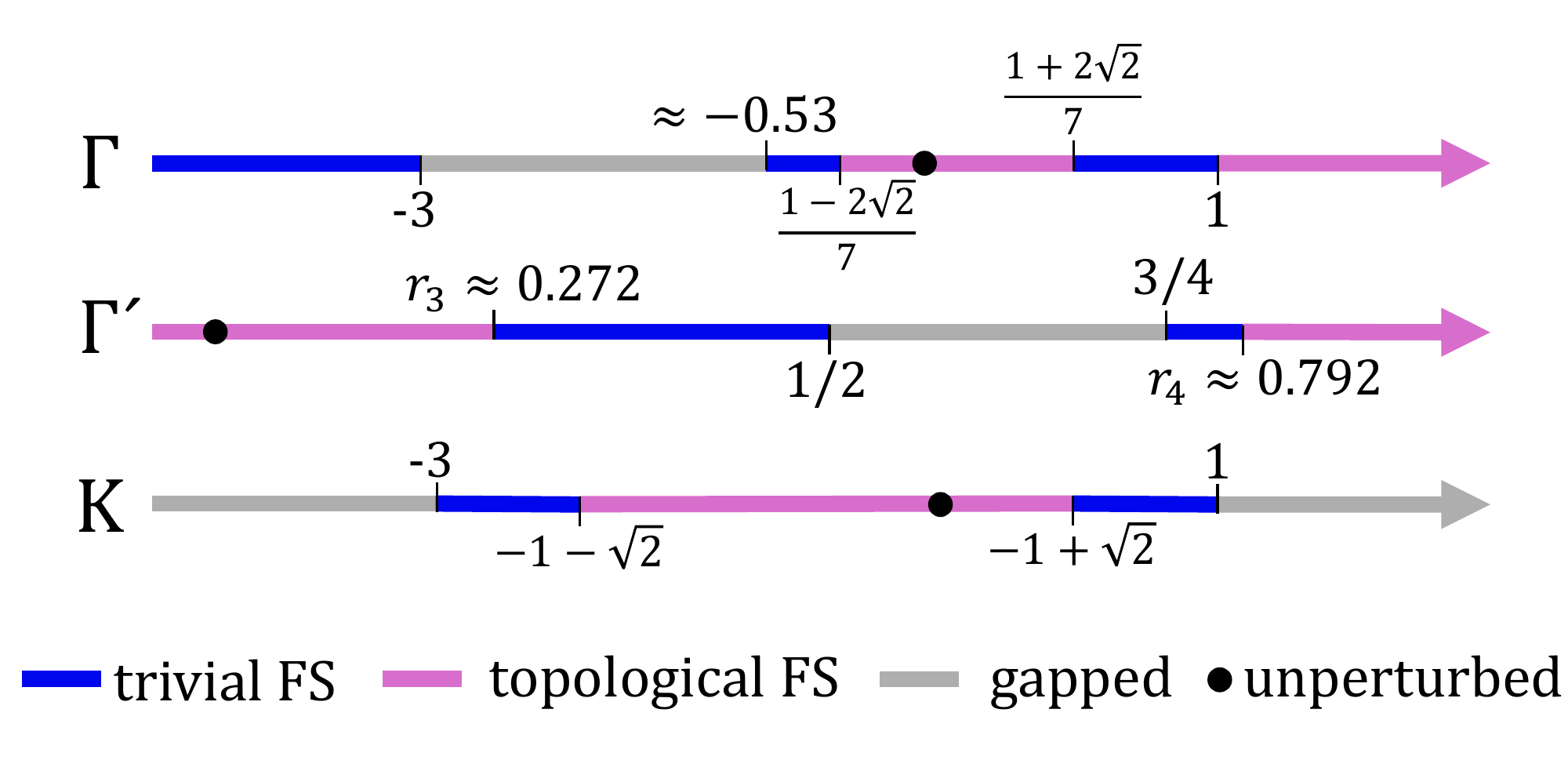}
    \caption{Phase diagram for NN perturbations on the hyperoctagon lattice, where only one of the TR invariant perturbations $\Gamma$, $\Gamma'$, and $K$ is changed.   $r_j$ denotes the $j$th root of $(1-12\Gamma'^2-8\Gamma'^3+4\Gamma'^4+16\Gamma'^5+16\Gamma'^6)$.}
    \label{fig:hyperoctagon_phase_diagram}
\end{figure}

There are three different scenarios for quadratic perturbations: (i) the system retains the topological FSs, (ii) the system retains the FSs, but they become topologically trivial, or (iii) the system becomes gapped. 
Note that (iii) can only take place if the FSs have already become trivial, i.e., there is no direct transition from a phase with topological FSs to a gapped one. 
All three scenarios are realized in extended regions in the phase diagram. 
Since the number of perturbations is rather large, we do not aim to compute a full phase diagram, but limit ourselves to phase diagrams where only one of the parameters is changed at the time. 
The NNN TR-breaking perturbation merely deforms the FSs without being able to affect the degeneracy or the topological charge. 
The onsite TR-breaking perturbation acts as a chemical potential---shrinking one of the FSs, expanding another, and keeping the third unchanged. 
It thus results in nested top. FSs---i.e., FSs sitting within each other---each with a topological charge of 2.
The TR-preserving perturbations, $\Gamma$, $\Gamma'$, and $K$, act non-trivially on the Majorana flavors and, thus, can induce phase transitions between the different scenarios described above. 
The phase diagrams for changing only one parameter, while keeping the others fixed to $0$, are shown in Fig.\ref{fig:hyperoctagon_phase_diagram}. 
Already, this simple analysis makes it clear that perturbing the system leads to a very rich phase diagram where all three scenarios are realized for a variety of parameter regimes.

For certain cases, we find an additional surprising feature, namely a persistent 6-fold degeneracy at the P point. 
In particular, this is the case for ($\Gamma\neq 0$, $\text{K}=\Gamma'=0$), (K$\neq 0$, $\Gamma=\Gamma'=0$) and for ($\Gamma=\Gamma'\neq 0$, K$=0$).  Figure~\ref{fig:10_3A_BZ_perturbations} showcases the first two of the scenarios: the 9-fold degeneracy has only split to two distinct energies for (c) and (d), while for (e) it is split fully into three threefold degenerate crossings. 
Note that the 6-fold degeneracy occurs for the lowest 6 bands for $K>0$ in (c), whereas it occurs for bands 3-9 for $\Gamma>0$ in (d). 

The case  ($\Gamma=\Gamma'\neq 0$, K$=0$) can be explained by the special rotation discussed in Sec.~\ref{subsubsec:TRIperturbations} that maps the $\Gamma=\Gamma'$ model to an unperturbed model with rescaled coupling constants for the $x/y$ flavor on the one hand, and the $z$ flavor on the other hand. 
Consequently, two of the three flavors are degenerate throughout the BZ, thus leading to a 6-fold degeneracy at the $P$ point. 
For the other two examples, we do not have such a simple explanation. 

In contrast to the threefold degeneracy discussed earlier, the 6-fold degeneracy is not associated with a symmetry, but only occurs on fine-tuned lines. 
Any infinitesimal perturbation immediately splits the 6-fold degeneracy into two threefold ones.
It also has no protected topological charge. 
Rather, it should be interpreted as the sum of the charges of the two threefold degeneracies that are fine-tuned to occur at the same energy. 
As the latter can have charges $\pm 2$, the combined object can have charges $2+2=4$, $2-2=0$ or $-2-2=-4$.

\subsection{Hyperhoneycomb}\label{subsec:Hyperhoneycomb}

	\begin{figure}[t]
        \centering
        \includegraphics[width=0.85\linewidth]{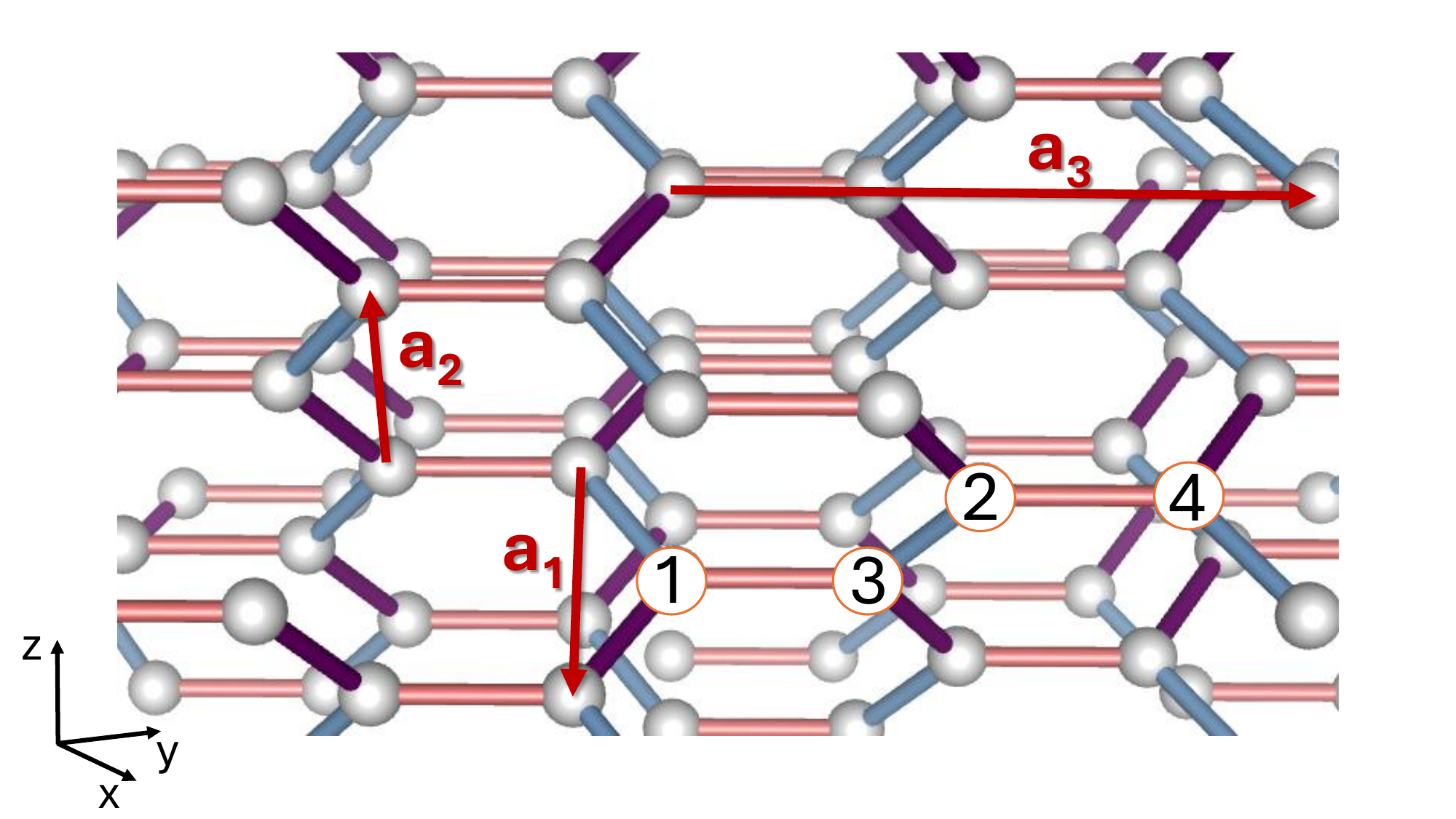}
        \caption{Hyperhoneycomb lattice with unit cell and translation vectors. Colors blue/purple/pink denote  $x$/$y$/$z$-bonds.}
        \label{fig:10_3B_lattice}
    \end{figure}
    
	The Kitaev model on the (10,3)b --- hyperhoneycomb --- lattice was introduced in \cite{mandal2009exactly}, and has since then been studied in a large variety of works, the earliest among them being Refs.~\cite{kimchi2014three,lee2014heisenberg,nasu2014finite}. 
   	It can be visualized as $xy$-zigzag chains that are coupled by the $z$ bonds.  
    It plays a central role in the experimental realization of Kitaev physics, as the Iridium atoms in $\beta$-Li$_2$IrO$_3$ arrange in a hyperhoneycomb lattice---making this our most prominent candidate for realizing 3D Kitaev spin liquids in materials \cite{takayama2015hyperhoneycomb,takagi2019concept}.

\begin{figure*}[t]
        \centering
        \includegraphics[width=\linewidth]{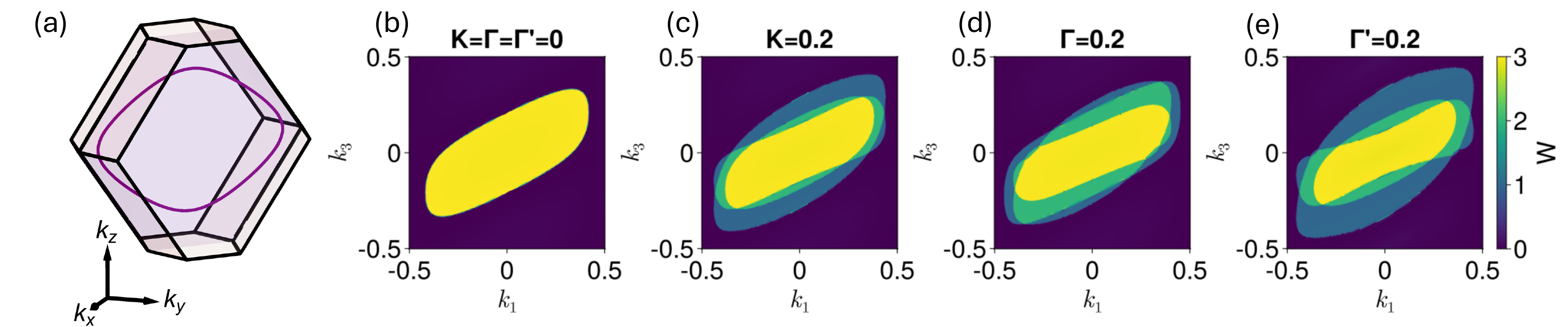}
        \caption{(a) BZ for the hyperhoneycomb lattice with the nodal lines of the unperturbed SOL. The nodal lines are threefold degenerate and lie in the $k_x+k_y=0$ plane. The remaining plots show the winding numbers, W, for the hyperhoneycomb lattice without perturbations (b), and for switching on one of the perturbations, (c) $K=0.2$, (d) $\Gamma=0.2$, and (e) $\Gamma'=0.2$.}
        \label{fig:10_3B_BZ_nodal_winding}
    \end{figure*}
	In the following, we will provide the necessary information about the lattice and the corresponding Hamiltonian to understand the results. 
	For a more thorough description of the lattice, its properties, and symmetries,  we refer the reader to the more in-depth discussion in Ref.~\cite{obrien_classification_2016}. 
	\subsubsection{The lattice and symmetries}
	We can describe the lattice by a four-site unit cell with site positions 
	\begin{align}
		\bf r_1 &= (0,0,0),&
		\bf r_2 &= (1,2,1),\nonumber\\
		\bf r_3 &= (1,1,0),&
		\bf r_4 &= (2,3,1)
	\end{align}
	with lattice translation vectors 
	\begin{align}
		\bf a_1 &=(-1,1,-2),& \bf a_2 &= (-1,1,2),& \bf a_3 &=(2,4,0). 
	\end{align}
	The lattice with our choice of unit cell and lattice translation vectors is illustrated in Fig.~\ref{fig:10_3B_lattice}. 
	The corresponding reciprocal lattice vectors read as 
	\begin{align}
		\bf q_1&=\left(-\frac{2\pi}{3},\frac{\pi}{3},-\frac{\pi}{2}\right),&
	    \bf q_2&=\left(-\frac{2\pi}{3},\frac{\pi}{3},\frac{\pi}{2}\right),\nonumber\\
		\bf q_3&=\left(\frac{\pi}{3},\frac{\pi}{3},0\right),  
	\end{align}
	and the Brillouin zone is visualized in Fig.~\ref{fig:10_3B_BZ_nodal_winding} (a). 
	Note that the (10,3)b lattice does not possess threefold rotation symmetry but only a 2-fold rotation around the z-bonds mapping $x$ and $y$ bonds onto each other.
   It also possesses inversion symmetry. 
   These symmetries strongly constrain the allowed nodal structures of the Majorana Hamiltonian.

\subsubsection{Unperturbed spin-orbital liquid}
For the unperturbed Hamiltonian, the SOL on the hyperhoneycomb lattice consists of three identical copies, \eqref{eq:10.3b_M}, of the corresponding Kitaev spin liquid---one for each itinerant Majorana flavor.
	Using the same gauge as in \cite{obrien_classification_2016}, the Hamiltonian for each Majorana flavor can be written as 
\begin{align}
    \begin{aligned}
        M(\bm{k})=
        \begin{pmatrix}
            &0 &0 &iJ_z &iA_{13} \\
            &0 &0 &iA_{2} &iJ_z \\
            &-iJ_z &-iA_{2}^* &0 &0 \\
            &-iA_{13}^* &-iJ_z &0 &0
        \end{pmatrix},
    \end{aligned}
    \label{eq:10.3b_M}
\end{align}
where $A_{13}=e^{-2\pi ik_3}(J_x+e^{2\pi ik_1}J_y)$, $A_{2}=J_x+J_y e^{2\pi i k_2}$.

The detailed analysis of the projective symmetries of the latter, see Ref.~\cite{obrien_classification_2016},  shows that the Kitaev spin liquid on (10,3)b  lattice hosts a nodal line, which (for $J_x = J_y$) is located on the $k_x+k_y=0$ plane, see Fig.~\ref{fig:10_3B_BZ_nodal_winding}(a). 
Thus, the unperturbed SOL exhibits a threefold degenerate nodal line on this high-symmetry plane, due to the emergent $\mathrm{SO}(3)$ flavor symmetry. 
We can most easily identify such nodal lines by computing the winding number---a $\mathbb{Z}$ topological invariant for systems with chiral symmetry in odd dimensions.
Fig.~\ref{fig:10_3B_BZ_nodal_winding}(b) shows the winding number for the isotropic, unperturbed SOL to be equal to $W= 3$ inside the nodal line, as expected from the $\mathrm{SO}(3)$ flavor symmetry. 

As long as TR symmetry is preserved, the nodal lines are stable against any (infinitesimal) symmetry-allowed quadratic perturbation. 
If one of the Ising couplings becomes dominant, the nodal lines shrink and disappear when  $J_z>J_x+J_y$(or cyclic permutations of $x$, $y$, $z$). 
Breaking TR, on the other hand,  immediately gaps the nodal line into pairs of WPs. 
The latter are pinned to $E=0$ as long as inversion symmetry is preserved, and they are stable against small symmetry-preserving perturbations:  they can only be removed by annihilation with opposite chirality, which requires a finite perturbation strength.

\subsubsection*{Effect of perturbations}
\begin{figure}[t]
        \centering
        \includegraphics[width=\linewidth]{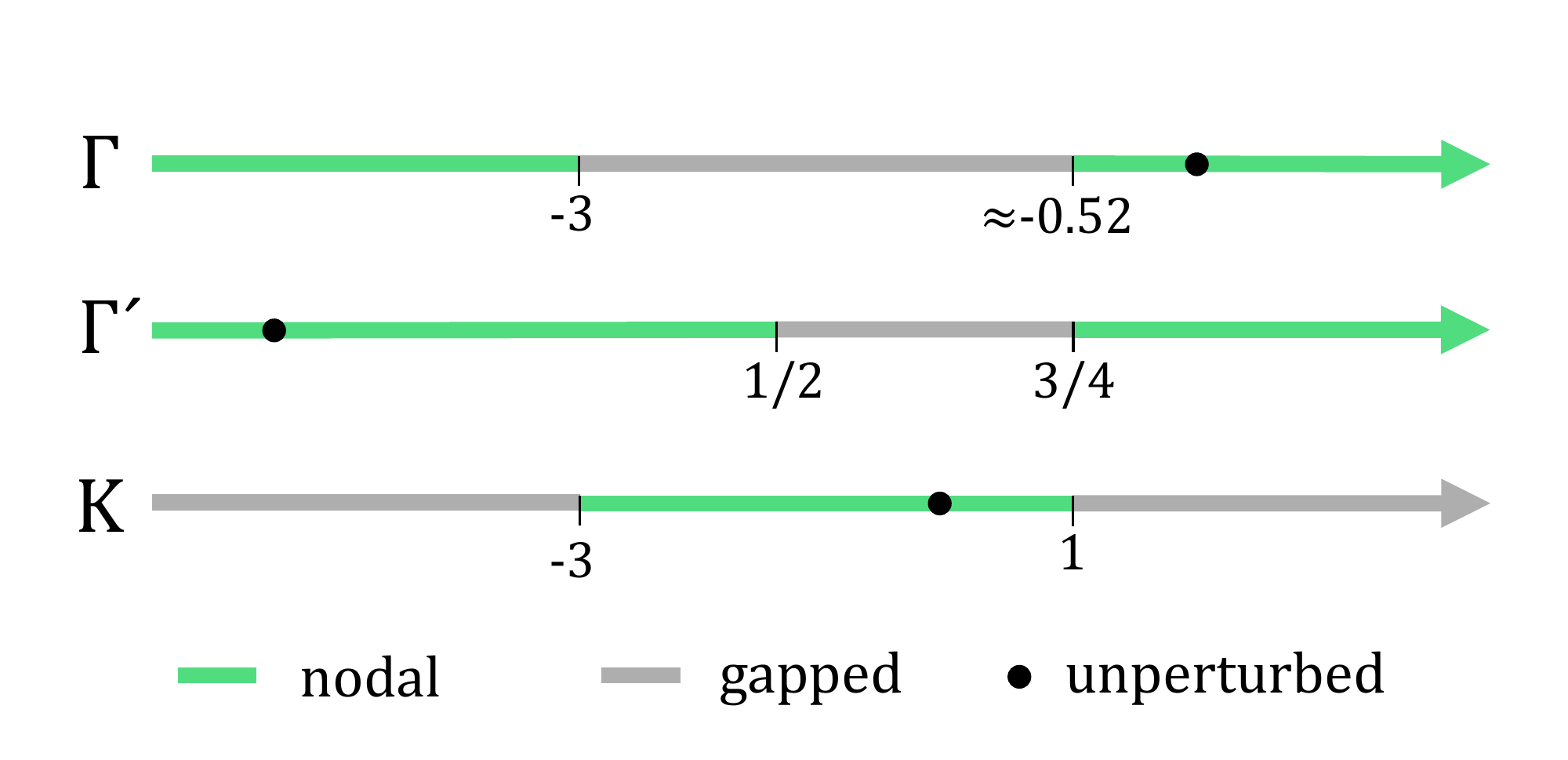}
        \caption{Phase diagram of NN perturbations on the hyperhoneycomb  lattice,  where only one of the TR invariant perturbations $\Gamma$, $\Gamma'$, and $K$ is changed. }
        \label{fig:10_3B_lattice_phase_diagram}
\end{figure}
We now discuss how the perturbations introduced in Sec.~\ref{subsec:pertubations} modify the Majorana band structure. For simplicity,  we restrict ourselves to the isotropic point, $J_x=J_y=J_z=1$. 

\paragraph{Time-reversal invariant perturbations}
The NN perturbations---$K$, $\Gamma$, and $\Gamma'$---preserve TR symmetry and therefore cannot immediately gap the nodal lines. 
Instead, they lift the threefold degeneracy by splitting the nodal line into multiple distinct ones. 
Generically, the threefold degenerate nodal line is split into three separate ones. 
Consequently, when computing the winding number, we can find regions for all possible values---0,1,2,3---as visualized in Fig.~\ref{fig:10_3B_BZ_nodal_winding} (c)-(e). 
Since none of the perturbations break the twofold rotation symmetry, even the split nodal lines remain in the $k_x+k_y=0$  plane.
Increasing $K$ causes the nodal lines to shrink; finally, for $K>1$ or $K<-3$, the system becomes fully gapped, as shown in Fig.~\ref{fig:10_3B_lattice_phase_diagram}. 
In contrast, the system continues to harbor nodal lines for large values of $|\Gamma|$ and $|\Gamma'|$, even though the system can be gapped in intermediate parameter regimes, see Fig.~\ref{fig:10_3B_lattice_phase_diagram}. 

\paragraph{Time-reversal-breaking perturbations}
As expected from the Kitaev spin liquid on the hyperhoneycomb lattice, breaking TR qualitatively changes the nodal structure. 
The NNN term $\kappa$ gaps the nodal line into pairs of WPs. 
For the unperturbed SOL, each Majorana flavor contributes a WP of charge $\pm 1$, thus in the absence of further perturbations the WPs carry charge $\pm 3$. 
This is, however, fine-tuned: generic NN perturbations will split these into multiple WPs of charge $\pm 1$, which can move independently from each other.  

\begin{figure}[t]
        \centering
        \includegraphics[width=0.7\linewidth]{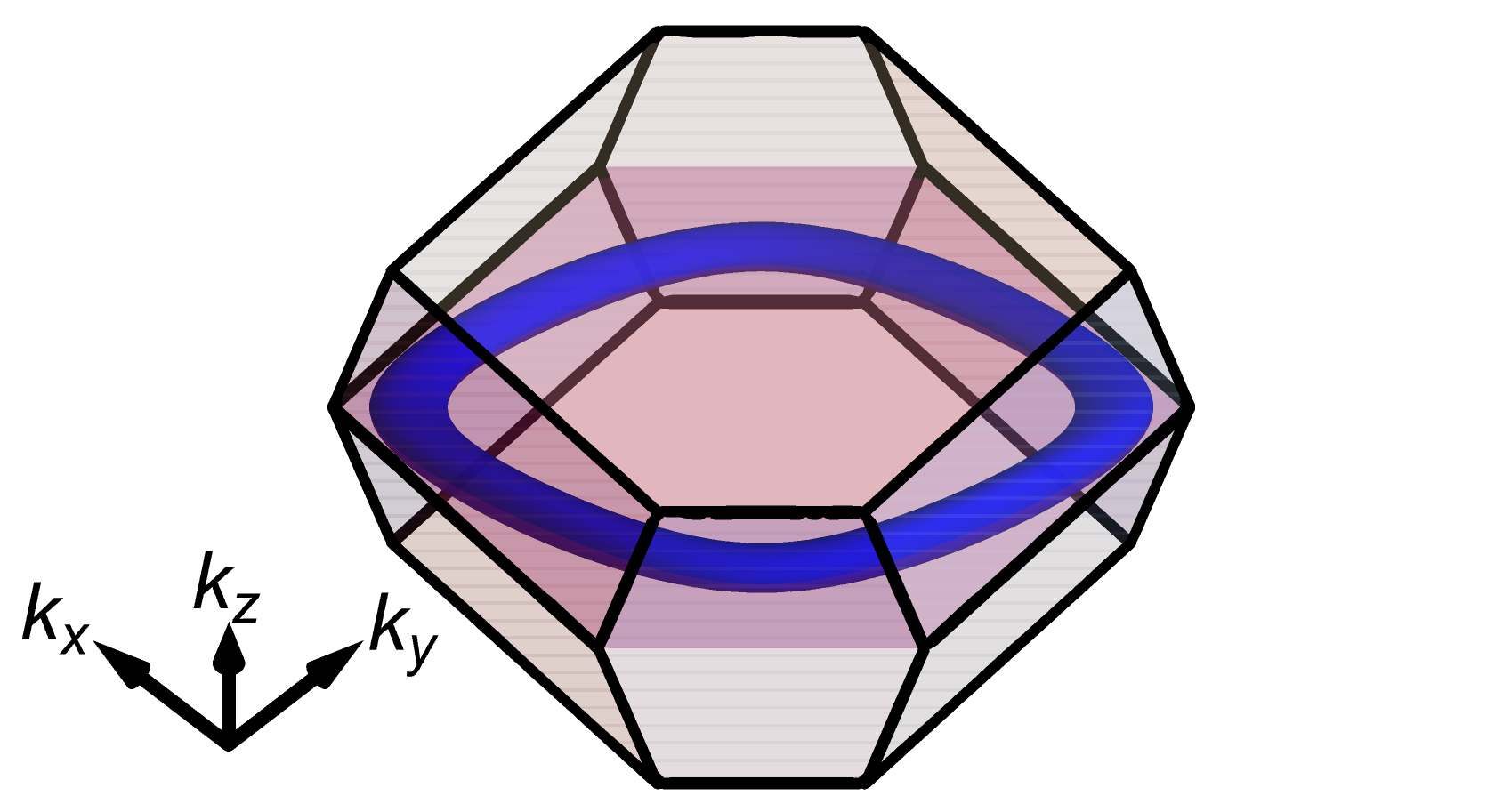}
        \caption{Adding the on-site perturbation $\mathbf{h}$ to the SOL on the hyperhoneycomb lattice results in a doubly degenerate tubular FS around the nodal line, here with $h_x=0.2$.}
        \label{fig:10_3B_tube}
\end{figure}

The on-site perturbation $\bf h$ does not gap the nodal line.
Instead, it acts as a `chemical potential' for the Majorana flavors---in complete analogy to the two-dimensional case discussed in \cite{chulliparambil_microscopic_2020}. 
One of the nodal lines will be shifted to positive energy $E$, another is moved to negative energy $E$, while the third remains at zero energy.  
This results in tubular FSs surrounding the nodal line, see Fig.~\ref{fig:10_3B_tube}.  
The resulting FSs are doubly degenerate due to inversion symmetry. 

\subsection{Hyperhexagon (8,3)b}\label{subsec:hyperhexagon}
\subsubsection{The lattice and symmetries}
\begin{figure}[t]
        \centering
        \includegraphics[width=0.85\linewidth]{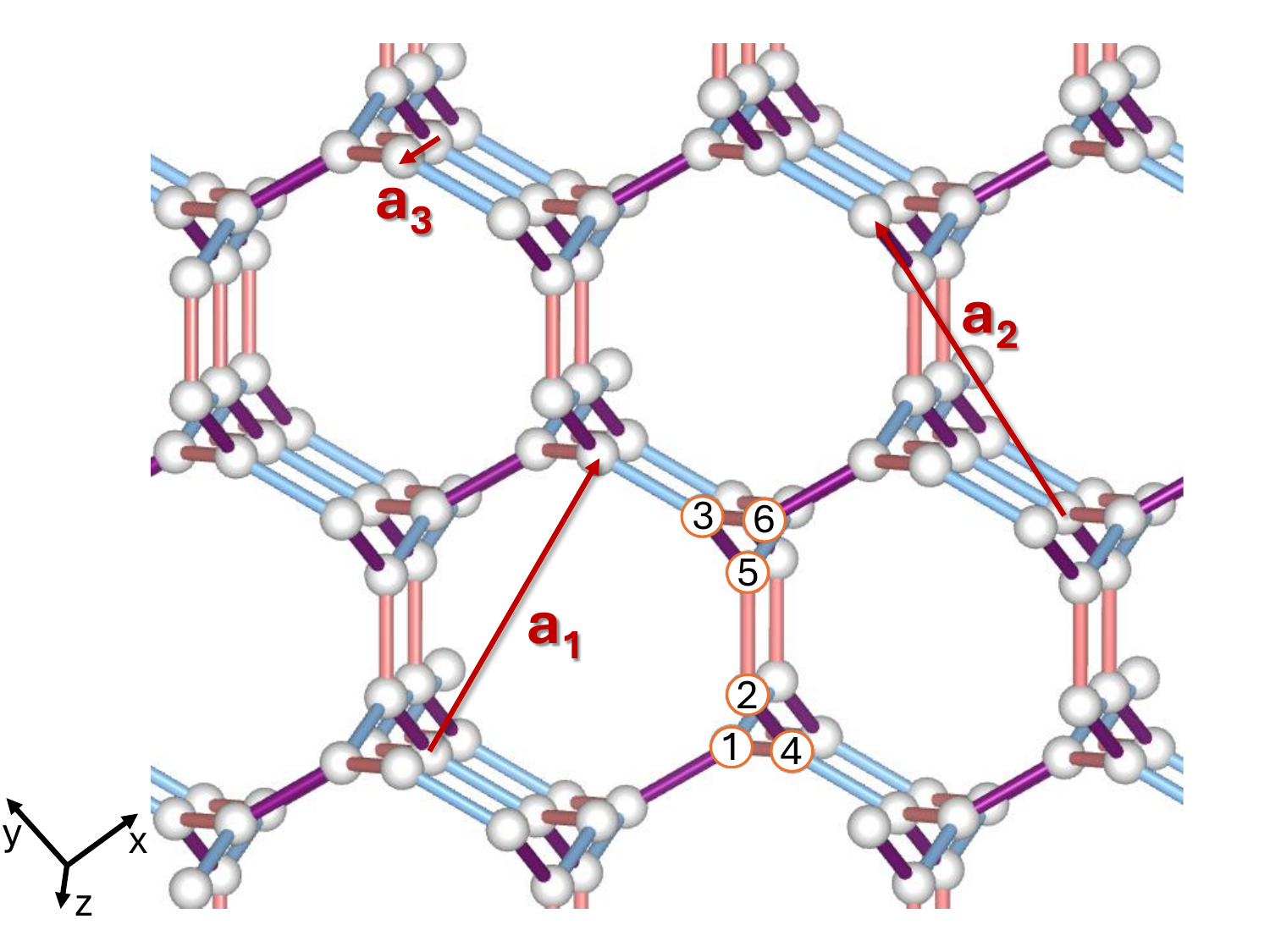}
        \caption{Unit cell and translation vectors for the Hyperhexagon lattice. Colors blue/purple/pink denote  $x$/$y$/$z$-bonds.}
        \label{fig:8_3b_lattice}
    \end{figure}

The (8,3)b or hyperhexagon lattice is our final example of SOLs on three-coordinated lattices. 
It can be visualized as a honeycomb lattice, where the original lattice sites are replaced by counter-rotating triangular spirals, see Fig.~\ref{fig:8_3b_lattice}. 
    It has a six-site unit cell, with sites located at \cite{obrien_classification_2016}
\begin{align}
    \mathbf{r}_1& =\left(\frac{1}{10},\frac{1}{2\sqrt{3}},\frac{1}{5}\sqrt{\frac{2}{3}} \right),& 
    \mathbf{r}_2 &=\left(\frac{1}{5},\frac{\sqrt{3}}{5},\frac{\sqrt{6}}{5} \right),\nonumber\\ 
    \mathbf{r}_3 &=\left(\frac{3}{10},\frac{11}{10\sqrt{3}},\frac{4}{5}\sqrt{\frac{2}{3}} \right), &
    \mathbf{r}_4 &=\left(\frac{1}{5},\frac{2}{5\sqrt{3}},\frac{2}{5} \sqrt{\frac{2}{3}}\right), \nonumber\\
    \mathbf{r}_5 &=\left(\frac{3}{10},\frac{3\sqrt{3}}{10},\frac{\sqrt{6}}{5} \right),& \mathbf{r}_6 &=\left(\frac{2}{5},\frac{1}{\sqrt{3}},\sqrt{\frac{2}{3}} \right),
\end{align}
and lattice translation vectors given by 
\begin{align}
    \mathbf{a}_1&=\left(\frac{1}{2},\frac{1}{2\sqrt{3}},\frac{1}{5}\sqrt{\frac{2}{3}}\right),& \mathbf{a}_2&=\left(0,\frac{1}{\sqrt{3}},\frac{2}{5}\sqrt{\frac{2}{3}}\right), \nonumber\\
    \mathbf{a}_3&=\left(0,0,\frac{\sqrt{6}}{5} \right),   
\end{align}
The corresponding reciprocal lattice vectors are 
\begin{align}
    \mathbf{q}_1&=(4\pi,0,0), & \mathbf{q}_2&=(-2\pi,2\sqrt{3}\pi,0),\nonumber\\ 
    \mathbf{q}_3&=\left(0,-\frac{4\pi}{\sqrt{3}},5\sqrt{\frac{2}{3}}\pi\right).
\end{align}
The lattice possesses inversion symmetry as well as threefold rotation symmetry that cyclically permutes the $x$-, $y$-, and $z$-bonds.
As a consequence, we restrict all Ising couplings to be equal.
\footnote{Note that the lattice has two types of bonds that are not connected by any symmetry---those constituting the spirals and those connecting them. Without loss of generality, we will set the coupling constants identical for both types. }

\begin{figure}[t]
	\centering
	\includegraphics[width=\linewidth]{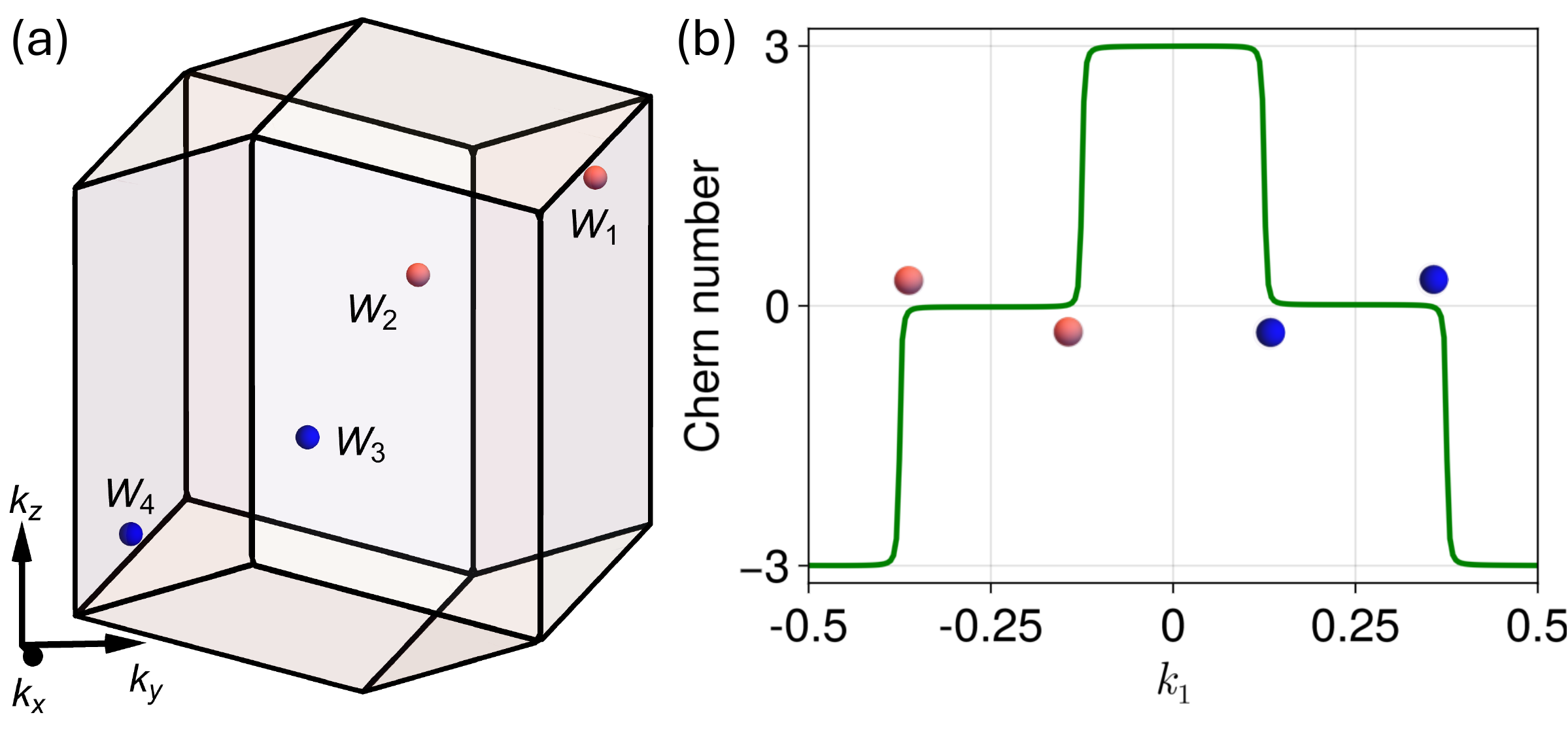}
	\caption{(a) WP positions in the BZ for the unperturbed SOL on the hyperhexagon lattice. The negatively charged WP are depicted in blue, while the positive ones are pink, with the corresponding Chern numbers shown in (b). 
	}
	\label{fig:8_3b_BZ}
\end{figure}

\begin{figure*}[t!]
	\centering
	\includegraphics[width=\linewidth]{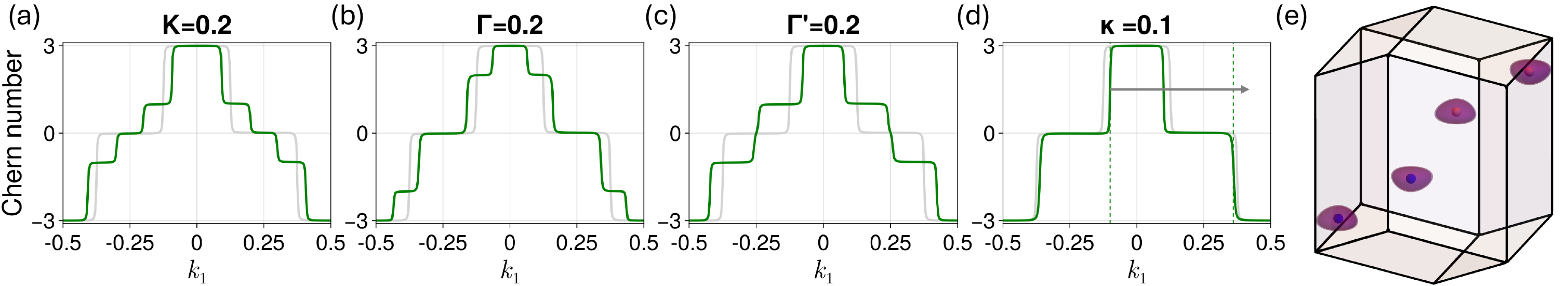}
	\caption{(a)-(d) Chern numbers for different perturbations on a SOL on the hyperhexagon lattice, with isotropic couplings J. The green line shows the Chern numbers of the perturbed model, while the unpertubed SOL is shown in gray.   (e) The on-site perturbation $\bm{h}$ splits the threefold degeneracy of the SOL on the hyperhexagon lattice, resulting in doubly degenerate topological FSs around the original WPs. The FSs in the figure are for $h_x=0.2$.
	}
	\label{fig:8_3b_perturbations}
\end{figure*}
\subsubsection{Unperturbed spin-orbital liquid}
For the unperturbed Hamiltonian, each Majorana flavor realizes the same band structure as in the Kitaev model on the hyperhexagon lattice.
Using the same conventions as in \cite{obrien_classification_2016}, the Hamiltonian describing a single flavor of the SOL on the hyperhexagon lattice is given by 
\begin{align}
    \begin{aligned}
        M(\bm{k})= i
        \begin{pmatrix}
            &0 &-A_3 &0 &J_z &0 &-A_{13} \\
            &A_3^* &0 &0 &- J_y  &J_z  &0 \\
            &0 &0 &0 &-A_2 &-J_y &J_z  \\
            &-J_z  &J_y  &A_2^* &0 &0 &0 \\
            &0 &-J_z &J_y  &0 &0 &A_3 \\
            &A_{13}^* &0 &-J_z  &0 &-A_3^* &0
        \end{pmatrix},
    \end{aligned}
\end{align}
where $A_{3}=J_{x} e^{-2\pi i k_{3}}$, $A_{13}=J_{y} e^{-2\pi i(k_1+k_3) }$, $A_{2}=J_{x} e^{2\pi i k_2}$.
Since lattice translations connect the A to the B sublattice, TR symmetry is implemented projectively with a non-vanishing translation in momentum space: relating states at $\mathbf{k}$ with their TR partners at $-\mathbf{k}+\mathbf{k_0}$, where $\mathbf{k_0}=(\mathbf{q}_1+\mathbf{q}_3)/2$.
Such a scenario usually leads to stable FSs, were it not for the fact that the hyperhexagon also possesses inversion symmetry that is implemented trivially, i.e., relates states at $\mathbf{k}$ to those at $-\mathbf{k}$. 
Together with particle-hole symmetry, it enforces that every energy $E(\mathbf{k})$ has a partner $-E(\mathbf{k})$, a symmetry that prohibits stable FSs. 
As a consequence, the only stable zero-energy states that the system can harbor are WPs. 
As long as TR is unbroken, these have to occur in multiples of four---each pair separated by $\mathbf{k}_0$ in momentum space. 
In the SOL, this structure is triplicated.
Consequently, for the unperturbed SOL, we find WPs of charge  $+3$ at 
$W_1=(5/8)\mathbf{q}_1+(3/4)\mathbf{q}_2+(3/8)\mathbf{q}_3$ and $W_2=-W_1+\mathbf{k_0}$. Their negatively charged partners are located at $W_4=-W_1$ and $W_3=-W_2$,  see Fig.~\ref{fig:8_3b_BZ}

\subsubsection{Effect of perturbations}

\paragraph{Time-reversal invariant perturbations}
The NN perturbations break the $\mathrm{SO}(3)$ flavor symmetry, thus allowing the charge 3 WPs to split into three separate ones that can move through the Brillouin zone individually.
This is the generic behavior that occurs when switching on $K$, $\Gamma$ and $\Gamma'$, visible in the Chern number plot \ref{fig:8_3b_perturbations}(a)-(c). 
The original jump of 3 in the Chern number is broken down to jumps of 1 and 2, reflecting the split of a 3-fold degenerate WP into three individual ones. 
The jumps of 2 does not imply that there is a residual degeneracy, but occurs since  the three split WPs are related to each other by the 3-fold rotation symmetry. 
\footnote{We note that in certain parameter regimes---in particular for $\Gamma'<0.5$, $K=\Gamma=0$---the system can exhibit ring-like low-energy features that, however, \emph{are not} nodal lines. 
Rather, the low-energy features originate from the presence of 6 chargeless zero-modes on the ring, with a very flat, but non-zero dispersion connecting them. }
Only at the fine-tuned points discussed in Sec.~\ref{subsec:TRIgeneral}, e.g., $\Gamma=\Gamma'=1/2$,  does the system deviate from the generic behavior, due to the appearance of doubly degenerate zero-energy bands.  

In order to gap the system, WPs of opposite charge need to recombine, which requires a finite perturbation strength. 
When considering only the $K$ perturbation, the Weyl spin liquid phase persists in the region $-2.52< K < 0.52$, while it is gapped outside. 
For $\Gamma$ and $\Gamma'$,  even large perturbations stabilize a Weyl spin liquid phase, even though the system can become gapped in an intermediate regime---as happens, e.g., for $-3.64<\Gamma<-.318$.

\paragraph{Time-reversal-breaking perturbations}
The NNN perturbations break the perfect nesting condition, i.e., the positions of WPs 1 and 2 (or 3 and 4) are no longer related by $\mathbf{k}_0$,  but can move independently of each other. 
In the presence of both NN and NNN perturbations, all WPs can move individually, apart from the trivial particle-hole doubling, which states that each WP at $\mathbf{k}$ has an oppositely charged partner at $-\mathbf{k}$. 
In the Chern number plot in Fig.~\ref{fig:8_3b_perturbations}(d),  WPs of opposite charge are no longer related by the vector $\mathbf{k}_0$, indicated by the gray arrow. 

The on-site term $\bm h$ acts as a chemical potential for the WPs, in complete analogy to the hyperhoneycomb lattice discussed previously. 
It splits the threefold degeneracy by shifting one of the zero-energy modes to positive energy $E$, another to negative energy $E$, while the third remains at zero. 
This happens for each of the WP positions, thus the system exhibits doubly degenerate topological FSs around the original WP positions, see Fig.~\ref{fig:8_3b_perturbations}(e). 
The double degeneracy is protected by inversion symmetry.

	\section{Four-coordinated lattices}\label{sec:Fourcoordinated}
\begin{table*}[t]
\centering
\caption{Overview table of possible SOL behavior on the four-coordinated lattices under perturbations. }
\label{tab:summary_fourcoord_tight}
\begin{tabular}{p{3.6cm}|p{2.9cm}|p{3.2cm}|p{3.2cm}|p{4.3cm}}
\hline\hline
\textbf{Lattice} &
\textbf{Unperturbed} &
\textbf{TR-preserving (\(\bar\Gamma\))} &
\textbf{TR-breaking (\(h_z\))} &
\textbf{TR-breaking (\(\kappa\))} \\
\hline
Chiral square-octagon &
Top. FS &
Top. FS&
Nested top. FS&
Deformed top. FS \\
\hline
Layered honeycomb &
Twofold degenerate nodal line &
Split nodal lines &
Tubular FS  &
 WPs (if mirror symmetry is broken) \\
\hline
\hline
\end{tabular}
\end{table*}

We now turn to the discussion of the SOL on four-coordinated lattices, focusing on the $\nu=2$  realization with two itinerant Majorana flavors. In contrast to the three-coordinated case---where the unperturbed SOL can be viewed as multiple decoupled copies of a Kitaev spin liquid---four-coordinated lattices provide a genuinely distinct setting in which the flavor structure is essential.  The interplay of lattice symmetries with a minimal flavor sector imposes distinct constraints on how nodal manifolds can split, shift, and recombine under solvable perturbations.
By analyzing two representative four-coordinated lattices, we provide a controlled characterization of these constraints and use them to extend the classification of SOLs.
An overview of the possible behaviors under perturbations is provided in Table~\ref{tab:summary_fourcoord_tight}. 

	\subsection{General comments}\label{subsec:Fourcoordinated_general}

     \subsubsection{Perturbations}

For four-coordinated lattices, the most direct analogue of the TR-breaking NNN term takes the form
\begin{equation} \label{eq:NNN_fourcoord}
    \tilde{\mathcal H}_\kappa^{(2)}=\kappa\sum_{\langle ijk\rangle_{\alpha\beta}}\epsilon^{\alpha\beta\gamma\delta} u_{ij}^\alpha u_{jk}^{\beta} \left(   i c_i^x c_k^x + i c_i^y c_k^y \right),
\end{equation}
where $\gamma$ and $\delta$ denote the two bond types that are not traversed by the path $\langle ijk \rangle$, always ordered so that $\gamma<\delta$. 
The totally antisymmetric Levi--Civita tensor $\epsilon^{\alpha\beta\gamma\delta}$ is 
defined as usual on the ordered set $(x,y,z,w)$:
\begin{align}
    \epsilon^{\alpha\beta\gamma\delta} =
    \begin{cases}
        +1 & \text{for even permutations of } (x,y,z,w),\\
        -1 & \text{for odd permutations of } (x,y,z,w),\\
        0  & \text{otherwise}.
    \end{cases}
\end{align}
This perturbation is chosen to be identical for both flavors and, thus, preserves $\mathrm{SO}(2)$ symmetry. 
Alternative implementations---e.g., with a relative minus sign between the flavors as used in \cite{ryu_three-dimensional_2009}---lead to the same qualitative physics and will not be discussed further. 

The remaining quadratic perturbations break $\mathrm{SO}(2)$ symmetry. 
The on-site term 
\begin{align}
\mathcal H^{(2)}_h&= -h_z \sum_{j} \sigma^z_j\otimes \textbf{1},\nonumber\\
\tilde{\mathcal H}^{(2)}_h&= h_z \sum_{j} i c_j^x c_j^y,
\end{align}
acts, in complete analogy to the three-coordinated case, as a chemical potential, while the nearest neighbor term 
\begin{align} \label{eq:fourcoord_GammaBar}
   \mathcal H^{(2)}_{\bar \Gamma}&= \sum_{\langle ij\rangle_\gamma}\bar \Gamma_\gamma (\sigma^x_i\sigma^y_j+\sigma^y_i \sigma^x_j)\otimes\tau_i^\gamma\tau_j^\gamma,\nonumber\\
     \tilde{\mathcal H}^{(2)}_{\bar \Gamma} &=\sum_{\langle ij\rangle_\gamma}-i u_{ij}^\gamma \bar \Gamma_\gamma (c^x_ic^y_j+c^y_ic^x_j),
\end{align}
mixes flavors. 
We can bring the $\bar \Gamma$ term into a simpler form by considering a  rotation in the spin space given by 
\begin{align}
    \sigma^x\rightarrow & \left(\tilde \sigma^x + \tilde \sigma^y\right)/\sqrt{2},\nonumber\\
    \sigma^y\rightarrow & \left(\tilde \sigma^y - \tilde \sigma^x\right)/\sqrt{2},
\end{align}
which in flavor space becomes 
\begin{align}
    (c_i^x, c_i^y)\rightarrow \frac{1}{\sqrt{2}}(c_i^x + c_i^y, -c_i^x + c_i^y). 
\end{align}
In terms of the rotated spins/Majoranas, the perturbation reads 
\begin{align}\label{eq:four_rotated_Gamma}
   \mathcal H^{(2)}_{\bar \Gamma}&= \sum_{\langle ij\rangle_\gamma}\bar \Gamma_\gamma \left(\tilde\sigma^x_i\tilde\sigma^x_j-\tilde\sigma^y_i\tilde\sigma^y_j\right)\otimes\tau_i^\gamma\tau_j^\gamma,\nonumber\\
   \tilde{\mathcal H}^{(2)}_{\bar \Gamma} &=\sum_{\langle ij\rangle_\gamma}-i u_{ij}^\gamma \bar \Gamma_\gamma (\tilde c^x_i\tilde c^x_j-\tilde c^y_i\tilde c^y_j), 
\end{align}
i.e., it has become diagonal in the rotated flavor index and shifts the Kitaev coupling on bond $\gamma$ from $J_\gamma$ to  $J_\gamma-\bar \Gamma_\gamma$ for the new $x$ flavor, while the coupling for the new $y$ flavor becomes $J_\gamma\rightarrow J_\gamma+\bar \Gamma_\gamma$. 
Both TR-breaking perturbations $\kappa$ and $h_z$ are diagonal in flavor space, and unaffected by the spin rotation. 
More specifically, $\kappa$ stays diagonal in the rotated flavors, while the $h_z$ perturbation remains off-diagonal, still mixing the rotated flavors.

\subsubsection{Structure of perturbed Hamiltonian in momentum space}
	The structure of the perturbed Hamiltonian for a four-coordinated lattice is similar to a three-coordinated lattice, if not simpler, as we now have two instead of three itinerant Majoranas. 
    As we did for the three-coordinated lattices, we can apply a Fourier transform and combine all the Majoranas into a single spinor $\psi$ 
\begin{align}
    \psi(\mathbf{k}) &= (c_{\mathbf{k},1}^x, \ldots,c_{\mathbf{k},d}^x,c_{\mathbf{k},1}^y, \ldots, c_{\mathbf{k},d}^y)
\end{align}
where $d$ denotes the size of the unit cell for a given lattice (the lattices considered here all have $d=4$).  The general matrix representation of the perturbed SOL is then
\begin{align}
		\tilde{\mathcal{H}}&=\sum_{\mathbf{k}}
		\psi(-\mathbf{k})^\dagger 
		H(\mathbf{k})
	\psi(\mathbf{k})\nonumber
\end{align}
which can be expressed in terms of $d\times d$ matrices $M_{\alpha\beta}$ as 
\begin{align} \label{eq:perturbedH_fourcoord}
	\begin{aligned}
		H(\mathbf{k})=
		\begin{pmatrix}
			&M_{xx}(\mathbf{k}) &M_{xy}(\mathbf{k}) \\
			&M_{xy}^\dagger(\mathbf{k}) &M_{yy}(\mathbf{k})  \\
		\end{pmatrix}.
	\end{aligned}
\end{align}
The matrices on the diagonal are identical, $M_{xx}(\mathbf{k})=M_{yy}(\mathbf{k})$, and describe just the $\mathrm{SO}(2)$ invariant Majorana Hamiltonian. The NN coupling and the on-site TR breaking term mix the two Majorana flavors and thus appear in the off-diagonal block $M_{xy}(\mathbf{k})$. 

We can bring this Hamiltonian into a mostly diagonal form by employing the spin rotation discussed above. 
In matrix form, this transformation can be written as
\begin{align}\label{eq:block_trafo}
    U=\frac{1}{\sqrt{2}}\begin{pmatrix}
        &I_4 &I_4\\
        &-I_4 &I_4
    \end{pmatrix},
\end{align}
where $I_4$ is a $4\times 4$ identity matrix. 
After employing this transformation, only the on-site term $h_z$ remains off-diagonal.

\subsection{Chiral square-octagon}\label{subsec:chiralSquareOctagon}

\begin{figure}[t!]
\centering
    \includegraphics[width=\linewidth]{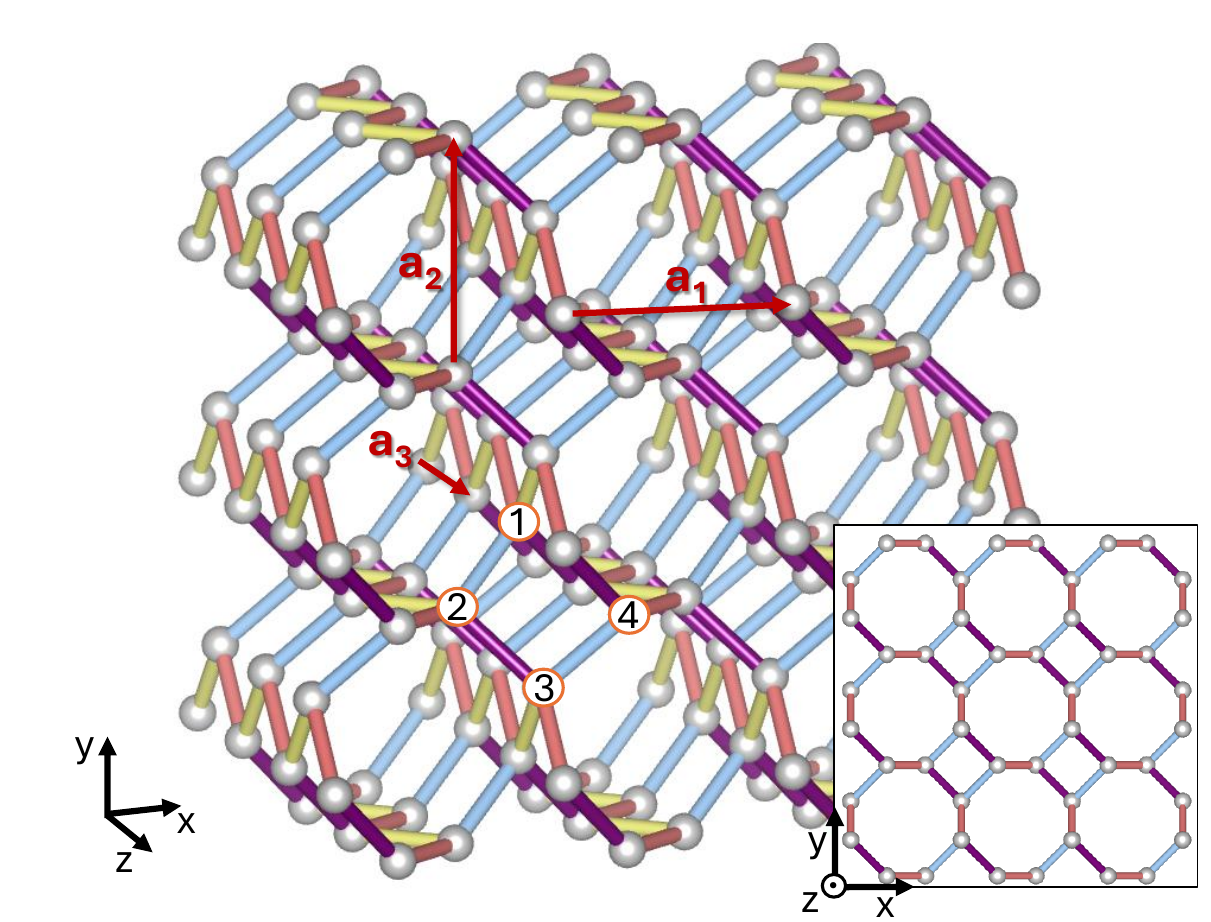}
    \caption{Chiral square-octagon lattice with the unit cell and translation vectors. Colors blue/purple/pink/yellow denote $x$/$y$/$z$/$w$ bonds.}
    \label{fig:chiral_square-octagon_lattice}
\end{figure}

    \subsubsection{The lattice and symmetries}
\begin{figure}[t]
    \centering
    \includegraphics[width=\linewidth]{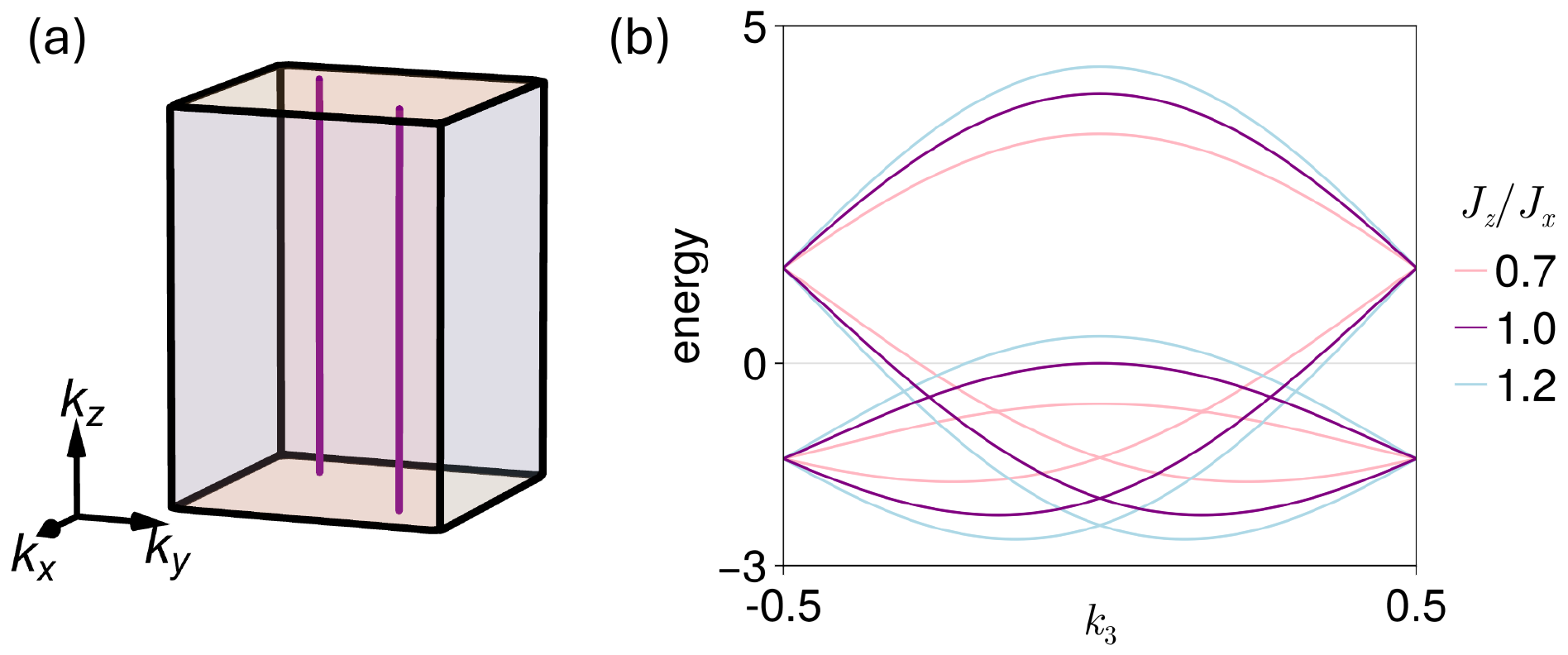}
    \caption{(a) BZ for the chiral square-octagon lattice, marked are also the $(1/4,1/4,k_3)$ and $(-1/4,-1/4,k_3)$ lines. (b) Energy spectrum along the $(-1/4,-1/4,k_3)$ line for different coupling ratios $f=J_z/J_x$ ($J_y=J_x$ and $J_w=J_z$).
    }
    \label{fig:ChSqOctagon_BZ_spectrum}
\end{figure}

    The chiral square-octagon can be visualized by co-rotating square spirals that are connected by zig-zag chains, see Fig.~\ref{fig:chiral_square-octagon_lattice}. 
    In its most symmetric form, its space group is 91 with Wickoff position 4a; see, e.g., the discussion on net 5 in Ref.~\cite{okeefe1992uninodal}. 
    It has a four-site unit cell, with sites located at 
\begin{align}
    \mathbf{r}_1&=(0,a,0),& \mathbf{r}_2&=(-a,0,\frac{1}{4\sqrt{2}}),\nonumber\\
     \mathbf{r}_3&=(0,-a,\frac{1}{2\sqrt{2}}),& \mathbf{r}_4&=(a,0,\frac{3}{4\sqrt{2}}),
\end{align}
where $a=\frac{1}{8}(8-\sqrt{29})$ ensures that all bonds are of equal length.  The lattice vectors are given by 
\begin{align}
    \mathbf{a}_1&=(1,0,0),& \mathbf{a}_2&=(0,1,0), &\mathbf{a}_3&=(0,0,\frac{1}{\sqrt{2}}), 
\end{align}
which yields the following reciprocal vectors 
\begin{align}
    \mathbf{q}_1&=(2\pi,0,0),& \mathbf{q}_2&=(0,2\pi,0),&  \mathbf{q}_3&=(0,0,2\sqrt{2}\pi).
\end{align}
The BZ of the lattice is shown in Fig.~\ref{fig:ChSqOctagon_BZ_spectrum}(a) together with the lines along which we will plot the energy spectrum. 

The chiral square-octagon has a fourfold screw rotation around $\hat z$, i.e., along the square spirals, as well as a twofold rotation around $\hat y$ and several twofold screw rotations. 
Note, however, that despite the various rotation symmetries, the bonds constituting the square spirals (blue/purple corresponding to $x$ and $y$ bonds) are not symmetry-related to those of the zig-zag chains connecting the spirals (pink/yellow corresponding to $z$ and $w$ bonds).
Thus, we can and will allow for different coupling strengths for the corresponding Ising interactions on those bonds. 

\begin{figure}[t]
    \centering
    \includegraphics[width=\linewidth]{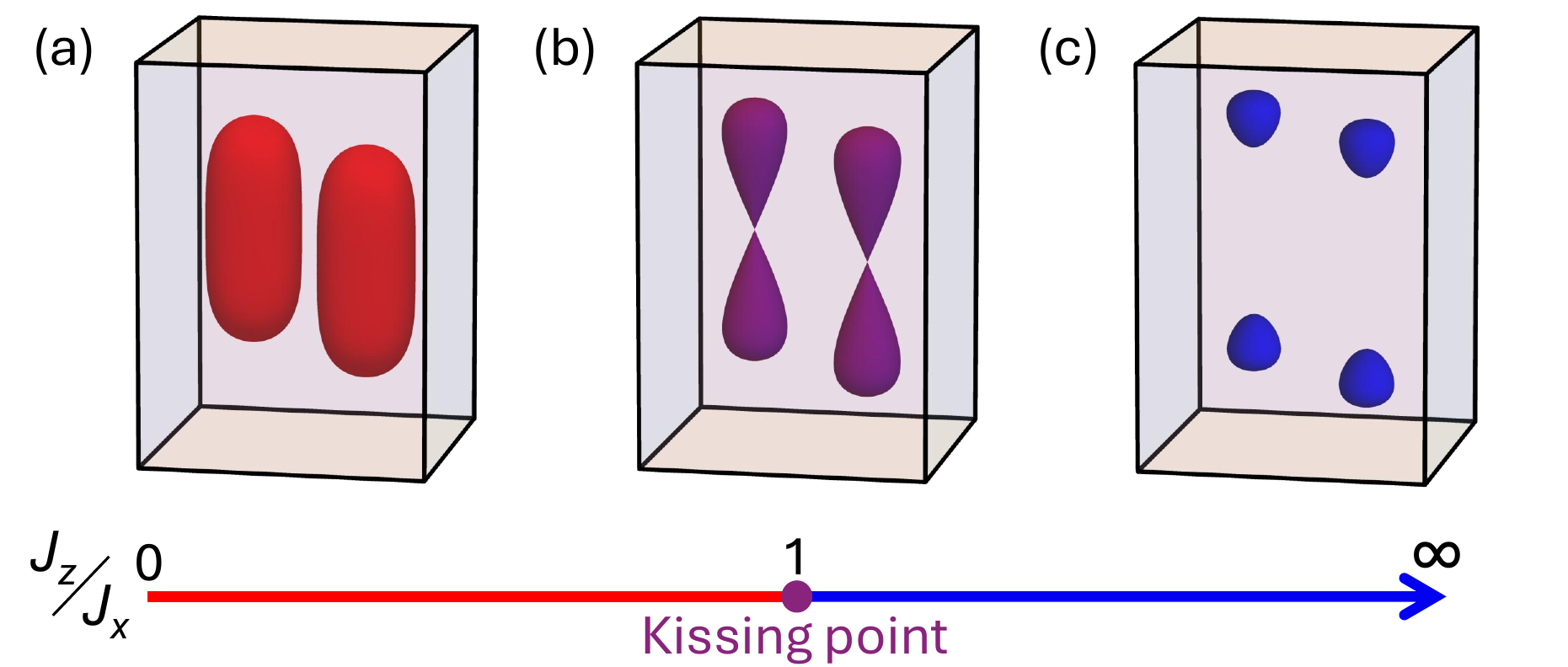}
    \caption{FSs for the chiral square-octagon lattice with different coupling ratios $f=J_z/J_x$ (with $J_y=J_x$ and $J_w=J_z$), (a) $f=0.7$, (b) $f=1$ and (c) $f=1.2$. For $|f|<1$ the model exhibits two FSs (red), for $|f|>1$, the model has 4 FSs (blue), and at $|f|=1$, the FSs touch (purple). }
    \label{fig:ChSqOctagon_BZ_Fermisurfaces}
\end{figure}
\begin{figure*}[t]
	\centering
	\includegraphics[width=0.9\linewidth]{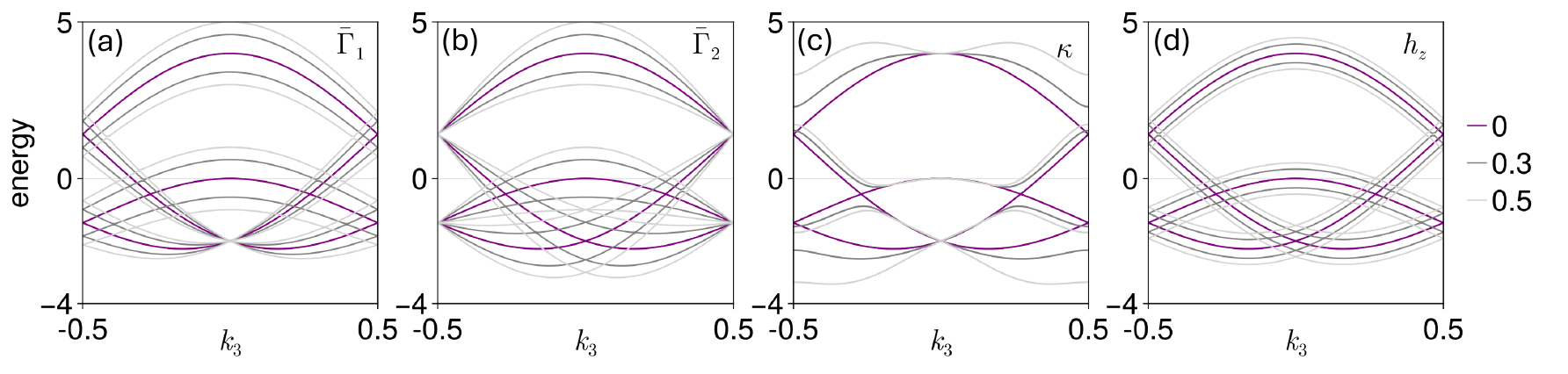}
	\caption{Energy spectrum along the $(-1/4,-1/4,k_3)$ line for the chiral square-octagon lattice with $f=1$, for different perturbations.}
	\label{fig:ChSqOctagon_spectrum_perturbations}
\end{figure*}
\begin{figure}[b]
    \centering
    \includegraphics[width=\linewidth]{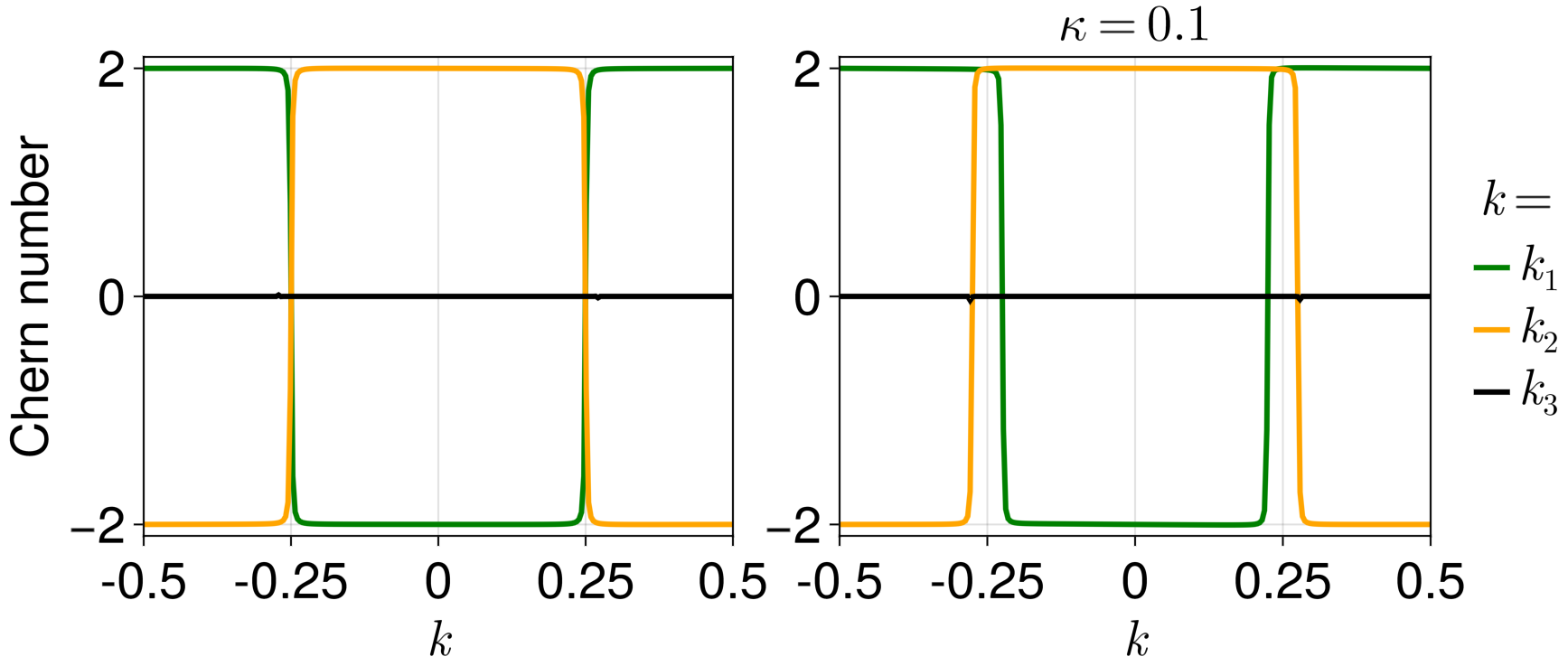}
    \caption{(left) Chern number for the TR invariant SOL  on the chiral square-octagon lattice. Since the WPs are fixed to the lines $k_1=k_2=\pm 1/4$, the Chern number jumps by 4 at $k=1/4$, independent of the actual parameters.  (right) Chern number when adding a small NNN perturbation, $\kappa=0.1$, to the unperturbed SOL with $f=1$. The WPs have moved away from the lines $k_1=k_2=\pm 1/4$. 
	}
    \label{fig:ChSqOctagon_BZ_chern}
\end{figure}

\subsubsection{Unperturbed spin-orbital liquid}
The chiral square-octagon lattice has 8 fundamental loops of length 6 per unit cell---though not all independent---and 1 of length 8, see Appendix~\ref{subsec:Loops_volume_constraints} for further details and pictures. 
While Lieb's theorem cannot be applied to this lattice,  our numerical simulations for small flux unit cells suggest that the  0 flux configuration gives the lowest ground state energy. 
Choosing a gauge configuration that ensures eigenvalues $+1$ for all the Wilson loops of length 6, see Appendix~\ref{subsubsec:ChiralSquareOctagon_appendix_H},  each of the two Majorana flavors of the unperturbed SOL is described by the Hamiltonian 
\begin{align}
    \begin{aligned}
        M(\bm{k})= i
        \begin{pmatrix}
            &0 &-J_x &A_{23} &J_y A_3  \\
            &J_x &0 &-J_y &A_{13} \\
            &-A_{23}^* &J_y &0 &-J_x \\
            &-J_y A_3^* &-A_{13}^* &J_x &0
        \end{pmatrix}
    \end{aligned}
\end{align}
where $A_{23}=-e^{2\pi ik_2}(J_z e^{-2\pi i k_3}+J_w)$, $A_{13}=e^{-2\pi ik_1}(J_w e^{-2\pi i k_3}+J_z)$, and $A_3=e^{-2\pi i k_3}$.
Rotation symmetry constrains the model to obey $J_x=J_y$ and $ J_z=J_w$, with only their ratio $f=J_z/J_x$ as a remaining tuning parameter. 
If not stated otherwise, this constraint on coupling constants is assumed in the following.

Since lattice translations map A sublattice sites to B sublattice sites, and vice versa, TR symmetry is implemented projectively with $\mathbf{k}_0=(\mathbf{q}_1 + \mathbf{q}_2)/2$. 
This implies that the stable zero-energy modes live on FSs. 
Indeed, for any value of $f$,  the system exhibits topological FSs: 2 for $|f|<1$ and 4 for $|f|>1$, with touching FSs at $|f|=1$, see Fig.~\ref{fig:ChSqOctagon_BZ_Fermisurfaces} for a visualization of the FSs for positive $f$. 
Even though the surfaces become very small for large $f$, they cannot disappear since they are protected by the enclosed WPs, located at $(k_1,k_2,k_3)=(\frac{1}{4},\frac{1}{4},\pm k_3) $ (both charge $+2$) and $(-\frac{1}{4},-\frac{1}{4},\pm k_3)$ (both charge $-2$),  where the value of $k_3$ depends on $f$. 
The energy spectrum along the line $k_1=k_2 =-\frac{1}{4}$, alternatively $(k_x,k_y,k_z)=(-\pi/2,-\pi/2,k_z)$, is shown in Fig.~\ref{fig:ChSqOctagon_BZ_spectrum}(b). 
For $f=1.2$ (blue line), the special line cuts through two FSs surrounding the WPs at $k_1=k_2=-\frac{1}{4}$, $k_3 \approx \pm 0.31$. 
For $f=0.7$ (pink line), the WP have moved closer together to $k_1=k_2=-\frac{1}{4}$, $k_3\approx \pm 0.16$, with a common FS surrounding both. 
For $f=1$ (purple line), the band touching $E=0$ at $k_3=0$ marks the kissing point in Fig.~\ref{fig:ChSqOctagon_BZ_Fermisurfaces}.  
The presence of the WPs can also be shown by the Chern number plot in Fig.~\ref{fig:ChSqOctagon_BZ_chern}. 
Since the WP are fixed to the two lines with $(k_1,k_2)=\pm \frac{1}{4}(1,1)$, the Chern number plot looks identical for any value of $f\neq 0$, though for very small $|f|$, numerics can become problematic due to the small band gaps.

\subsubsection{Effect of perturbations}
\begin{figure}[b]
    \centering
    \includegraphics[width=\linewidth]{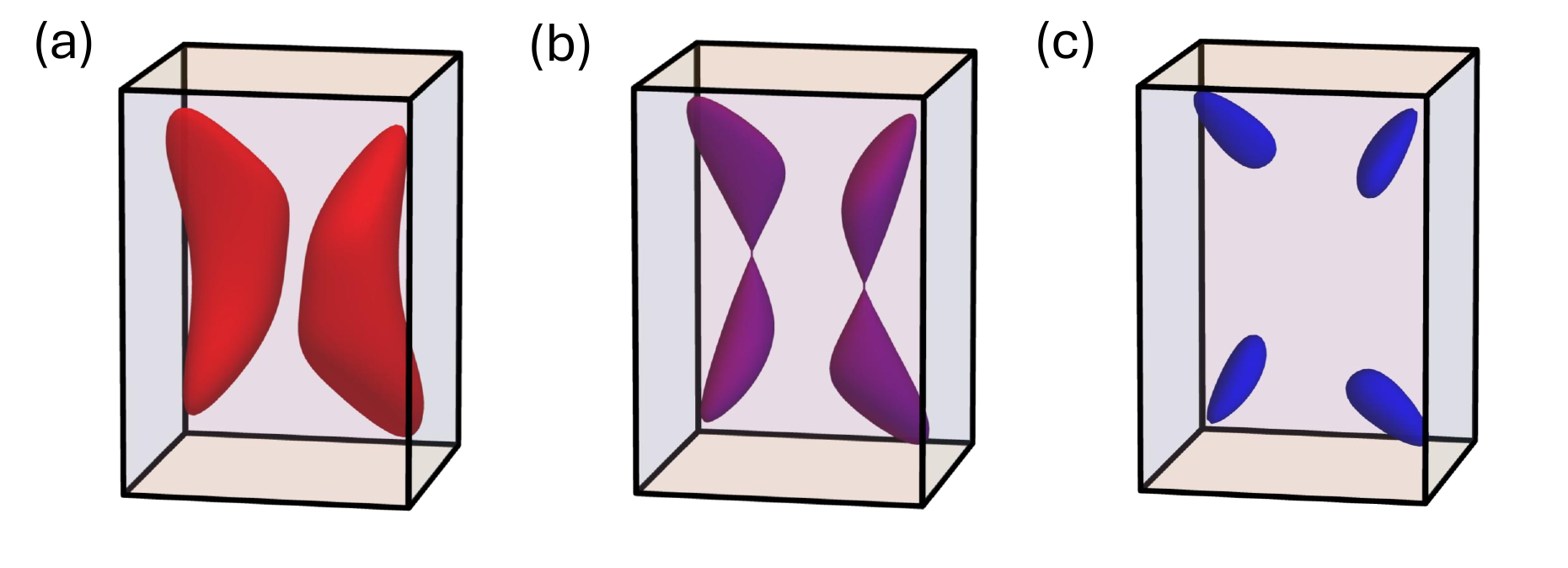}
    \caption{FSs for the chiral square-octagon lattice with $\kappa=0.2$, for (a) $f=0.7$, (b) $f=1$, and (c) $f=1.2$.}
    \label{fig:ChSqOctagon_FS_perturbations}
\end{figure}
We now proceed with discussing the effect of perturbations, starting with the NN TR-preserving perturbations. 
The chiral square-octagon has two symmetry-inequivalent bonds. 
We denote the perturbation strength of the nearest-neighbor perturbation \eqref{eq:fourcoord_GammaBar} as $\bar \Gamma_1$ for the $x$ and $y$-bonds, while the one on the $z$ and $w$-bonds is denoted by $\bar \Gamma_2$.

\begin{figure*}[t]
	\centering
	\includegraphics[width=0.9\linewidth]{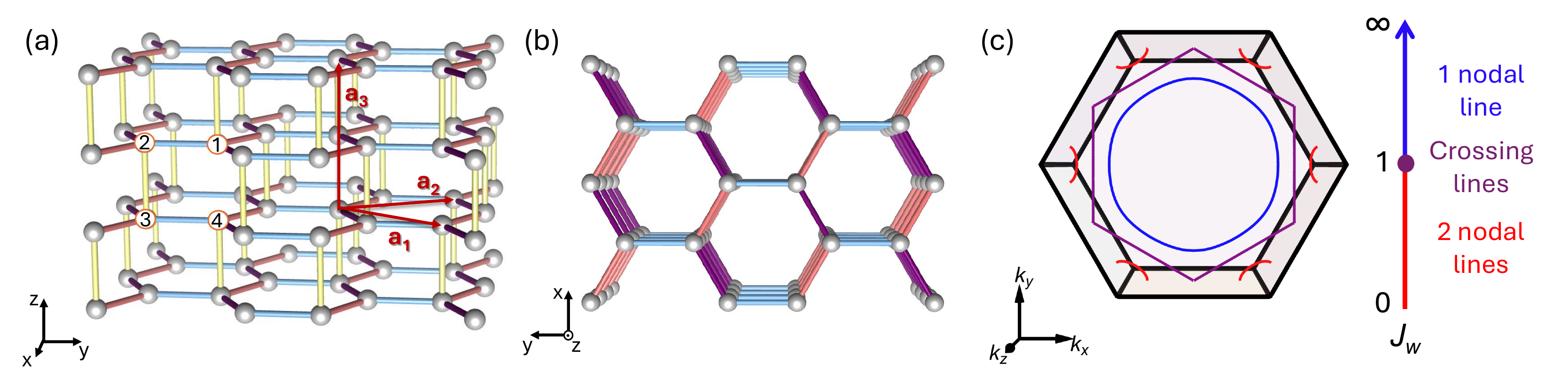}
	\caption{(a)-(b) Layered honeycomb lattice. Colors blue/purple/pink/yellow denote $x$/$y$/$z$/$w$ bonds. (c) Brillouin zone for the layered honeycomb lattice with nodal lines for different interlayer couplings $J_w$ ($J_x=J_y=J_z=1$), $J_w=0.5$ (red), $J_w=1$ (purple), and $J_w=1.5$ (blue). For $J_w<1$ there are 2 nodal lines, at $J_w=1$ the lines cross, and for $J_w>1$ there is one nodal line. The nodal lines lie in the $k_z=0$ plane.}
	\label{fig:LayHoney}
\end{figure*}

Due to the projective implementation of TR symmetry---relating momenta $\mathbf{k}$ to momenta $-\mathbf{k}+\mathbf{k}_0$ with $\mathbf{k}_0=(\mathbf{q}_1+\mathbf{q}_2)/2$---the FSs obey a perfect nesting condition---translation by $\mathbf{k}_0$ maps the FSs onto each other, clearly visible in Fig.~\ref{fig:ChSqOctagon_BZ_Fermisurfaces}. 
In addition, TR  pins the enclosed WPs to the two TR-invariant lines $\pm (1/4,1/4,k_3)$, as visualized in Fig.~\ref{fig:ChSqOctagon_spectrum_perturbations}(a) and (b), which show the energy spectrum along $k_1=k_2=-1/4$. 
For all perturbation strengths, one can detect band crossings, which show the position of the WPs on this line. 
Because the positive and negative WPs are confined to disconnected lines, generic TR-preserving perturbations cannot bring them together to annihilate.
Thus, the system harbors topological FSs for any of the symmetry-allowed, quadratic, TR-preserving perturbations.

The on-site perturbation $h_z$---despite breaking TR---cannot remove the WPs from the special lines either.
Instead, it splits the bands, thus enlarging some of the topological FSs while shrinking others, yielding nested topological FSs. 
The splitting of the bands is evident from the energy dispersion in Fig.~\ref{fig:ChSqOctagon_spectrum_perturbations}(d) and the shifted $E=0$ crossings indicated the respective sizes of the nested FSs. 

In contrast, the NNN TR perturbation breaks the perfect nesting condition and allows the WP to move away from the special lines. 
Consequently, Fig.~\ref{fig:ChSqOctagon_spectrum_perturbations} (c) does not show any band crossings.  
This opens up, in principle, the possibility of removing the topological charge of the FSs---thus turning them into trivial ones---and eventually gapping the system. 
In practice, however, the topological FSs persist even for sizeable 
$\kappa$ \footnote{In our numerical simulations, we considered values up to $\kappa=5$.}, although their detailed geometry is strongly deformed, as shown in Fig.~\ref{fig:ChSqOctagon_FS_perturbations} for three different choices of $f$.
In conclusion, none of the symmetry-allowed perturbations we consider here is able to drive the system into a different QSL regime.

\subsection{Layered honeycomb}\label{subsec:LayeredHoneycomb}
\subsubsection{The lattice and symmetries}

The layered honeycomb lattice is easiest visualized as AA stacked honeycomb layers (i.e., all sites of the layers sit exactly above one another), where the A  sites of a given layer connect to the A sites of the layer above, whereas the B sites connect to the B sites below. 
The resulting lattice is again bipartite. 
The most important symmetries of this lattice include threefold rotation around the z-axis, inversion symmetry (with inversion centers in the center of the hexagonal plaquettes as well as the middle of the $x$, $y$, or $z$ bonds), and mirror planes in between the honeycomb layers. The layered honeycomb lattice has a four-site unit cell. We choose our conventions such that all bonds have the same length, which yields site positions
\begin{align}
    \mathbf{r}_1&=(0,0,0), &\mathbf{r}_2&=(0,1,0),\nonumber\\
    \mathbf{r}_3&=(0,1,1),& \mathbf{r}_4&=(0,0,1), 
\end{align}
and lattice translation vectors
\begin{align}
    \mathbf{a}_1&=\left(\frac{\sqrt{3}}2,\frac{{3}}2,0\right),&
    \mathbf{a}_2&=\left(-\frac{\sqrt{3}}2,\frac{{3}}2,0\right),\nonumber\\
    \mathbf{a}_3&=(0,0,2).
\end{align}
The corresponding reciprocal lattice vectors are given by 
\begin{align}
    \mathbf{q}_1&=2\pi\left(\frac{1}{\sqrt{3}},\frac{1}{3},0\right),& \mathbf{q}_2&=2\pi\left(-\frac{1}{\sqrt{3}},\frac{1}{3},0\right),\nonumber\\
    \mathbf{q}_3&=\pi(0,0,1).
\end{align}
Figure~\ref{fig:LayHoney} illustrates the lattice, as well of the most natural choice of assigning Ising couplings to the different bonds.
The layered honeycomb lattice has some similarities to another four-coordinated lattice, the diamond lattice, which is discussed in Appendix~\ref{subsec:deformedLayeredHoneycomb}.
    
\subsubsection{Unperturbed spin-orbital liquid}
The layered honeycomb has 8 fundamental loops of length 6 per unit cell. 
These are not all independent, as we have three volume constraints per unit cell. The fundamental loops as well as the constraints are visualized in Appendix~\ref{subsubsec:Loops_LayeredHoneycomb}. 
While the mirror plane between the honeycomb layers ensures that 6 of the 8 loops carry 0-flux (according to Lieb's theorem), there is no mirror plane that constrains the flux of the original honeycomb plaquettes \footnote{The mirror planes of the honeycomb lattice become glide-mirrors for the layered honeycomb. The latter are of no use in Lieb's theorem.  }.
However, our numerical studies strongly suggest that the ground state resides in the 0-flux sector of these plaquettes as well. 
In the following, we will fix a gauge, see Appendix~\ref{subsec:explicitHamiltonians}, that yields 0-flux for all the 6-loops. 

For an unperturbed SOL on the layered honeycomb lattice, the Hamiltonian for a single Majorana flavor can then be written as 
\begin{align}
    \begin{aligned}
        M(\bm{k})=i
        \begin{pmatrix}
            &0 &A &0 &J_w e^{2\pi i k_3}  \\
            &-A^* &0 &-J_w &0 \\
            &0 &J_w &0 &A^* \\
            &-J_w e^{-2\pi i k_3} &0 &- A &0
        \end{pmatrix}
    \end{aligned}
\end{align}
    where $A = J_x + J_ye^{2\pi i k_1} + J_ze^{2\pi i k_2}$. 
    The combination of TR and sublattice symmetry of the Hamiltonian then allows us to rewrite it in terms of off-diagonal blocks
    \begin{align}
    \tilde{M}(\bm{k})&=
            \begin{pmatrix}
                &0 &Q \\
                &Q^\dagger &0\\
            \end{pmatrix},&
        Q (\bm{k})&= i
        \begin{pmatrix}
            &A &J_w e^{2\pi i k_3} \\
            &J_w &A^*
        \end{pmatrix}.
    \end{align}
    The gapless modes can be found by taking $\det{Q}=0$. For $J_w\neq0$, the unperturbed model exhibits nodal lines that lie in the $k_3=0$ plane and are described by
\begin{align}\label{eq:layHoney_zeromodes}
    \begin{aligned}
	 J_w^2 -|A|^2=0. \\
    \end{aligned}
\end{align}

If we fix the intralayer couplings to 1 and study the behavior of the nodal lines while varying $J_w$, we find three different behaviors. 
When $J_w<1$, the unperturbed SOL exhibits 2 nodal lines, while $J_w>1$ exhibits a single nodal line, and for $J_w=1$ it has crossing lines. Figure~\ref{fig:LayHoney}(c) shows the nodal lines in the Brillouin zone for $J_w=0.5, 1, 1.5$. 
If $J_w$ goes to zero, we have essentially returned to the original Kitaev model, as the layers are no longer coupled. 
The two nodal lines then contract into what look like two gapless points in the $k_3=0$ plane, but are actually nodal lines perpendicular to the $k_3=0$ plane. 
If we instead vary $J_x$ and set $J_y=J_z=J_w=1$, we get a similar result as for the $J_w$ case. 
There are two nodal lines when the coupling is smaller than 1, and a single nodal line when the coupling is larger than 1. 
This also holds for the case where you vary $J_y$ or $J_z$, as they are related to $J_x$ by threefold rotation symmetry.

The unperturbed SOL model exhibits twofold degenerate nodal lines, since we have two itinerant Majorana flavors. 
Each of the nodal lines carries a winding number with $|W|=1$, yielding a total winding number of $|W|=2$, see Fig.~\ref{fig:LayHoney_pert_windingnumbers}(d). 
Intimately connected with a non-vanishing winding number are topologically protected zero-energy surface states---called drumhead states---when imposing open boundary conditions.

\subsubsection{Effect of perturbations}
    \begin{figure}[t]
        \centering
        \includegraphics[width=\linewidth]{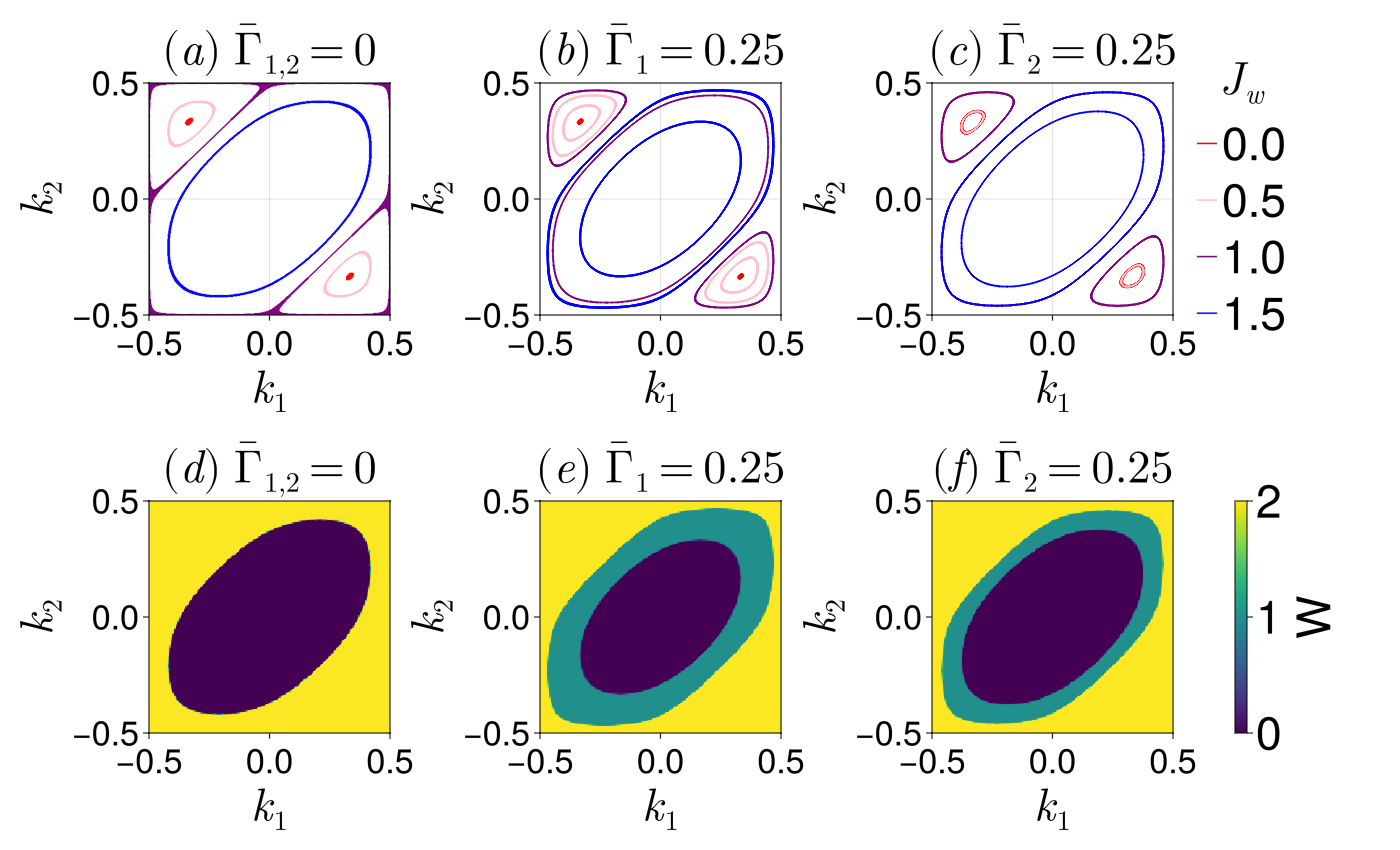}
        \caption{Upper row shows the nodal lines of a SOL on the layered honeycomb lattice for different interlayer couplings $J_w$, where (a) is unperturbed, (b) $\bar{\Gamma}_1=0.25$, and (c) $\bar{\Gamma}_2=0.25$. Lower row shows the winding numbers, W, of a SOL on the layered honeycomb lattice, with $J_w=1.5$, where (d) is unperturbed, (e) $\bar{\Gamma}_1=0.25$, and (f) $\bar{\Gamma}_2=0.25$. }
        \label{fig:LayHoney_pert_windingnumbers}
    \end{figure} 

The layered honeycomb lattice distinguishes intralayer $(x,y,z)$ bonds from the interlayer $w$ bonds. 
Accordingly, the symmetry-allowed nearest-neighbor perturbations reduce to two independent couplings: $\bar\Gamma_1$ on the $(x,y,z)$ bonds and $\bar\Gamma_2$ on the $w$ bond. 
Either perturbation breaks the twofold flavor degeneracy and generically splits each nodal line into two, while the nodal structure remains constrained to the $k_3=0$ plane as long as mirror symmetry (with the mirror plane in-between the honeycomb layers, visualized in Fig. ~\ref{fig:LayHoney_variation}) is preserved.
Figures~\ref{fig:LayHoney_pert_windingnumbers}(b) and (c) show the split nodal lines as a function of $(k_1,k_2)$, while (e) and (f) visualize the corresponding winding numbers in the presence of NN perturbations.

The fate of the nodal lines under TR breaking depends crucially on spatial symmetries. 
One might expect that breaking time reversal immediately converts nodal lines into point nodes; however, for the layered honeycomb, the nodal lines persist under generic TR-breaking perturbations so long as the mirror symmetry between layers
remains intact. 
In this case, the nodal lines no longer carry a winding number, and the associated drumhead surface states disappear.

Since inversion symmetry (in combination with particle-hole) enforces particle-hole symmetry for \emph{fixed} $\mathbf{k}$---meaning that each state at energy $E(\mathbf{k})$ has an inversion partner at $-E(\mathbf{k})$---there are now several scenarios when breaking additional lattice symmetries. 
If mirror and TR symmetry are broken simultaneously, the nodal lines are gapped to pairs of WPs. 
These lie at $E=0$ as long as inversion is preserved, and the system shows a symmetry-protected Weyl spin liquid phase. 
If all three symmetries---TR, mirror, and inversion---are broken, the  WPs move away from $E=0$, and the system harbors topological FSs instead. 
On the other hand, breaking inversion alongside TR symmetry, but keeping the mirror symmetry intact,  will replace the nodal lines with tubular FSs. 
In contrast to the situation for the SOL on the hyperhoneycomb lattice, these are not symmetry-protected to be doubly-degenerate. 
For a closer discussion on the mirror/inversion symmetry of the layered honeycomb lattice, see Appendix~\ref{subsec:broken_mirrorLayeredHoneycomb}.

	\section{Discussion}\label{sec:Conclusions}

A central feature of the Kitaev spin liquid is the fractionalization of microscopic spins into two emergent degrees of freedom: itinerant Majorana fermions and a static $\mathbb{Z}_2$ gauge field. 
While the gauge field is generically gapped, the Majorana sector can host robust gapless excitations, giving rise to a Majorana metal ground state.
In Kitaev’s original honeycomb model, this gapless phase consists of TR-protected Dirac points, whereas in three dimensions, the structure of gapless phases becomes substantially richer, allowing for topological FSs, symmetry-protected nodal lines, and WPs. 

In this work, we studied SOLs, which extend the Kitaev construction by hosting multiple itinerant Majorana flavors. 
Unlike conventional KSLs, SOLs can be defined on both three- and four-coordinated lattices, leading to qualitatively different classes of Majorana metals. 
On three-coordinated lattices, the resulting band structures closely resemble multiple coupled copies of the 3D Kitaev spin liquid, with additional possibilities arising from flavor mixing. 
In contrast, four-coordinated lattices have no direct KSL analogue and introduce an entirely new setting in which two-flavor Majorana bands give rise to distinct gapless structures.
We investigate how these differences manifest in the topology of the Majorana band structures across a set of representative three-dimensional lattices, identifying the nodal structures that arise robustly in each case and analyzing their evolution under symmetry-breaking perturbations.

Our analysis establishes a simple organizing principle for three-dimensional SOLs: once the ground-state flux sector is fixed, the low-energy physics is governed by free Majorana bands whose nodal manifolds---FSs, nodal lines, and WPs---are constrained by lattice geometry and symmetry, while the flavor structure controls degeneracies, splittings, and how the system can be gapped. 
For three-coordinated lattices, the unperturbed SOL reduces to three degenerate copies of the corresponding 3D Kitaev model, providing a direct bridge between the known classification of 3D KSLs and the richer phenomenology enabled by flavor mixing.
Interestingly, flavor-mixing terms offer multiple distinct pathways for lifting degeneracies and reshaping the nodal structure.

Turning to four-coordinated lattices ($\nu = 2$ Majorana flavors), we enter a setting that is largely unexplored beyond the original diamond-lattice example and closely related constructions \cite{ryu_three-dimensional_2009}.
These models are not naturally interpretable as multiple copies of a known 3D Kitaev spin liquid. 
Instead, they provide the minimal three-dimensional setting in which flavor is intrinsic and does not simply act as a degeneracy. 
Generic solvable perturbations naturally generate flavor anisotropies and flavor mixing, turning the classification problem into one of two-component Majorana band topology coupled to a static $\mathbb Z_2$ gauge field.
Four-coordinated lattices thus provide the simplest solvable 3D spin liquid in which topology and nodal structure are shaped by an intrinsically two-component Majorana sector, rather than by replication of a single-flavor Kitaev band.
By extending the analysis to four-coordinated lattices, our results therefore address a more basic question than which symmetry protects which node: which features of 3D Majorana metals are universal across solvable fractionalized phases? 
In this sense, the four-coordinated models serve as a complementary benchmark class for the broader SOL landscape in 3D.

A central assumption throughout is that we work in the ground-state flux sector appropriate to each lattice.
While we argue that this is appropriate in the vicinity of the unperturbed model, strong perturbations can stabilize alternative flux sectors (including flux-crystal patterns) and lead to even richer physics. 
Addressing this requires sophisticated numerical simulations and lies outside the scope of the present work.
Second, we have focused on solvable perturbations that preserve exact tractability in the Majorana description. 
This enables a controlled and systematic study of how symmetry-breaking terms reshape Majorana metals, but it does not address interaction-driven instabilities that may arise from generic non-solvable terms. 
The latter are expected to be particularly relevant near extended FSs, where residual interactions could in principle drive symmetry breaking or pairing-like instabilities in the Majorana sector \cite{hermanns2015spin}.
Finally, while three-dimensional Kitaev-type models admit a finite-temperature regime with dilute vison loops (and thus thermal stability absent in 2D), we do not attempt to map out finite-temperature phase diagrams here.

\section{Outlook}
Our work suggests several concrete directions for future research.
 A systematic numerical study could determine when flavor mixing and symmetry breaking favor alternative flux patterns, and whether new flux-crystal phases emerge in 3D SOLs beyond the Lieb sector. 
Incorporating generic symmetry-allowed terms would allow one to test the stability of the Majorana metals found here---in particular Majorana FSs---against interaction effects, and to explore whether partially gapped descendant phases (e.g., nodal-line phases) arise generically.
 The symmetry protection mechanisms identified here (winding-protected nodal lines versus mirror-protected nodal lines, Weyl spin liquid phases, and tubular FSs) naturally motivate a systematic study of surface spectra and experimentally relevant probes such as dynamical structure factors, Raman response, and thermal transport, with special attention to how flavor splittings reorganize low-energy thresholds. 
Finally, extending the present ``representative lattices'' approach to other lattices ---e.g., some with higher coordination numbers---would clarify which nodal manifolds and symmetry-protection mechanisms are generic in 3D SOLs, and whether additional multi-flavor phenomena—beyond those accessible for $\nu=2,3$ appear naturally.

In summary, by combining exact solvability with genuinely three-dimensional lattice geometry, we have mapped out a broad set of stable Majorana metals in 3D spin–orbital liquids and clarified how symmetry and flavor mixing organize their topology and perturbative evolution. 
This provides a flexible starting point for exploring more realistic settings where exact solvability is lost, while retaining a clear diagnostic language---based on nodal manifolds, symmetry protection, and flavor structure---for classifying emergent fractionalized metals in three dimensions.

{\bf Acknowledgements---}
The authors acknowledge helpful discussions with Lukas K\"onig. 
This project was partially funded by the Knut and Alice Wallenberg Foundation as part of the Wallenberg Academy Fellows
project and the Swedish Research Council under project number 2025-04091. 
Figures for most of the 3D lattices were made using the VESTA software~\cite{momma_vesta_2011}, while the figures displaying Chern numbers, winding numbers and  energy spectra utilize the Makie package~\cite{DanischKrumbiegel2021}.

\section{Appendix}

\subsection{Gauge-fixing and explicit Hamiltonians}\label{subsec:explicitHamiltonians}
This appendix includes the gauge choices for each of the lattices, as well as the explicit sub-matrices for the Hamiltonians with perturbations. As a reminder (from Eq. \eqref{eq:perturbedH}), the perturbed momentum Hamiltonian for the three-coordinated lattices can be written in terms of $d\times d$ matrices $M_{\alpha\beta}$ as 
\begin{align}
	\begin{aligned}
		H(\mathbf{k})=
		\begin{pmatrix}
			&M_{xx}(\mathbf{k}) &M_{xy}(\mathbf{k}) &M_{xz}(\mathbf{k}) \\
			&M_{xy}^\dagger(\mathbf{k}) &M_{yy}(\mathbf{k}) &M_{yz}(\mathbf{k}) \\
			&M_{xz}^\dagger(\mathbf{k}) &M_{yz}^\dagger(\mathbf{k}) &M_{zz}(\mathbf{k})
		\end{pmatrix},
	\end{aligned}
\end{align}
where $d$ is the number of sites in the unit cell. $x$, $y$, and $z$ denote the three different flavors of Majoranas. Similarly, for four-coordinated lattices, the perturbed Hamiltonians can be written as
\begin{align}
	\begin{aligned}
		H(\mathbf{k})=
		\begin{pmatrix}
			&M_{xx}(\mathbf{k}) &M_{xy}(\mathbf{k}) \\
			&M_{xy}^\dagger(\mathbf{k}) &M_{yy}(\mathbf{k})  \\
		\end{pmatrix}.
	\end{aligned}
\end{align}
The matrices on the diagonal describe the unperturbed Majoranas, as well as perturbations that do not mix different Majorana flavors. The off-diagonal blocks describe perturbations that mix Majorana flavors.

\subsubsection{Hyperoctagon}\label{subsubsec:Hyperoctagon_appendix_H}
For the hyperoctagon, or (10,3)a, lattice, we use the same reference gauge as in \cite{obrien_classification_2016}, that is,
\begin{align}
    \begin{aligned}
        u_{12}^x&=-1, & u_{13}^y&=-1, & u_{32}^z&=+1, \\
        u_{34}^x&=-1, & u_{24}^y&=+1, & u_{14}^z&=-1.
    \end{aligned}
    \label{gauge_10_3a}
\end{align}
For the unperturbed Hamiltonian, the matrices on the diagonal, $M_{xx}(\mathbf{k})=M_{yy}(\mathbf{k})=M_{zz}(\mathbf{k})=M(\mathbf{k})$, take the form
\begin{align}
    \begin{aligned}
        M(\bm{k})=
        \begin{pmatrix}
            &0 &-iA_2 &-iJ_y&-i A_1 \\
            &iA_2^* &0 &-iJ_z &iJ_y  \\
            &iJ_y  &iJ_z  &0 &-iA_3 \\
            &i A_1^* &-iJ_y &iA_3^* &0
        \end{pmatrix},
    \end{aligned}
\end{align}
where $A_1=J_z e^{-2\pi i k_1}$,  $A_2=J_x e^{-2\pi i k_2}$, and $A_3=J_xe^{-2\pi i k_3}$. The $K$ perturbation does not mix any Majorana flavors, but rather shifts one of the couplings for each respective Majorana flavor. In $M_{xx}(\mathbf{k})$ it shifts the couplings on the $x$-bonds, $J_x\rightarrow J_x+K$. Similarly, in $M_{yy}(\mathbf{k})$ the $y$-bonds are shifted, $J_y\rightarrow J_y+K$, and in $M_{zz}(\mathbf{k})$ the $z$-bonds are shifted, $J_z\rightarrow J_z+K$. The $\Gamma$, $\Gamma'$, and $\mathbf{h}$ perturbations do, however, mix the Majorana flavors. The off-diagonal matrices take the form 
\begin{align}
    \begin{aligned}
        &M_{xy}(\bm{k})=\\
        &i \begin{pmatrix}
            &h_z &\Gamma'e^{-2\pi i k_2} &\Gamma'&\Gamma e^{-2\pi i k_1} \\
            &-\Gamma'e^{2\pi i k_2} &h_z &\Gamma &-\Gamma'  \\
            &-\Gamma'  &-\Gamma  &h_z &\Gamma'e^{-2\pi i k_3} \\
            &-\Gamma e^{2\pi i k_1} &\Gamma' &-\Gamma'e^{2\pi i k_3} &h_z
        \end{pmatrix},\\
                &M_{xz}(\bm{k})=\\
        &i \begin{pmatrix}\\
            &-h_y &\Gamma'e^{-2\pi i k_2} &\Gamma &\Gamma'e^{-2\pi i k_1} \\
            &-\Gamma'e^{2\pi i k_2} &-h_y &\Gamma' &-\Gamma  \\
            &-\Gamma  &-\Gamma'  &-h_y &\Gamma'e^{-2\pi i k_3} \\
            &-\Gamma'e^{2\pi i k_1} &\Gamma &-\Gamma'e^{2\pi i k_3} &-h_y
        \end{pmatrix},\\
        &M_{yz}(\bm{k})=\\
        &i \begin{pmatrix}
            &h_x &\Gamma e^{-2\pi i k_2} &\Gamma'&\Gamma'e^{-2\pi i k_1} \\
            &-\Gamma e^{2\pi i k_2} &h_x &\Gamma' &-\Gamma'  \\
            &-\Gamma'  &-\Gamma'  &h_x &\Gamma e^{-2\pi i k_3} \\
            &-\Gamma'e^{2\pi i k_1} &\Gamma' &-\Gamma e^{2\pi i k_3} &h_x
        \end{pmatrix}.
    \end{aligned}
\end{align}
The next-nearest neighbor perturbation $\kappa$ appears in the matrices on the diagonal as $M_{xx}^\kappa(\bm{k})=M_{yy}^\kappa(\bm{k})=M_{zz}^\kappa(\bm{k})= M^\kappa(\bm{k})$, where 
\begin{align}
    \begin{aligned}
        M^{\kappa}(\mathbf{k})&=i\kappa
        \begin{pmatrix}
            &0 &A &B &C \\
            &-A^* &0 &C^* &D \\
            &-B^* &-C &0 &A \\
            &-C^* &-D^* &-A^* &0
        \end{pmatrix},
    \end{aligned}
\end{align}
with $A=-1- e^{-2\pi i k_1}$, $B=- e^{-2\pi i k_2}- e^{2\pi i(k_3-k_1)}$, $C=- e^{-2\pi i k_2}- e^{-2\pi i k_3}$, and $D= e^{2\pi i (k_2-k_1)}+ e^{-2\pi i k_3}$.

\subsubsection{Hyperhoneycomb}\label{subsubsec:Hyperhoneycomb_appendix_H}
For the hyperhoneycomb, or (10,3)b, lattice, we use the same reference gauge as in \cite{obrien_classification_2016}, that is,
\begin{align}
    \begin{aligned}
        u_{23}^x&=+1, & u_{14}^y&=+1, & u_{13}^z&=+1, \\
        u_{14}^x&=+1, & u_{23}^y&=+1, & u_{24}^z&=+1.
    \end{aligned}
    \label{gauge_10_3b}
\end{align}
For the unperturbed Hamiltonian, the matrices on the diagonal, $M_{xx}(\mathbf{k})=M_{yy}(\mathbf{k})=M_{zz}(\mathbf{k})=M(\mathbf{k})$, then take the form
\begin{align}
    \begin{aligned}
        M(\bm{k})=
        \begin{pmatrix}
            &0 &0 &iJ_z &iA_{13} \\
            &0 &0 &iA_{2} &iJ_z \\
            &-iJ_z &-iA_{2}^* &0 &0 \\
            &-iA_{13}^* &-iJ_z &0 &0
        \end{pmatrix},
    \end{aligned}
\end{align}
where $A_{13}=e^{-2\pi ik_3}(J_x+e^{2\pi ik_1}J_y)$, $A_{2}=J_x+J_y e^{2\pi i k_2}$. The $K$ perturbation simply shifts one of the couplings, which coupling depends on the Majorana flavor, i.e., for $M_{xx}(\mathbf{k}): J_x\rightarrow J_x+K$, $M_{yy}(\mathbf{k}):J_y\rightarrow J_y+K$, and $M_{zz}(\mathbf{k}):J_z\rightarrow J_z+K$. The $\Gamma$, $\Gamma'$, and $\mathbf{h}$ perturbations, on the other hand, mix the Majorana flavors. The off-diagonal matrices take the form
\begin{align} 
\begin{aligned}
    &M_{xy}(\mathbf{k}) = \\
&i \begin{pmatrix}
h_z &0 &-\Gamma &-\Gamma'(\alpha+\beta_3) \\
0 &h_z &-\Gamma'(1+\beta_2) &-\Gamma \\
\Gamma &\Gamma'(1+\beta_2^*) &h_z &0 \\
\Gamma'(\alpha^*+\beta_3^*) &\Gamma &0 &h_z
\end{pmatrix},\\
&M_{xz}(\mathbf{k}) = \\
&i \begin{pmatrix}
-h_y &0 &-\Gamma' & -\Gamma \alpha-\Gamma'\beta_3 \\
0 &-h_y &-\Gamma \beta_2-\Gamma' &-\Gamma' \\
\Gamma' &\Gamma \beta_2^*+\Gamma' &-h_y &0 \\
\Gamma \alpha^*+\Gamma' \beta_3^* &\Gamma' &0 &-h_y
\end{pmatrix}, \\
&M_{yz}(\mathbf{k}) = \\
&i \begin{pmatrix}
h_x &0 &-\Gamma' & -\Gamma \beta_3-\Gamma'\alpha \\
0 &h_x &-\Gamma-\Gamma'\beta_2 &-\Gamma' \\
\Gamma' &\Gamma+\Gamma'\beta_2^* &h_x &0 \\
\Gamma \beta_3^*+\Gamma'\alpha^* &\Gamma' &0 &h_x
\end{pmatrix},
\end{aligned}
\end{align}

where $\alpha=e^{2\pi i(k_1 - k_3)}$, $\beta_3=e^{-2\pi i k_3}$ and $\beta_2=e^{2\pi i k_2}$. 

The next-nearest neighbor perturbation $\kappa$ appears in the matrices on the diagonal as $M_{xx}^\kappa(\bm{k})=M_{yy}^\kappa(\bm{k})=M_{zz}^\kappa(\bm{k})= M^{\kappa}(\bm{k})$, where 
\begin{align}
    \begin{aligned}
        M^\kappa(\mathbf{k})=i\kappa
        \begin{pmatrix}
            &A_{1} &B &0 &0 \\
            &-B^* &A_{2} &0 &0 \\
            &0 &0 &-A_{2} &-B \\
            &0 &0 &B^* &-A_{1}
        \end{pmatrix},
    \end{aligned}
\end{align}
with 
 $A_{1}=2i\sin(2\pi k_1)$, $A_{2}=2i\sin(2\pi k_2)$, and $B=-1+e^{-2\pi i k_2}+e^{-2\pi i k_3}-e^{2\pi i (k_1-k_3)}$.

\subsubsection{Hyperhexagon}\label{subsubsec:Hyperhexagon_appendix_H}
For the hyperhexagon, or (8,3)b, lattice, we use the same reference gauge as in \cite{obrien_classification_2016}, that is,
\begin{align}
    \begin{aligned}
        u^{x}_{43}&=+1, & u^y_{42}&=+1,&  u^z_{14}&=+1, \\
        u^x_{21}&=+1,& u^y_{53}&=+1,& u^z_{25}&=+1, \\
        u^x_{56}&=+1,& u^y_{61}&=+1,& u^z_{36}&=+1.
    \end{aligned}
    \label{gauge_8_3b}
\end{align}
For the unperturbed Hamiltonian, the matrices on the diagonal, $M_{xx}(\mathbf{k})=M_{yy}(\mathbf{k})=M_{zz}(\mathbf{k})=M(\mathbf{k})$, then take the form
\begin{align}
    \begin{aligned}
        M(\bm{k})= i
        \begin{pmatrix}
            &0 &-A_3 &0 &J_z &0 &-A_{13} \\
            &A_3^* &0 &0 &- J_y  &J_z  &0 \\
            &0 &0 &0 &-A_2 &-J_y &J_z  \\
            &-J_z  &J_y  &A_2^* &0 &0 &0 \\
            &0 &-J_z &J_y  &0 &0 &A_3 \\
            &A_{13}^* &0 &-J_z  &0 &-A_3^* &0
        \end{pmatrix},
    \end{aligned}
\end{align}
where $A_{3}=J_{x} e^{-2\pi i k_{3}}$, $A_{13}=J_{y} e^{-2\pi i(k_1+k_3) }$, $A_{2}=J_{x} e^{2\pi i k_2}$. 
The $K$ perturbation shifts one of the couplings, which coupling depends on the Majorana flavor, i.e., for $M_{xx}(\mathbf{k}): J_x\rightarrow J_x+K$, $M_{yy}(\mathbf{k}):J_y\rightarrow J_y+K$, and $M_{zz}(\mathbf{k}):J_z\rightarrow J_z+K$. The Majorana flavor-mixing perturbations, $\Gamma$, $\Gamma'$, and $\mathbf{h}$ are given by the off-diagonal matrices, 

\begin{align}
    \begin{aligned}
        &M_{xy}(\bm{k})=\\
        &i \begin{pmatrix}
        h_z & \Gamma'\alpha_3 & 0 & -\Gamma & 0 & \Gamma'\alpha_{13} \\
        -\Gamma'\alpha_3^* & h_z & 0 & \Gamma' & -\Gamma & 0 \\
        0 & 0 & h_z & \Gamma'\alpha_2 & \Gamma' & -\Gamma \\
        \Gamma & -\Gamma' & -\Gamma'\alpha_2^* & h_z & 0 & 0 \\
        0 & \Gamma & -\Gamma' & 0 & h_z & -\Gamma'\alpha_3 \\
        -\Gamma'\alpha_{13}^* & 0 & \Gamma & 0 & \Gamma'\alpha_3^* & h_z \\
        \end{pmatrix},\\
        &M_{xz}(\bm{k})=\\
        &i \begin{pmatrix}
        -h_y & \Gamma'\alpha_3 & 0 & -\Gamma' & 0 & \Gamma \alpha_{13} \\
        -\Gamma'\alpha_3^* & -h_y & 0 & \Gamma & -\Gamma' & 0 \\
        0 & 0 & -h_y & \Gamma' \alpha_2 & \Gamma & -\Gamma' \\
        \Gamma' & -\Gamma & -\Gamma' \alpha_2^* & -h_y & 0 & 0 \\
        0 & \Gamma' & -\Gamma & 0 & -h_y & -\Gamma' \alpha_3 \\
        -\Gamma \alpha_{13}^* & 0 & \Gamma' & 0 & \Gamma'\alpha_3^* & -h_y \\
        \end{pmatrix},\\
        &M_{yz}(\bm{k})=\\
        &i \begin{pmatrix}
        h_x & \Gamma \alpha_3 & 0 & -\Gamma' & 0 & \Gamma' \alpha_{13} \\
        -\Gamma \alpha_3^* & h_x & 0 & \Gamma' & -\Gamma' & 0 \\
        0 & 0 & h_x & \Gamma \alpha_2 & \Gamma' & -\Gamma' \\
        \Gamma' & -\Gamma' & -\Gamma \alpha_2^* & h_x & 0 & 0 \\
        0 & \Gamma' & -\Gamma' & 0 & h_x & -\Gamma \alpha_3 \\
        -\Gamma' \alpha_{13}^* & 0 & \Gamma' & 0 & \Gamma \alpha_3^* & h_x \\
        \end{pmatrix},
    \end{aligned}
\end{align}
where $\alpha_{13}=e^{-2\pi i (k_1 + k_3)}$, $\alpha_2=e^{2\pi i k_2}$, and $\alpha_3=e^{-2\pi i k_3}$. The next-nearest neighbor perturbation $\kappa$ shows up in matrices on the diagonal as $M_{xx}^\kappa(\bm{k})=M_{yy}^\kappa(\bm{k})=M_{zz}^\kappa(\bm{k})= M^\kappa(\bm{k})$, where 
\begin{align}
    \begin{aligned}
        &M^\kappa(\mathbf{k})=i\kappa\\
        &\begin{pmatrix}
            &0 &-1 &A &B &C &0 \\
            &1 &0 &D &-B^* &0 &C \\
            &-A^* &-D^* &0 &0 &-B^* &B \\
            &-B^* &B &0 &0 &-D &-A \\
            &-C^* &0 &B &D^* &0 &1 \\
            &0 &-C^* &-B^* &A^* &-1 &0
        \end{pmatrix},
    \end{aligned}
\end{align}
with $A=e^{-2\pi i k_2}+e^{-2\pi i (k_1+k_3)}$, $B=e^{-2\pi ik_3}$, $C=-e^{-2\pi ik_1}+e^{-2\pi i k_3}$, and $D=e^{-2\pi i k_2}-1$.

 \subsubsection{Chiral square-octagon}\label{subsubsec:ChiralSquareOctagon_appendix_H}
 The chiral square-octagon lattice has eight loop operators of length 6, see Appendix~\ref{subsubsec:Loops_ChiralSquareOctagon}. Lieb's theorem does not apply to the chiral square-octagon lattice, as it lacks mirror planes. We will, however, still assume that the ground state is given by a zero-flux configuration (loop operators have eigenvalue +1). We need a convention for the bonds that ensures that all loops of length six have positive value, while also being translationally invariant (i.e., we do not wish to enlarge the unit cell). 
The current convention on the bonds is the following: for blue, purple, and yellow bonds ($J_x/J_y/J_w$), the bond operator is positive in the direction of increased $z$ component; for the pink bonds ($J_z$), on the other hand, the opposite convention is chosen. In other words, the reference gauge is
\begin{align}
\begin{aligned}
    u_{21}^x=+1, \quad u_{43}^x=+1, \quad u_{32}^y=+1, \quad u_{14}^y=+1, \\
    u_{42}^z=-1, \quad u_{13}^z=-1, \quad u_{31}^w=+1, \quad u_{24}^w=+1.
    \end{aligned}
    \label{eq:chiral_square_octagon_gauge}
\end{align}
For the unperturbed Hamiltonian, the matrices on the diagonal, $M_{xx}(\mathbf{k})=M_{yy}(\mathbf{k})=M(\mathbf{k})$, then take the form
\begin{align}
    \begin{aligned}
        M(\bm{k})= i
        \begin{pmatrix}
            &0 &-J_x &A_{23} &J_y A_3  \\
            &J_x &0 &-J_y &A_{13} \\
            &-A_{23}^* &J_y &0 &-J_x \\
            &-J_y A_3^* &-A_{13}^* &J_x &0
        \end{pmatrix},
    \end{aligned}
\end{align}
where $A_{23}=-e^{2\pi ik_2}(J_z e^{-2\pi i k_3}+J_w)$, $A_{13}=e^{-2\pi ik_1}(J_w e^{-2\pi i k_3}+J_z)$, and $A_3=e^{-2\pi i k_3}$. The $\Bar{\Gamma}_1$ and $\bar{\Gamma}_2$ perturbations mix the Majorana flavors on the $x$/$y$-bonds and $z$/$w$-bonds, respectively. $h_z$ mixes the two Majorana flavors on a site. This is described by the following matrix 
\begin{align}
    \begin{aligned}
        M_{xy}(\bm{k})= i
        \begin{pmatrix}
        h_z & \Bar{\Gamma}_1 & -\Bar{\Gamma}_2 \alpha_{23} & -\Bar{\Gamma}_1 \alpha_3 \\
        -\Bar{\Gamma}_1 & h_z & \Bar{\Gamma}_1 & -\Bar{\Gamma}_2 \alpha_{13} \\
        \Bar{\Gamma}_2 \alpha_{23}^* & -\Bar{\Gamma}_1 & h_z & \Bar{\Gamma}_1 \\
        \Bar{\Gamma}_1\alpha_3^* & \Bar{\Gamma}_2 \alpha_{13}^* & -\Bar{\Gamma}_1 & h_z
    \end{pmatrix},
    \end{aligned}
\end{align}
where $\alpha_{23}=-e^{2\pi ik_2}(e^{-2\pi i k_3}+1)$, $\alpha_{13}=e^{-2\pi ik_1}(e^{-2\pi i k_3}+1)$, and $\alpha_3=e^{-2\pi i k_3}$.
The next-nearest neighbor perturbation $\kappa$ shows up in the matrices on the diagonal, where $M_{xx}^{NNN}(\bm{k})=M_{yy}^{NNN}(\bm{k})= M^\kappa(\bm{k})$, and
\begin{align}
    \begin{aligned}
        M^\kappa(\mathbf{k})=i\kappa
        \begin{pmatrix}
            B &A_{12}A_3 &A_{3} &- CA_3 \\
            -A_{12}^*A_{3}^* &-B &- CA_3 &-A_{3} \\
            -A_{3}^* & C^*A_3^* &-B &A_{12}^*A_{3} \\
             C^*A_3^* &A_{3}^* &-A_{12}A_{3}^* &B
        \end{pmatrix}
    \end{aligned},
\end{align}
 where $B=2i\sin{2\pi k_3}$,  $A_{12}=(e^{2\pi i k_1}-e^{2\pi i k_2})$, $A_{3}= 1-e^{-2\pi i k_3}$, $C=(e^{2\pi i k_2}-e^{-2\pi i k_1})$.

\subsubsection{Layered honeycomb}\label{subsubsec:LayeredHoneycomb_appendix_H}
The layered honeycomb lattice has 8 loop operators of length 6, see Appendix~\ref{subsubsec:Loops_LayeredHoneycomb}. 
Thus,  Lieb's theorem suggests that  the ground state has flux zero. We choose the gauge such that going from the A to the B sublattice is positive (where A=site 2,4 and B=site 1,3),
\begin{align}
    \begin{aligned}
        &u^x_{12}=+1, \quad u^y_{12}=+1, \quad u^z_{12}=+1, \quad u^w_{32}=+1, \\
        &u^x_{34}=+1, \quad u^y_{34}=+1, \quad u^z_{34}=+1, \quad u^w_{14}=+1.
    \end{aligned}
    \label{eq:gauge_layHoney}
\end{align}
This ensures that all loops have a positive eigenvalue, i.e., the flux is zero. The unperturbed Hamiltonian, i.e., the matrices on the diagonal, $M_{xx}(\mathbf{k})=M_{yy}(\mathbf{k})=M(\mathbf{k})$, then take the form
\begin{align}
    \begin{aligned}
        M(\bm{k})=i
        \begin{pmatrix}
            &0 &A &0 &J_w e^{2\pi i k_3}  \\
            &-A^* &0 &-J_w &0 \\
            &0 &J_w &0 &A^* \\
            &-J_w e^{-2\pi i k_3} &0 &- A &0
        \end{pmatrix},
    \end{aligned}
\end{align}
with $A = J_x + J_ye^{2\pi i k_1} + J_ze^{2\pi i k_2}$. The $\Bar{\Gamma}_1$ and $\bar{\Gamma}_2$ perturbations mix the Majorana flavors on the $x$/$y$/$z$-bonds and $w$-bonds, respectively. $h_z$ mixes the two Majorana flavors on a site. These perturbations are described by the matrix 
\begin{align}
    \begin{aligned}
        M_{xy}(\bm{k})=
        i \begin{pmatrix}
            &h_z &-\Bar{\Gamma}_1\alpha &0 &-\Bar{\Gamma}_2e^{2\pi i k_3}  \\
            &\Bar{\Gamma}_1\alpha^* &h_z &\Bar{\Gamma}_2 &0 \\
            &0 &-\Bar{\Gamma}_2&h_z &-\Bar{\Gamma}_1\alpha^* \\
            &\Bar{\Gamma}_2e^{-2\pi i k_3} &0 &\Bar{\Gamma}_1\alpha &h_z
        \end{pmatrix},
    \end{aligned}
\end{align}
where $\alpha = 1 + e^{2\pi i k_1} + e^{2\pi i k_2}$. 

\begin{figure*}[t]
	\centering
	\includegraphics[width=\linewidth]{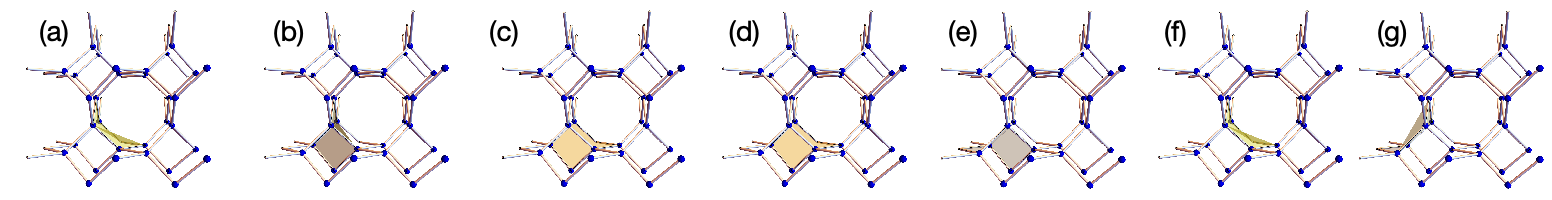}
	\caption{
		Visualizations of the volume constraints in the chiral square–octagon lattice.
		(a)-(c) showing one of the three-loop constraints and (d)-(g) showing one of the four-loop constraints. The other constraints of the same type are related by a 4-fold screw rotation.
	}
	\label{fig:chiral_sq_oct_vol_combined}
\end{figure*}

The next-nearest neighbor perturbation $\kappa$ shows up in the matrices on the diagonal, 
where $M_{xx}^{NNN}(\bm{k})=M_{yy}^{NNN}(\bm{k})=M^\kappa(\bm{k})$, and
\begin{align}
    \begin{aligned}
        M^\kappa(\mathbf{k})=i\kappa
        \begin{pmatrix}
            &A &0 &-BC &0\\
            &0 &-A &0 &B^*C \\
            &B^*C^* &0 &-A &0 \\
            &0 &-BC^* &0 & A
        \end{pmatrix},
    \end{aligned}
\end{align}
where $A=2i  (\sin{2\pi k_1}- \sin{2\pi k_2}- \sin{2\pi(k_1-k_2)})$, $B=(1-e^{2\pi i k_1}+e^{2\pi i k_2})$, and $C=(1- e^{2\pi i k_3})$.

    \subsection{Fundamental loops and volume constraints}\label{subsec:Loops_volume_constraints}
    \subsubsection{Chiral square-octagon}\label{subsubsec:Loops_ChiralSquareOctagon}
\begin{figure}
    \centering
    \includegraphics[width=\linewidth]{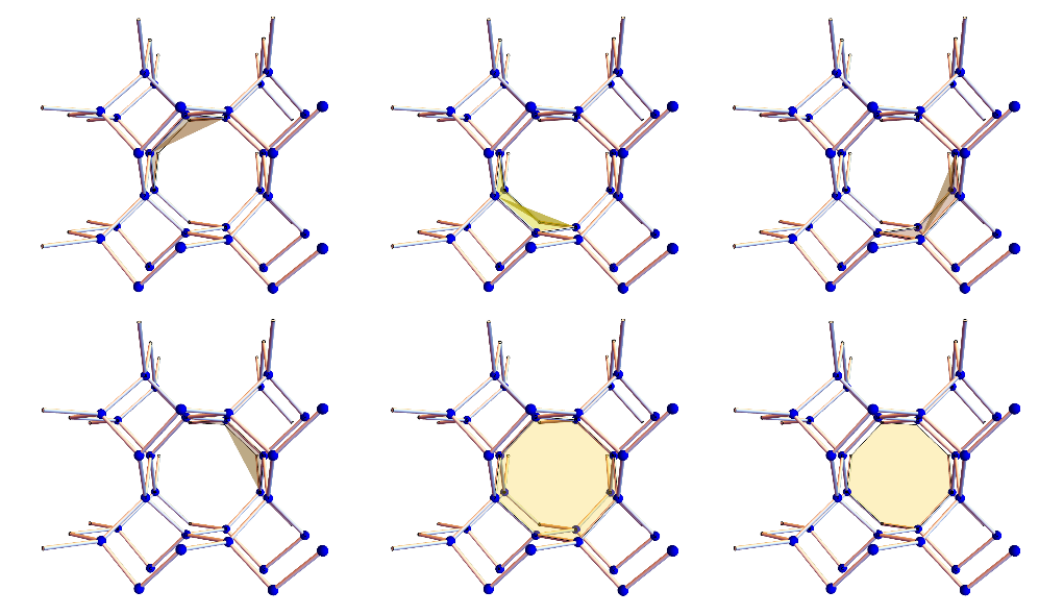}
    \caption{Visualization of the 6 loops constituting one of the volume constraints in the chiral square octagon lattice. }
    \label{fig:chiral_sq_oct_vol_2}
\end{figure}

The chiral square octagon has eight loops of length 6 and one loop of length 8 per unit cell. 
Four loops of length 6 can be visualized as transversing once around the square-spiral and moving back via the zig-zag chain to the starting point. 
Fig.~\ref{fig:chiral_sq_oct_vol_combined}(b), (c) and (e)  show three of these loops. 
Another four can be pictured as connecting two neighboring zig-zag chains; see, e.g., the upper panel of Fig.~\ref{fig:chiral_sq_oct_vol_2}. 
These loops are related to each other by four three-loop constraints and two four-loop constraints. 
One example of each type is shown in Fig.~~\ref{fig:chiral_sq_oct_vol_combined}: (a)-(c) show a three-loop constraint, while (d)-(g) a four-loop one. 
The rest of the constraints can be obtained by a 4-fold screw rotation around $\hat z$.

The loop of length 8 transverses the octagon of the chiral square octagon lattice. 
Also this loop is not linearly independent of the first four, since they are related by the volume constraint depicted in Fig.~\ref{fig:chiral_sq_oct_vol_2}. 
In total, we find two independent loops per unit cell, as expected.

    \subsubsection{Layered honeycomb}\label{subsubsec:Loops_LayeredHoneycomb}
\begin{figure}
    \centering
    \includegraphics[width=\linewidth]{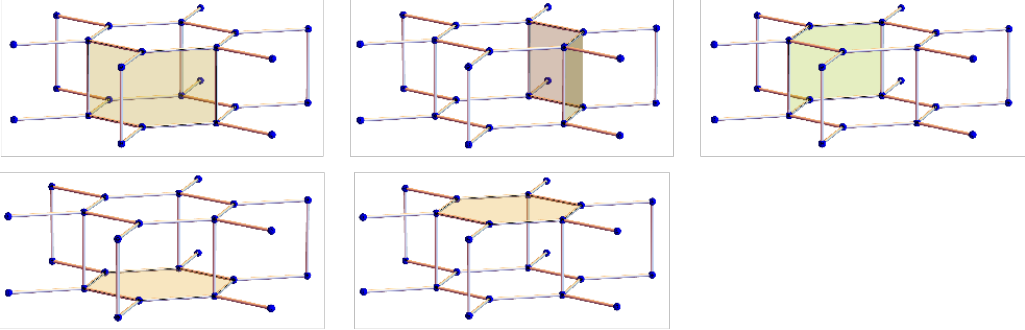}
    \caption{Visualization of the 5 loops constituting one of the volume constraints in the layered honeycomb lattice. }
    \label{fig:lay_hon_vol_1}
\end{figure}

\begin{figure}
    \centering
    \includegraphics[width=\linewidth]{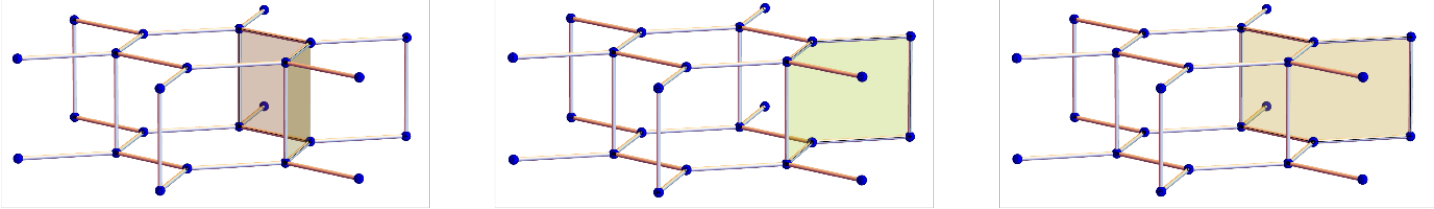}
    \caption{Visualization of the 3 loops constituting one of the volume constraints in the layered honeycomb lattice. }
    \label{fig:lay_hon_vol_2}
\end{figure}

The layered honeycomb lattice has 8 loops of length 6 per unit cell: two honeycomb plaquettes of the original honeycomb layer and six loops that connect the two layers, see Fig.~\ref{fig:lay_hon_vol_1} for three of these loops. The second three are obtained by a 6-fold glide rotation around the $\hat z$ axis. 
These loops are not linearly independent: there are two volume constraints of the type shown in Fig.~\ref{fig:lay_hon_vol_1} (the `top' of the first acts as the `bottom' of the second) and four of the type shown in Fig.~\ref{fig:lay_hon_vol_2}, one for each site of the unit cell. 
Thus, there are only 2 independent loops per unit, as expected. 

    \subsection{Deformed layered honeycomb}\label{subsec:deformedLayeredHoneycomb}
\begin{figure}[t]
        \centering
        \includegraphics[width=0.45\linewidth]{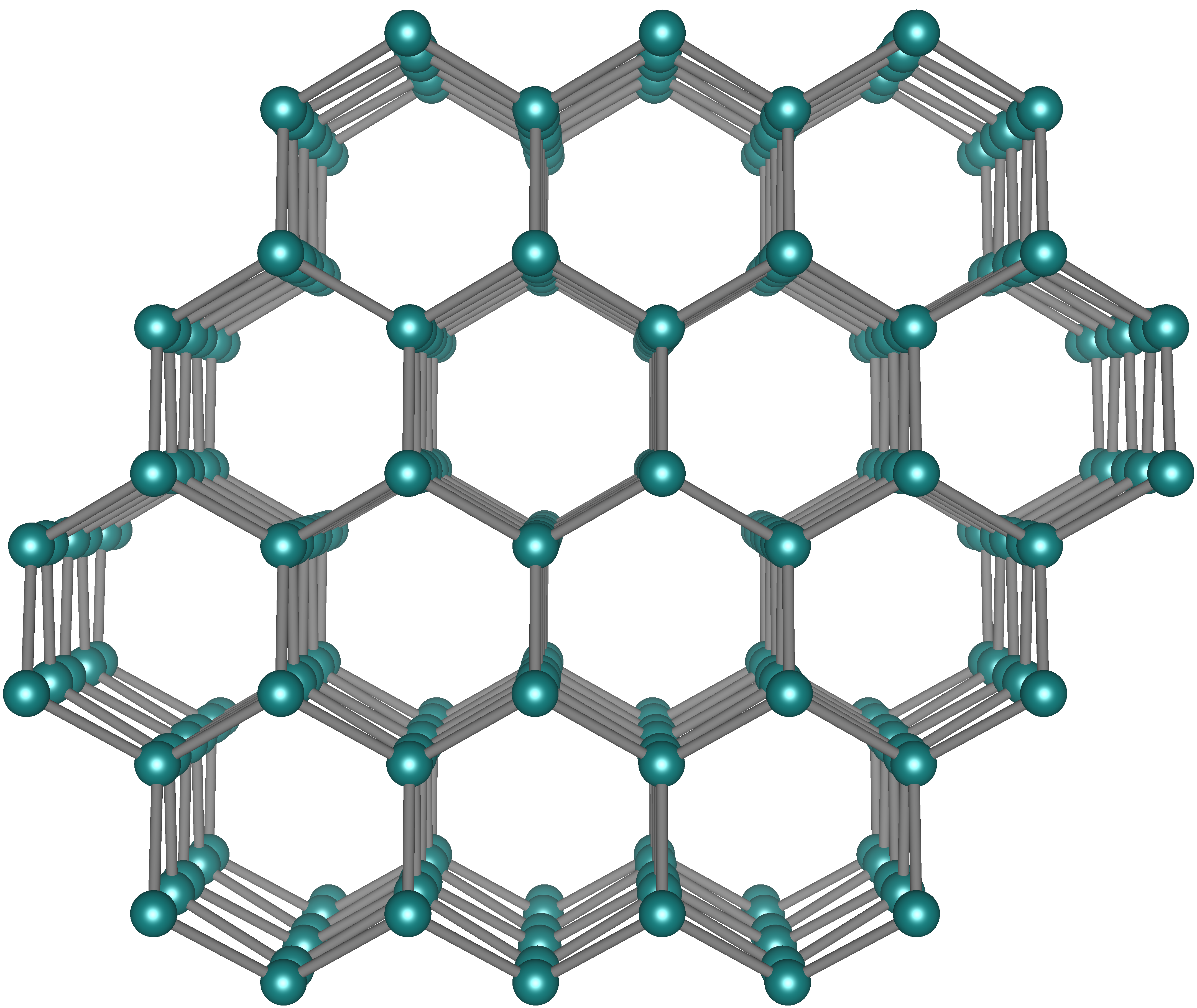}
        \includegraphics[width=0.45\linewidth]{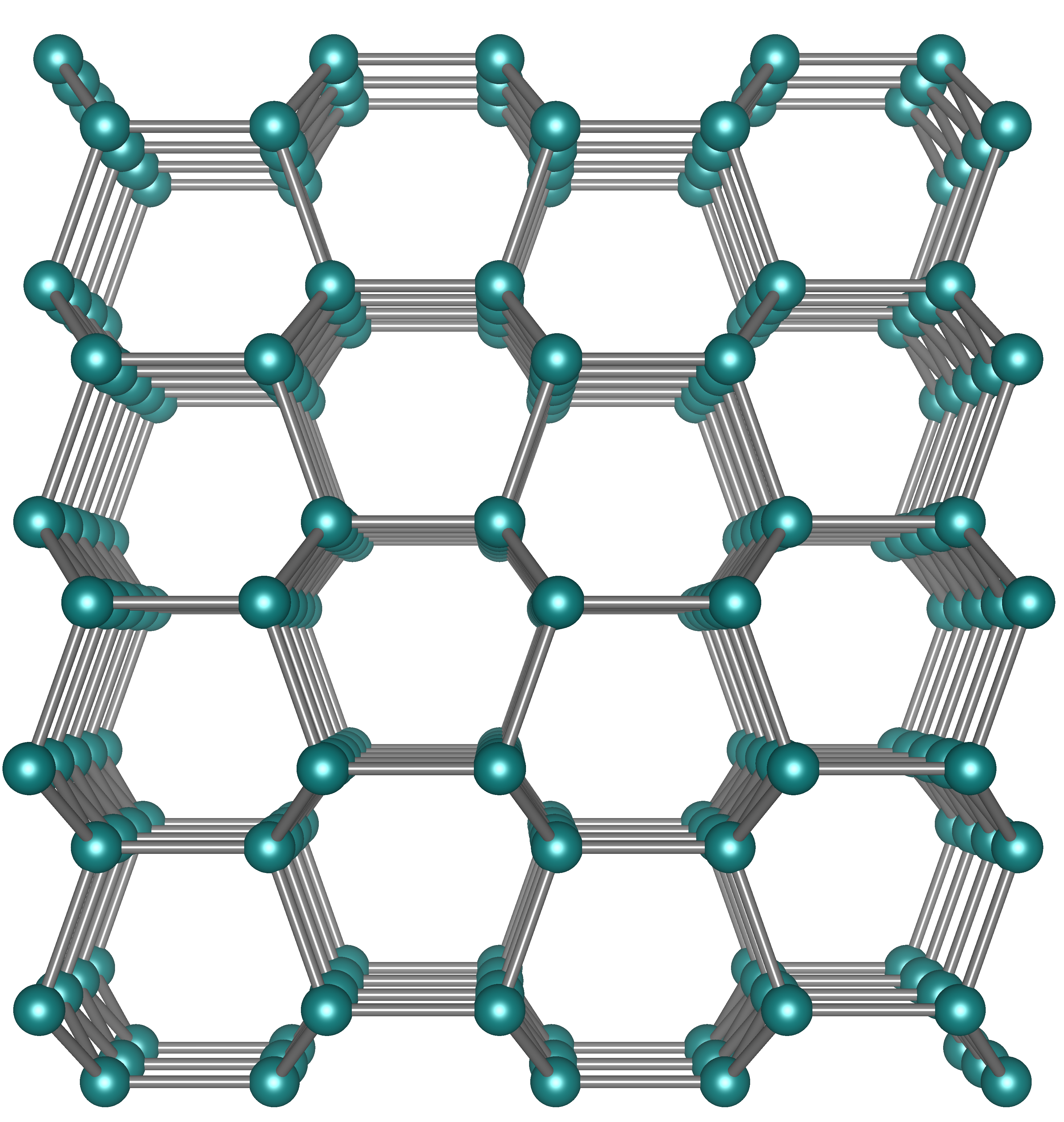}
        \caption{Deformed layered honeycomb lattice with equal bond angles of $\approx 109.47^\circ$. (left) view along the $z$-axis, (right) view along the ($xy$)-ladders. }
        \label{fig:LayHoney_diamond_angles}
    \end{figure}

Even though not obvious at first glance, the layered honeycomb lattice is locally similar to the diamond lattice---though the two lattices are distinct. 
While the layered honeycomb was obtained by connecting AA stacked honeycomb layers, the diamond can be thought of as connecting AB stacked layers, i.e., the A sites of a given layer sit above/below the B sites of the neighboring layers, while the B sites sit at the center of the hexagonal plaquettes of the neighboring layers. 
While the local motif---a site is surrounded by four others with equal bond length and equal angles---is the same, the connectivity differs. 
In particular, all bonds in the diamond lattice are related by symmetries, while for the layered honeycomb, only three of the four bonds are symmetry-related. 
In order to deform the layered honeycomb to achieve equal bond angles of $\arccos(-1/3) \approx109.47^\circ $, we choose site positions
\begin{align}
    \mathbf{r}_1&= \left(0,0,\frac{1}{\sqrt{8}}\right),& 
    \mathbf{r}_2&= \left({0,1,0}\right)\nonumber\\
    \mathbf{r}_3&= \left(0,1,\frac{5}{\sqrt{8}}\right),& 
    \mathbf{r}_4&= \left({0,0,\sqrt{2}}\right),
\end{align}
with slightly changed translation vectors (to keep all bonds at equal length)
\begin{align}
    \mathbf{a}_1 &= \left(\sqrt{3},3,0\right)/2,&
    \mathbf{a}_2 &= \left(-\sqrt{3},3,0\right)/2,\nonumber\\
    \mathbf{a}_3 &= \left(0,0,\sqrt{8}\right).
\end{align}
The resulting lattice has bond lengths of $\frac{3}{\sqrt{8}}$ and the same bond angles as the diamond lattice, see Figure~\ref{fig:LayHoney_diamond_angles}.

    \subsection{Layered honeycomb---broken mirror symmetry}
    \label{subsec:broken_mirrorLayeredHoneycomb}
    \begin{figure}[t]
        \centering
        \includegraphics[width=\linewidth]{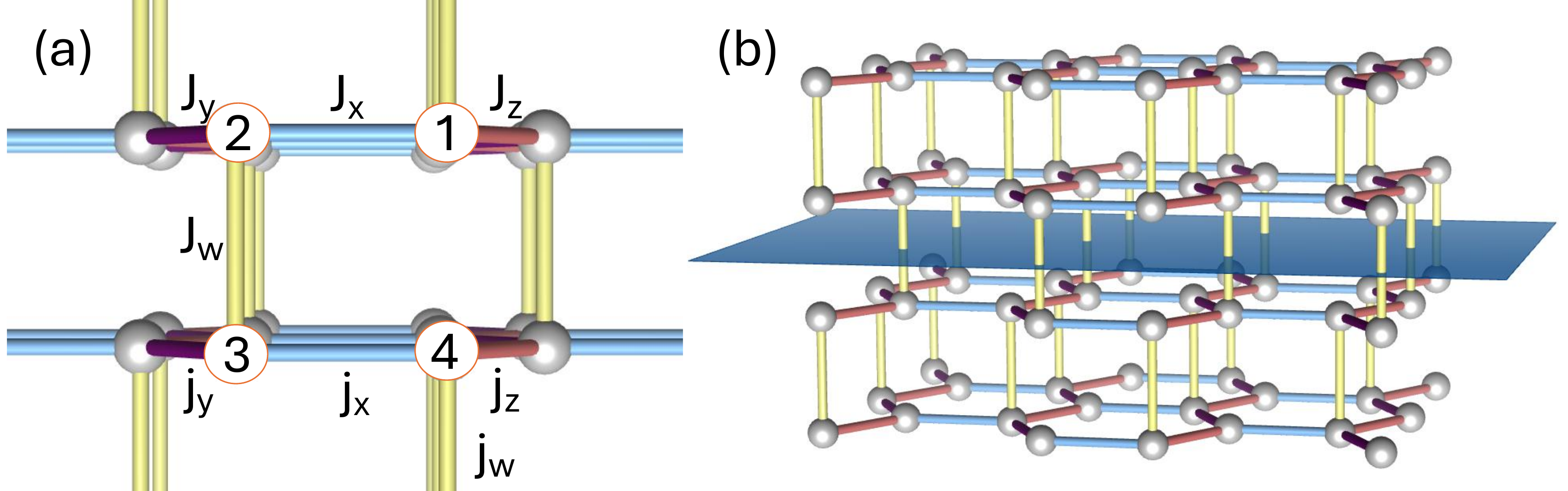}
        \caption{(a) Variation on the layered honeycomb lattice that breaks mirror/inversion symmetry and (b) the mirror plane between the layers of the layered honeycomb lattice.}
        \label{fig:LayHoney_variation}
    \end{figure}
    The nodal lines of the layered honeycomb lattice are protected not only by TR symmetry, but also by mirror symmetry between the honeycomb layers, see Figure~\ref{fig:LayHoney_variation}(b). If we break TR symmetry, but not mirror/inversion symmetry, the nodal lines persist, but they no longer carry any winding numbers, and thus no surface drumhead states.  
    
    We may instead consider what happens if we break mirror and/or inversion symmetry, but not TR symmetry. 
    Mirror symmetry can be broken by allowing the coupling constants in the two honeycomb layers to be different, see Figure~\ref{fig:LayHoney_variation}(a).
    We denote the coupling in the layer of sites 1 and 2 as $J_\alpha$, and the coupling in the layer of sites 3 and 4 as $j_\alpha$. 
    Inversion symmetry can be broken by choosing the couplings between layers differently,  $j_w$ between sites 1 and 4, and $J_w$ between sites 2 and 3.

The nodal lines at zero energy will remain as long as TR symmetry is intact. However, TR symmetry does not require that the nodal lines lie in the $k_3=0$ plane. We will see that this restraint, in fact, comes from the mirror symmetry. The condition for the nodal lines to lie in the $k_3=0$ plane can be found by first taking the perturbed Hamiltonian into a block-diagonal form. This can be done by applying the transformation in Eq. \eqref{eq:block_trafo}, which is just a rotation in the Majorana basis: $(c_i^x, c_i^y)\rightarrow \frac{1}{\sqrt{2}}(c_i^x+c_i^y,c_i^x-c_i^y)$. Thereafter, by applying sublattice symmetry, we can bring one of these $4\times 4$ blocks into an off-block diagonal form, with the $2\times 2$ block,
\begin{align}
i\begin{pmatrix}
 A_{\tilde{J}} & e^{2  \pi i k_3} \tilde{j}_w \\
 \tilde{J}_w & A_{\tilde{j}} \\
\end{pmatrix},
\label{eq:LayHoney_mirrorsym_offdiagblock}
\end{align}
where $A_{\tilde{J}}=(\tilde{J}_x+e^{2  \pi i k_1} \tilde{J}_y+e^{2  \pi i k_2} \tilde{J}_z)$ and $A_{\tilde{j}}=(\tilde{j}_x+e^{-2  \pi i k_1} \tilde{j_y}+e^{-2  \pi i k_2} \tilde{j}_z)$. $\tilde{J}_\alpha=J_\alpha-\Gamma_1$, $\tilde{j}_\alpha=j_\alpha-\Gamma_1$, are the couplings for the bonds $\alpha=x,y,z$ in the layers, and $\tilde{J}_w=J_w-\Gamma_2$, $\tilde{j}_w=j_w-\Gamma_2$, are the couplings between the layers. 
In this basis, $\Gamma_1$ and $\Gamma_2$ just shift the coupling constants. 
Finally, the zero modes can be identified by requiring that both the imaginary and real parts of the determinant of this block (Eq. \eqref{eq:LayHoney_mirrorsym_offdiagblock}) are zero. We then find that the nodal lines lie in the $k_3=0$ plane if both layers are identical ($J_\alpha=j_\alpha$), that is, if we have mirror symmetry. 

\nocite{Data3DSOLs}
\bibliography{references}
\end{document}